\documentclass{article}
\usepackage{amsmath}
\usepackage{arxiv}
\usepackage[english]{babel}
\usepackage{bm}
\usepackage{float}
\usepackage{graphicx}
\usepackage[colorlinks=true, allcolors=blue]{hyperref}
\usepackage[normalem]{ulem}
\usepackage{subcaption}

\title{A meshing framework for digital twins for extrusion based additive manufacturing}
\author{
        Lucas~Gallup\thanks{Corresponding Author: \texttt{lkgallu@sandia.gov}}\\
        Materials and Failure Modeling\\
		Sandia National Laboratories\\
		Albuquerque, NM 87123 \\
        \And
        Kevin N.~Long\\
		Materials and Failure Modeling\\
		Sandia National Laboratories\\
		Albuquerque, NM 87123 \\
        \And
        Devin J.~Roach\\
        Department of Mechanical, Industrial\\ and Manufacturing Engineering\\
        Oregon State University\\
        Corvallis, OR, 97331\\
        \And
        William D.~Reinholtz\\
		Advanced Materials Lab\\
		Sandia National Laboratories\\
		Albuquerque, NM 87123 \\
        \And
        Adam Cook\\
		Advanced Materials Lab\\
		Sandia National Laboratories\\
		Albuquerque, NM 87123 \\
        \And
        Craig M.~Hamel\thanks{Corresponding Author: \texttt{chamel@sandia.gov}}\\
		Materials and Failure Modeling\\
		Sandia National Laboratories\\
		Albuquerque, NM 87123 \\
}
\renewcommand{\shorttitle}{A meshing framework for digital twins for extrusion based additive manufacturing}

\begin{document}
\maketitle

\begin{abstract}
Additive manufacturing (AM) allows for manufacturing of complex three-dimensional geometries not typically realizable with standard subtractive manufacturing practices. Additionally, the internal microstructure of a selected material of a 3D printed component can have a significant impact on its mechanical, vibrational, and shock properties and allow for a richer design space when this microstructure is controllable. Due to the complex interactions of the internal geometry of an extrusion-based AM component , it is common practice to assume a homogeneous behavior or to perform characterization testing on the specific toolpath configurations being used. To avoid unnecessary testing or material waste, it is necessary to develop an accurate and consistent numerical simulation framework with relevant boundary value problems that can handle the complicated geometry of internal material microstructure present in AM components. Herein, a framework is proposed to directly create computational meshes suitable for finite element analysis (FEA) of the fine-scale features of parts generated from extrusion-based AM tool paths to maintain a strong linkage between process control and fine geometric structure fidelity and component properties, or process-structure-property-performance (PSPP) linkage. This mesh can be manually or automatically analyzed using standard FEA simulations such as quasi-static preloading, modal analysis, or thermal analysis. The framework allows an in-silico assessment of a target AM geometry where fine-scale features may greatly impact quantities of design interest such as in soft elastomeric lattices where toolpath infill can greatly influence the self contact of a structure in compression, which we will use as a motivating exemplar. This approach greatly reduces the waste of both time and resources consumed through traditional build and test design cycles for non-intuitive design spaces. It also further allows for the exploration of toolpath infill to optimize component properties beyond simple linear properties such as density and stiffness.
\end{abstract}

\section{Introduction}

Additive manufacturing (AM) is an increasingly powerful tool that can be used to create complex three-dimensional geometries that are not typically realizable using standard subtractive manufacturing techniques. The library of materials that AM can leverage is growing daily with most classes of engineering materials such as ceramics, metals, plastics, etc. being covered by one or more AM process, ~\cite{BOURELL2017659, NGO2018172, LI2019242}. Different classes of materials can often require different AM processing techniques such as selective laser sintering (SLS) for metals or fused filament fabrication (FFF) for thermoplastics. However, there is overlap between classes of materials and processes where certain classes of materials can be manufactured with multiple AM processes such as FFF and inkjet printing for plastics. An additional advantage of AM is the ability to design and manufacture parts with locally tailorable material properties and microstructures. Despite these capabilities, it is not always efficient in terms of cost, material, or time to design, print, and test components. Numerical simulation provides a solution to this, allowing for designs to be tested and optimized in-silico before being implemented in reality. Deep understanding and insight of a process is necessary in order to leverage it for tailored microstructure design.

Typically, research into numerical simulation of AM focuses on the printing process itself or residual stress imparted upon a part during and directly after printing. For example, the spreading of powder print medium during printing was modeled with the discrete element method ~\cite{NAN2019801} and the printing parameters of resin manufactured with digital light processing, DLP, were studied to reduce shrinkage and warping caused by curing, ~\cite{ZHANG2021101403}. However, simulation of fully realized AM components after manufacturing is less common due to the computational expense and difficulty in obtaining geometric representations of microstructures for some processes, such as SLS. For example, a viscoplastic constitutive model which considers internal gas porosity and fusion voids in laser powder bed printing was fitted to test components to understand the impact of random inclusions on the failure and plasticity of printed stainless steel, ~\cite{Johnson_2019}. Crack initiation and propagation behaviour was well predicted, however varied validation experiments resulted in the sporadic force-displacement curves. The authors found that the porosity distribution caused considerate variation in the force-displacement curves, which may explain the randomness sometimes associated with AM components. Similarly, the effect of synthetic microstructures common in 3D printed metals were investigated for simple loading cases, ~\cite{Rodgers_2018}. The authors found a clear correlation between the local microstructures and the elastic stress response where the variation was not impacted by the grain size, especially in comparison to a homogenized model. Likewise, a study of advancements in numerical simulation of additively manufactured metal suggests an emphasis on fatigue analysis, printable materials, and lattice structures, ~\cite{Gandhi_2022}. The authors found that the numerical simulations from the studies were typically in agreement with the respective experiments, and developing models with the structure-process-fatigue properties will support the reliability of AM parts in the future. These authors investigated the impact of microstructures inherent in additive methods like SLS, however, for extrusion and light-based AM processes, such as FFF or DLP, void dispersity and material microstructure are often not as significant of a problem. So, we directly leverage the processing parameters (e.g. toolpath or photomask information) to generate "as-printed" geometries using standard computational geometry approaches such as computer aided design (CAD) or various meshing strategies, with the aim of accurately representing printed microstructures in numerical simulation.

Since studies predominantly center on simulating 3D printing metallic mediums, it is necessary to focus on other printing methods and materials, such as FFF and polymers. Some effort has already been made in this respect; experimental compression tests of the direct ink writing (DIW) method were conducted and the results fed to an artificial neural network (ANN) with the goal of determining compressive response and mechanical properties, ~\cite{ROACH2021101950}. The ANN was able to accurately predict the behavior of printed specimens that were not used in the training data set. In conjunction, the authors used a genetic algorithm and the results of the ANN to predict the printing parameters necessary to achieve a desired stress-strain curve in compression. With respect to the microstructural geometry of a printed component, a numerical optimization framework was proposed to generate smooth gradient internal geometry of varying microstructure cells to achieve a desired elastic response, ~\cite{SCHUMACHER2015}. The optimization was used to recreate the desired properties by tiling the predetermined cells, with the unique feature of interpolating the structures to ensure a smoother transition between property regions. Experimental validation showed excellent agreement between the predicted tangent modulus and elastic modulus for tensile and compression tests respectively. 

Process parameters of FFF and polymers have also seen a some degree of focus. Two inks were studied to predict their free-flow behavior when extruded through a syringe nozzle, ~\cite{Yongqiang2022}. The authors simulated the continuous flow which achieved a relative diameter error of less than 10.22\%, and were able to use the model to determine optimal piston speeds for stable flow of the two inks. Low density, functionally graded foams were printed using grayscale digital light projection (DLP) printing to understand the impact of light intensity on the mechanical behavior of the printed resin, and to compare these to a graded foam printed with variable light intensity values, ~\cite{MONTGOMERY2021101323}. A model was derived for several print cases using drop test experiments performed on the foams that accurately predicted their energy absorption. The authors concluded by showing that the graded foam was able to absorb energy efficiently over a wider range of impact conditions than a uniform foam. Laser sintered polyamide 12, a nylon powder printing medium, was experimentally tested in both isothermal and non-isothermal tensile tests, ~\cite{ROSZAK2021103893}. The temperature dependent material parameters of the proposed Chaboche, ~\cite{CHABOCHE1989247}, and Gurson-Tvergaard-Needleman, ~\cite{TVERGAARD1984157}, models were fitted to the experimental results using an optimization algorithm; a numerical simulation found that isothermal simulations resulted in excellent matching to the experimental results and the non-isothermal simulation modeled the experiment with less that 12\% deviation. To reduce the necessary time and computational power needed to perform high fidelity simulations of thermal distribution during the fused deposition modeling (FDM) printing process, a Gaussian process-constrained general path model was proposed to predict the results of high fidelity finite element analysis from low fidelity simulations, ~\cite{WANG2020101211}. The general path model was trained using two different training data set types; firstly by using one layer from multiple designs as the training set, and secondly by using all layers in a single design as the training set. When compared to four common benchmark methods for performance prediction of the model, it was found that the root-mean-square-errors of the four benchmark methods were from 50-195\% higher than the proposed model. While these studies were able to predict various properties and responses of AM components and processes via experimentation, reduced-order modeling, or neural networks, their application are highly limited to the respective microstructures, materials, and use cases.

The simulation of AM components often requires multiple programs and software to handling respective data and file types; going from a geometry file, to a slicing software to generate G-code, to numerical simulation requires time and handling by the user that isn't always necessary. Some work has been done to streamline the simulation of 3D printed models. A visualizer program was proposed that can detect potential print failure conditions, ~\cite{ShyhKuang2018}. The authors reported several objects that, when using default printer settings, resulted in print anomalies; however, the proposed program was able to detect the potential anomalies and suggest different printing parameters. The suggested parameters resulted in a significantly better print quality, and allows for a useful way to quickly test prints before actual manufacturing. This form of digital twin proved useful for intuitively detectable print anomalies, however, it cannot predict print behavior due to physics and the potential failures therein. A custom processor was developed to translate G-code of additively manufactured models into input files for finite element modeling (FEM), ~\cite{VORISEK2023103279}. By emulating the G-code over a  high fidelity voxel grid, the approximate structure is recreated based on the printer motion commands. While the ideal size of the voxel grid for FEM was not determined, the processor was able to determine the print time of all tested models with in 2.1\% error and the mass of the printed models within 3.3\%. The thermal distribution during the FDM printing process from a G-code was examined with 3D voxels, ~\cite{gamdha2023geometricmodelingphysicssimulation}. The use of parallel adaptive octree meshes allow for the the rapid representation of the object and simulation of the transient heat distribution, and establishes a foundation for real-time digital twin representation. Nevertheless, the use of voxelization to simulate a complex process such as AM reduces the fidelity of the microstructure and fundamentally disconnects the digital twin from the PSPP linkage with the physical AM component.

In this paper, a modular framework for the simulation of digital twins produced via extrusion-based AM structures is proposed with the principle goal of reducing redundant procedures typical in simulation schemes. Our framework's main strength is the tight linkage between tool-path generation and computational mesh formation, thereby automating unnecessary steps of data transfer and parsing. Through this approach, testing AM components manufactured through extrusion-based techniques can be streamlined with common toolsets available to respective users for both single component studies as well as multi-component structures. 

Our paper is outlined as follows. In section ~\ref{sectionMethods} we outline the proposed framework and highlight the modularity and how it can be applied to differing software packages at each step. In section ~\ref{sectionResults} we show results for the applicaiton of our framework to exemplar problems in the design and evaluation of soft lattice meta-materials structures. We specifically utilize these structures because they serve as a valuable test suite for our meshing approach due to the severe geometric and material non-linearaties present from large deformation and wide-spread self contact; specifically, numerical simulation examples showcasing the applicability of the framework, such as one-dimensional compression, Eigenfrequency analysis, and print-parameter optimization. Finally, a conclusion is provided in Section \ref{sectionConclusion}.

\begin{figure}[H]
\centering
\includegraphics[width=\linewidth]{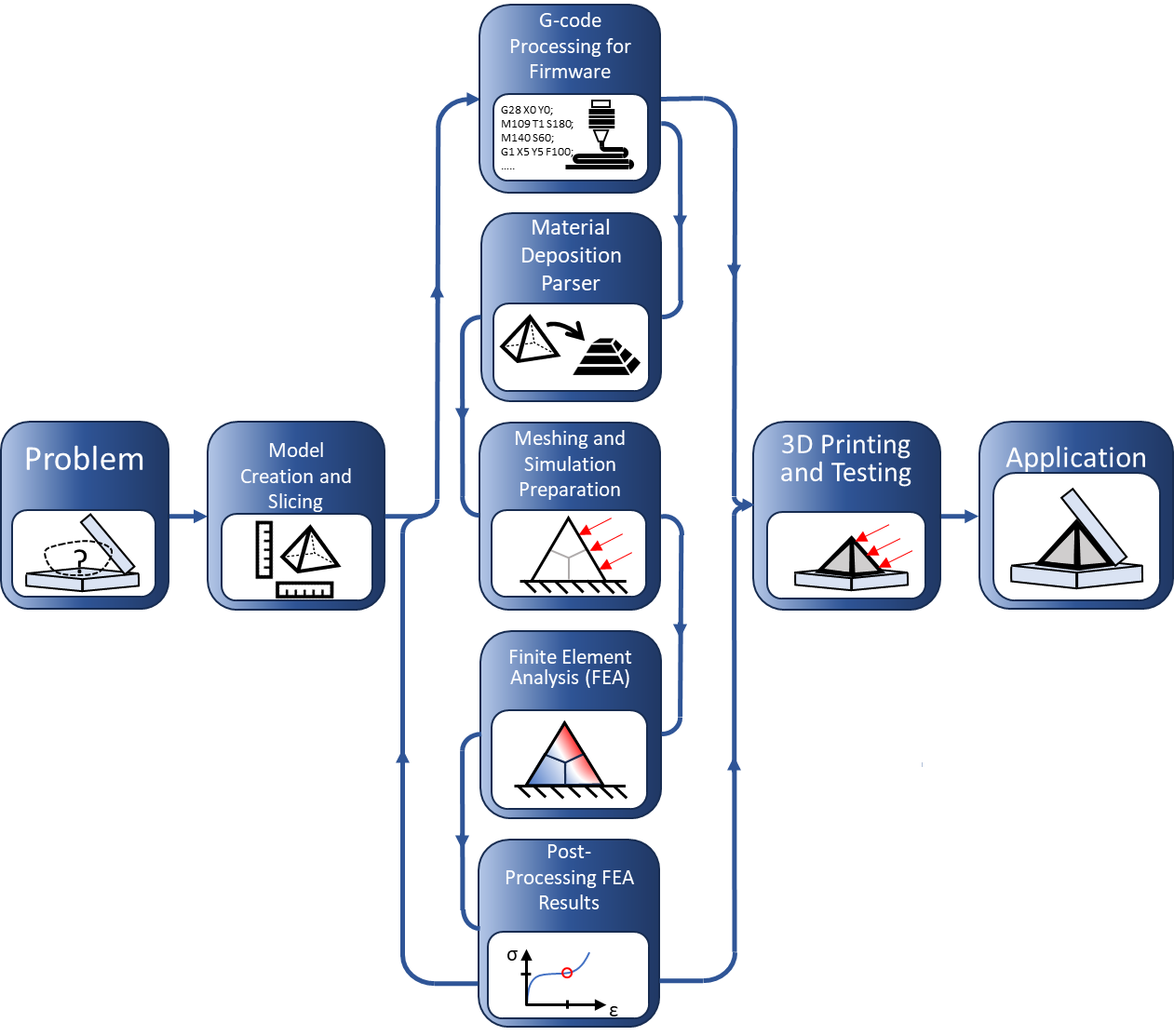} 
\caption{Illustration of our proposed framework. Each box represents a different stage in the generic execution of FEA on a target AM structure.}
\label{fig:WorkflowDiagram}
\end{figure}

\section{Methods}{\label{sectionMethods}}

The proposed framework has many different elements requiring deep interdisciplinary teaming to be successful, and lays out a blueprint for the incorporation of specific software packages (e.g. tool path generators, mesh generators, etc.) for the simplification of simulating additively manufactured  components. When an application and need for numerical simulation are identified, the process begins. The framework consists of six key steps: tool path generation for AM, tool path parsing and lexing to remove travel paths irrelevant to material deposition, mesh generation, boundary condition assignment, numerical simulation execution, and processing the results for quantities of interest.

The following subsections describe each step of the proposed framework for high fidelity numerical simulation of extrusion based, additively manufactured components. For the purpose of this work, the printing process we considered in detail was direct ink write (DIW) printing, using a soft polymer resin and mechanical meta-materials as exemplars. While the equipment and setting described references DIW printing explicitly, this framework can be applied to other extrusion based printing methods using other software and file types due to the object-oriented nature of our software stack. Each framework element is modular in the sense that different elements have the same user facing interface but can have vastly different backends.

\subsection{Toolpath Generation}

To begin, a geometry file is needed, such as an STL file, which describes the structure of the object in three dimensional space, ~\cite{iancu20183d}. While an STL file tends to be the most common geometry file format, several other AM file formats are becoming increasingly more popular. Among them including 3D Manufacturing Format (3MF), Additive Manufacturing File Format (AMF), and Standard for the Exchange of Product data model-Numerical Control (STEP-NC), which allow for the storage of more information regarding the object(s) being printed, such as color, material, and support material inclusion, ~\cite{Krueckemeier_2023}. The geometry file is processed into a series of commands that are readable by the printer using a  generation software, or slicer in common parlance; the object is discretized into stacked layers for sequential printing. There are many household and proprietary slicer softwares readily available. The toolpath generator then determines optimal toolpaths necessary for material deposition in each later. This is commonly output in a file format known as G-code but other more recent file formats such as STEP-NC are being developed to store the printing options as well, ~\cite{Krueckemeier_2023}. For the sake of this work, we restricted ourselves to G-code file formats for wider applicability across both modern and aging extrusion-based AM platforms. A key feature of tool path generators is the ability to customize tool path commands for the purpose of altering material deposition density, print time/speed, materials used, etc. With slicer software packages, we have fine grain control over AM microstructrual features through common tool path generation settings such as infill density, infill pattern, wall line counts, etc.

Confirming the accuracy and validity of numerical simulations through experimentation of the simulated component may require the generated G-code to be reconciled for the desired physical printing platform. As the framework is designed to be generalized for most extrusion based additive manufacturing, an intermediate step is used which can be edited for individual systems; in the case of this work, the more traditional FFF G-code is edited for our direct ink write (DIW) platform utilizing the Aerotech A3200 platform, ~\cite{AEROTECH}.

In the below, we want to emphasize that slicers are usually not a catch all for all tool path generation. We specifically seperate the "initial/generic" toolpath generation from specific platforms to make things more modular and play nice with as many software packages as possible. Then you can further motivate things with the DIW example below. The system we used for DIW printing requires several commands and file structures, such as headers and footers, that were not included; a module is implemented to add these necessary features for the printing platform firmware to process and print. One such step is converting the FFF filament extrusion to the firmware commands for ink extrusion, using a relationship similar to the following,

\begin{equation}
    V = Av
\end{equation}
where \textit{V} is the extruded volumetric flow rate (volume/time), \textit{A} is the area of the printer nozzle (area), and \textit{v} is the rate of linear material flow through the nozzle (length/time).

\subsection{Mesh Generation}

\begin{figure}[H]
    \centering
    \begin{subfigure}{1.0\textwidth}
        \includegraphics[trim={0pt 0pt 0pt 0pt}, clip=true, width=\linewidth]{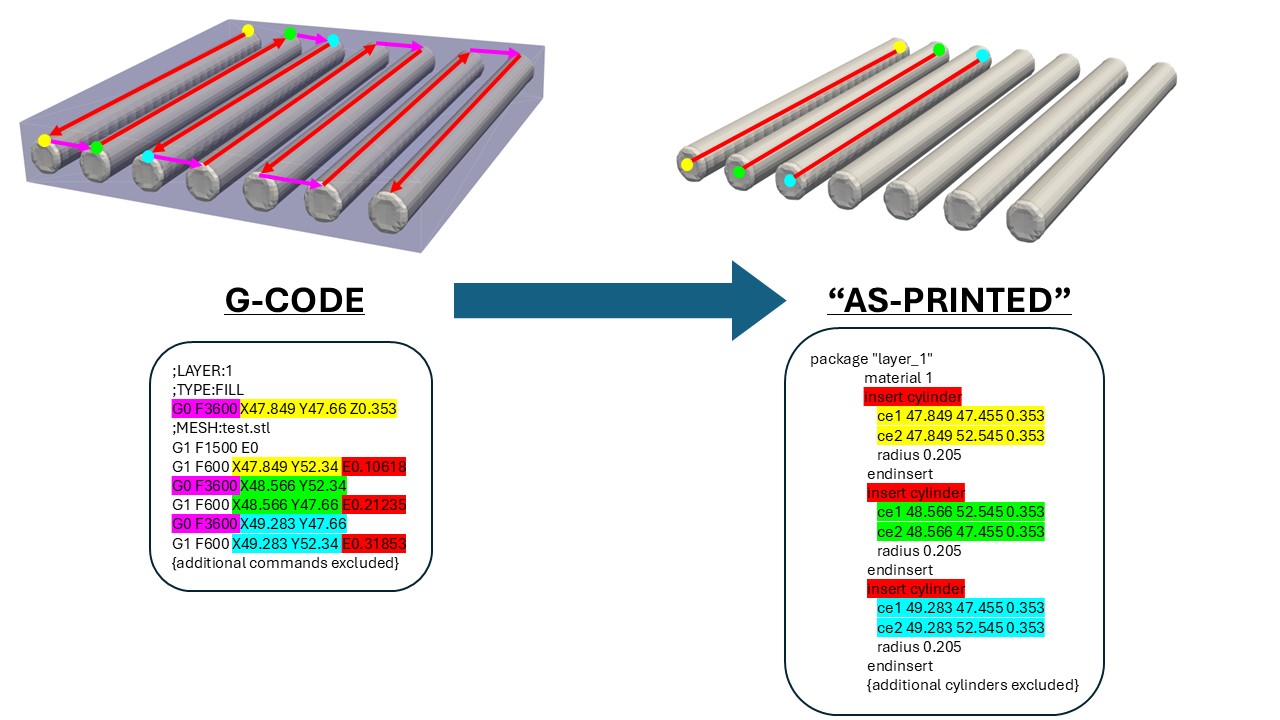} 
    \end{subfigure}
    \caption{Outline of the conversion from G-code to "as-printed" geometry. An STL geometry file (right, gray) is converted to G-code via a slicing software, and the G-code commands are processed as seen in the color-coded lines and text.}
    \label{fig:gcode_conversion}
\end{figure}

After the geometries are processed and the tool paths are generated, the typical design cycle would entail sending the G-code files to a printer, performing some test for design quantities of interest (QOI), and then iterating based on the outcomes of the test. This process can be exceedingly expensive in terms of time and material waste, and is typically guided more by intuition, experience, and empiricism. To alleviate this cost issue, it is highly desirable to create digital twins of appropriate fidelity for target AM structure designs while not sacrificing a strong PSPP linkage. In order to do this, computational discretizations of complex AM microstructures are necessary. This is by no means a trivial task; automation of this process is necessary even for the simplest of AM microstructures. Figure~\ref{fig:gcode_conversion} shows a generalized outline of the conversion of g-code to an "as-printed" geometry model. 

In preparation for meshing, relevant and irrelevant information is parsed from the G-code file; the former, the path of the tool head where material was actively being deposited, and the latter, tool head dwell time, movement that did not result in material deposition, tool head homing, etc.. Depending on the platform and technique in use, the exact method required for the parser logic to filter this information may be different. For example, with fused filament fabrication (FFF), it may be necessary to check for positive feed rate in G0/G1/G2/G3 commands. However, for the DIW method used as our exemplar, there were several options for checking for material deposition, such as a trigger on the pressure controller for the syringe or movement of the extrusion pump. Regardless, our framework allows for a modular definition of a material extrusion parser logic through object-oriented programming, via a python interface. 

The 3D printed representation file is passed to a meshing software to generate the elements and nodes necessary for finite element analysis (FEA). The meshing process posses another stage of customization that can greatly affect the efficiency and accuracy of simulations, depending on the size and shape of elements. The nodes and edges that will be used to apply boundary conditions in the FEA are also identified. 

In the provided examples, Sculpt, a powerful software that uses a unique, mesh-first method to define meshes is used to generate a robust hexahedral mesh of the provided geometry file, ~\cite{sculpt_2019, OWEN2017167}. The high quality meshes are automatically generated for FEA and computational fluid dynamics (CFD), with local and global element quality control. A Cartesian grid is overlaid on the geometry features, which are then carved, or sculpted, out of the overlay. The boundaries are smoothed to create the resulting hexahedral mesh, which are beneficial for numerical simulation due to their ability to pack volume more efficiently than other common element types. The exemplar geometries are provided in a Sandia National Laboratories legacy file format called a diatom which were generated during the noted G-code parsing stage.

Next, the file is passed to Cubit, a finite element mesh and geometry preparation software developed by Sandia National Laboratories, ~\cite{cubit_2024}. While Cubit is capable of generating the mesh of the geometry, it is used in this framework to prepare the all-hex mesh created by Sculpt for FEA. Here collections of nodes or element facets are defined for the application of defined boundary conditions. At this stage as well, the sculpted mesh can be passed into Cubit to integrate with other traditionally meshed objects, such as platens, housings, or other assemblies. This is useful for multi-component simulations, where one or more component-of-interest can be individually processed in this framework and integrated into larger structures.

\subsection{Finite Element Simulations}

The meshed AM structure is now ready for use in FEA. Our object-oriented programming interface handles user provided simulation template files, writing key information such as model parameters and settings for the given analysis. Boundary conditions, such as Dirichlet or Neumann, are applied to the relevant sets of nodes or element facets. Other key factors, such as contact control, solver settings, or time stepping control, and QOI's for post-processing are also defined for the simulation. The exact simulation software package utilized is customizable; users are able to define which FEA software is used and what kind of simulation is conducted, such as ANSYS, Autodesk, and Altair or uniaxial compression, thermal diffusion, or modal analysis, respectively. To utilize the framework for optimization, the variable parameters and objective function are also defined. After the simulation template is properly defined, it is passed to the respective FEA software and executed. 

In this study, Sierra/Solid Mechanics (Sierra/SM) or Sierra/Solid Dynamics (Sierra/SD) are utilized to conduct numerical simulation on AM components, ~\cite{sierra_sm_2024, sierra_sd_2024}. Sierra/SM is a solid mechanics code developed by Sandia National Laboratories with both explicit transient and implicit quasistatic capabilities, with a large variety of nonlinear material models ~\cite{osti_2430028}, robust element library, contact, and large deformation capabilities. Sierra/SD is a structural dynamics finite element analysis code capable of high fidelity modal, vibration, static, or shock analysis; its linear and nonlinear computational abilities are specifically designed for parallel use. Both Sierra/SM and Sierra/SD are executable via an input file, making the implementation of finite element analysis in the framework straightforward. For optimization analysis of designs, the independent toolpath generation parameters are defined as well as their constrained ranges; a necessary objective function, such as minimization of eigenmodes, is also defined. A template input file is open and edited using Python to the user specified boundary conditions and QOI's. Then, the input file is passed to Sierra, and the simulation is executed.

\subsection{Post-Processing Finite Element Analysis Results}

The specified QOI's from the simulation are stored in a results file, which will differ across different simulation software suites. The exact program, language, or toolkit used to process the resultant data is a user preference, and can include any number of software packages or languages, such as ParaView ~\cite{Paraview2005}, Python and its ecosystem's many packages, Matlab, etc. In cases where design optimization is used, the optimized results can be analyzed with respect to the initial configuration. Our proposed framework allows for simple wrapping for seamless integration of different software packages. 

The results of the Sierra simulations conducted in this study are stored in two forms; a heartbeat file and an exodusII file, a Sandia National Laboratories developed data file format, ~\cite{Schoof_1994}. The heartbeat file is a file that contaion global output data requested by the user in a column separated value (CSV) file, such as time step, global load, and displacement. Mechanical properties, such as the Elastic Modulus and Plateau Stress can be reduced from this global data using various Python packages, such as Numpy, Pandas, Matplotlib, and Scipy, \cite{harris2020array, Hunter:2007, reback2020pandas, 2020SciPy-NMeth}. The exodus file contains the element and node coordinates, the connectivity, and result variables such as nodal displacements and element averaged Cauchy stresses. Programs such as Paraview were used to visualize the specimen deformed shapes in the case mechanics problems, and the frequency response in dynamic problems, providing a better understanding of the model response.

\section{Results}\label{sectionResults}

Our proposed framework is adaptable to the user's desired slicer, parser, mesh method, and numerical simulation software. In addition to the software specified in Section \ref{sectionMethods}, our framework was configured to produce an exemplar soft elastomer, using a custom-built Direct Ink Write printer commanded by an Aerotech motino control interface. The following shows the results of model conversion from an input geometry STL file to the simulation of the digital twin representation of the AM structure, as well as several basic applications of our proposed framework. First, the impact of varying several common printing parameters are explored; then, the results of meshing the representative model with Sculpt are discussed. Next, a series of numerical simulations showcasing the capabilities of this framework are investigated. For solid mechanics, 1D compression is simulated to a high degree of strain, and relevant mechanical information is derived from the generated results. For solid dynamics, an examination of the eigen solutions for these complex geometries is given. Lastly, an example of optimization of the first eigenmode is based on varying printer parameters is conducted. All three implementations used the same Neo-Hookean material model to represent the printed material, Table~\ref{NeoHookean}.

\begin{table}[H]
\centering
\begin{tabular}{ | c | c | c | }
    \hline
    Density ($kg/m^3$) & Shear Modulus (\textit{MPa}) & Bulk Modulus (\textit{MPa}) \\
    \hline
    \hline
    1130 & 2.18 & 920.0 \\
    \hline
\end{tabular}
\caption{Neo-Hookean model parameters used in the provided examples.}
\label{NeoHookean}
\end{table}

\subsection{Effect of Infill Parameters}{\label{subsectionObjectSlicing}}

\begin{figure}[H]
    \centering
    \begin{subfigure}{0.45\textwidth}
        \includegraphics[trim={600pt 0pt 600pt 0pt}, clip=true, width=\linewidth]{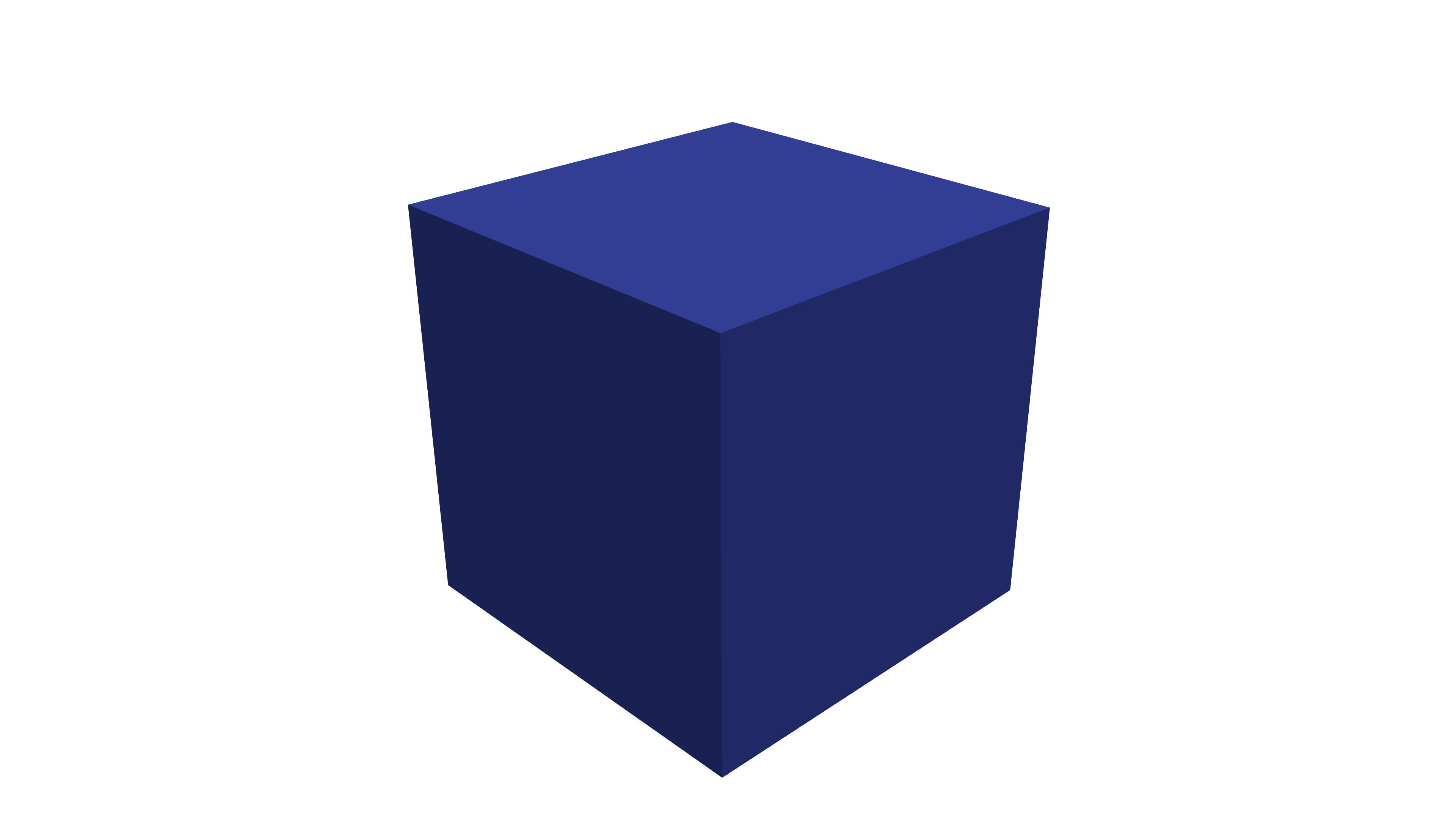} 
        \caption{}
        \label{fig:infillSpacinga}
    \end{subfigure}
    \begin{subfigure}{0.45\textwidth}
        \includegraphics[trim={600pt 0pt 600pt 0pt}, clip=true, width=\linewidth]{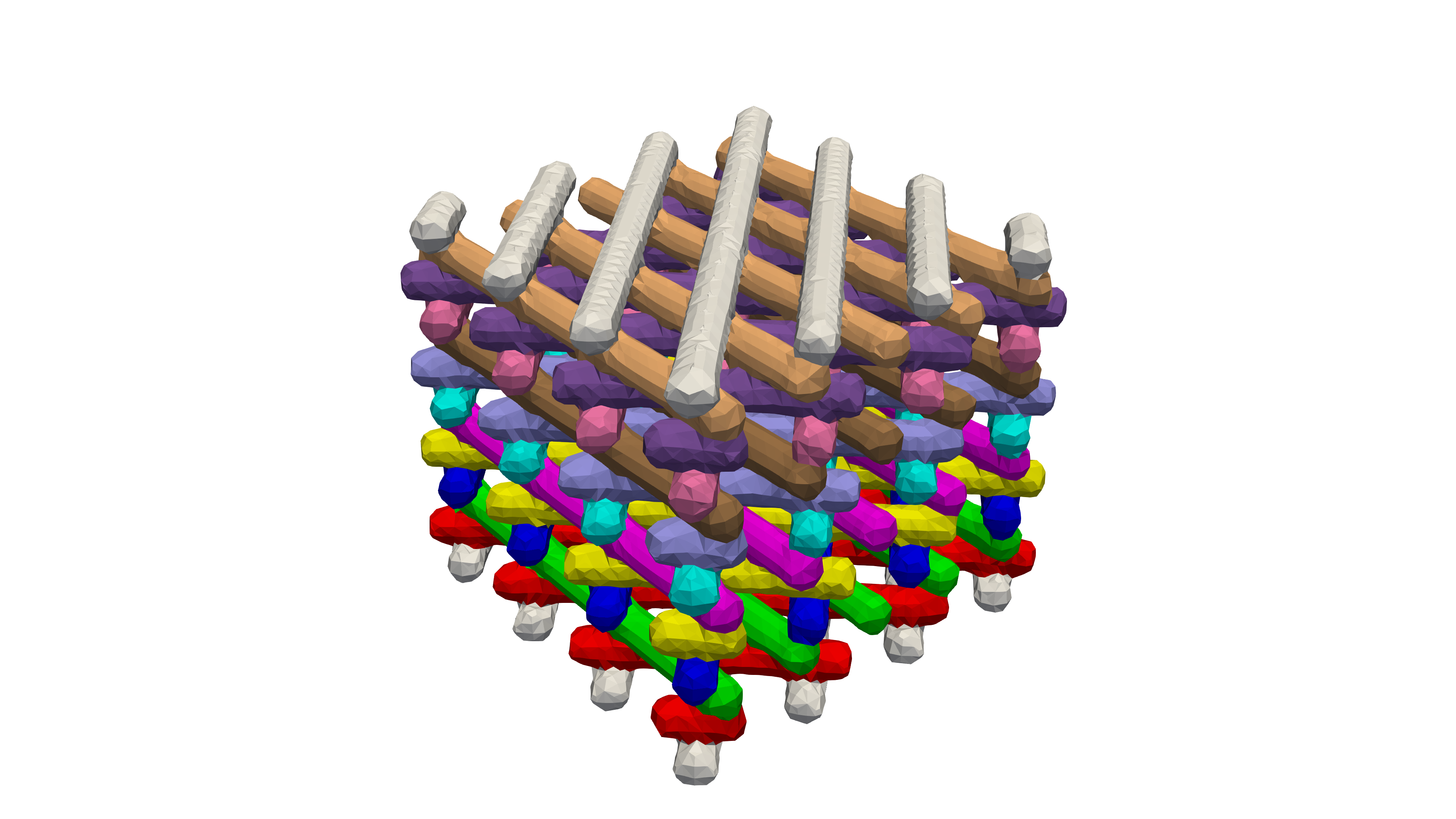} 
        \caption{}
        \label{fig:infillSpacingb}
    \end{subfigure}
    \begin{subfigure}{0.45\textwidth}
        \includegraphics[trim={600pt 0pt 600pt 0pt}, clip=true, width=\linewidth]{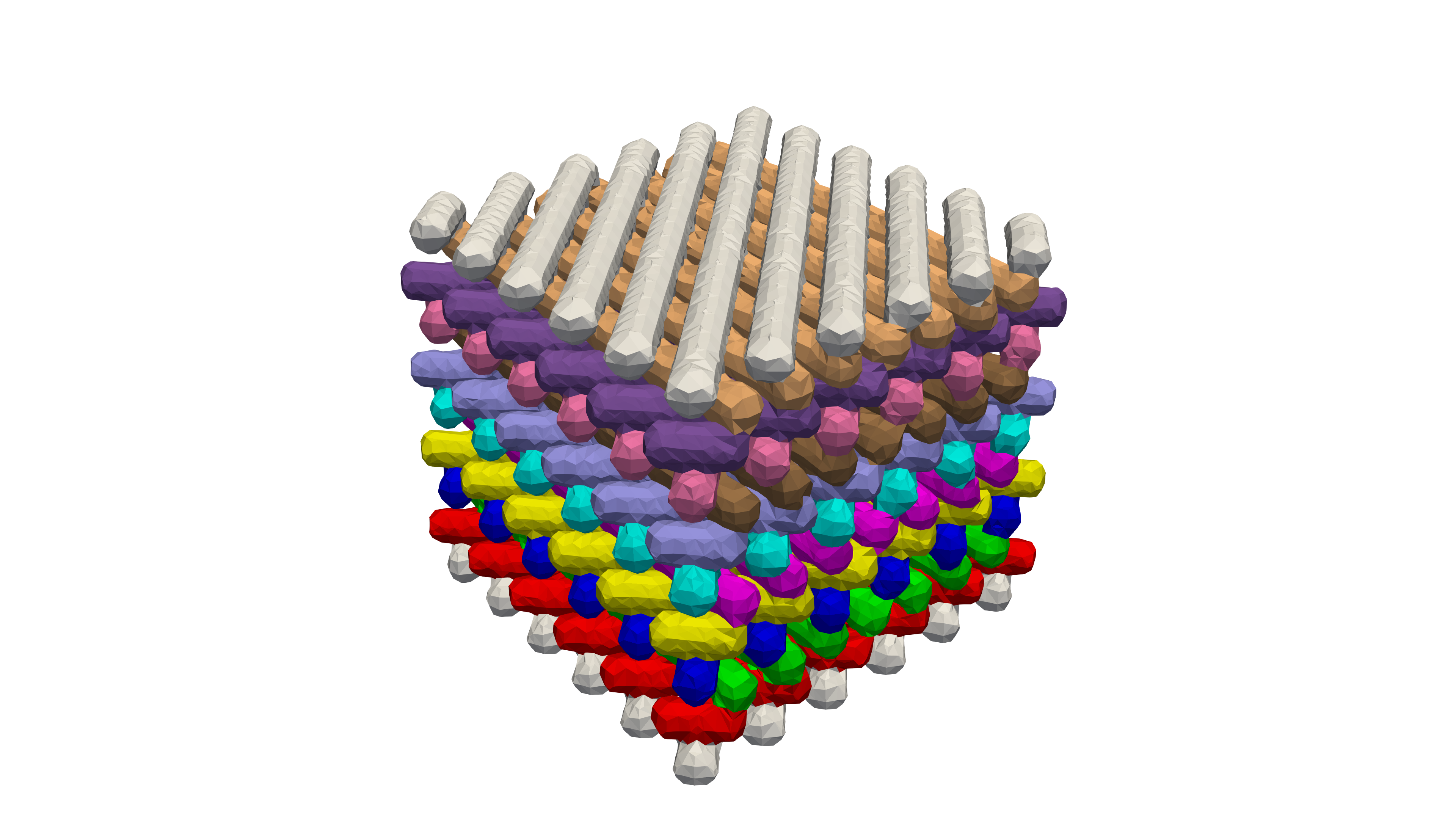}
        \caption{}
        \label{fig:infillSpacingc}
    \end{subfigure}
    \caption{Isometric meshed cube with varying infill spacing; \ref{fig:infillSpacinga} Geometry file, \ref{fig:infillSpacingb} sparser spacing, \ref{fig:infillSpacingc} denser spacing.}
    \label{fig:infillSpacing}
\end{figure}

Since the proposed framework integrates a slicing software, many of the common 3D printing parameters, such as infill density, geometry, and angle, are readily customizable from the start. Infill spacing, or conversely infill density, has an influence on the mechanical properties and behavior of an AM structure and should be considered, \cite{QAMARTANVEER2022100}. Increasing the infill spacing will result in fewer toolpaths, which typically lowers the stiffness of the AM structure. In terms of simulation, the infill spacing impacts the quantity and connectivity of toolpaths that will experience contact; decreasing infill spacing, for example, creates more toolpaths that result in more connections and contacts to other toolpaths, thus increasing the computational load and complexity. The proposed framework is capable of handling a variety of infill spacing options, from very dense to significantly sparse, as shown in Figure~\ref{fig:infillSpacing}. Limitations to the degree of infill spacing used should be based on the relevant material and printer; too sparse spacing may result in a toolpath being insufficiently supported, while too dense spacing may cause material buildup and misshaped prints due to over-extrusion. It should be noted that herein when referencing infill spacing, this represents the distance between printed lines of material in terms of multiples of the filament diameter; e.g. an infill spacing of 2.0 results in the distance between print lines being 2.0x the filament diameter.

\begin{figure}[H]
    \centering
    \begin{subfigure}{0.45\textwidth}
        \includegraphics[trim={600pt 0pt 600pt 0pt}, clip=true, width=\linewidth]{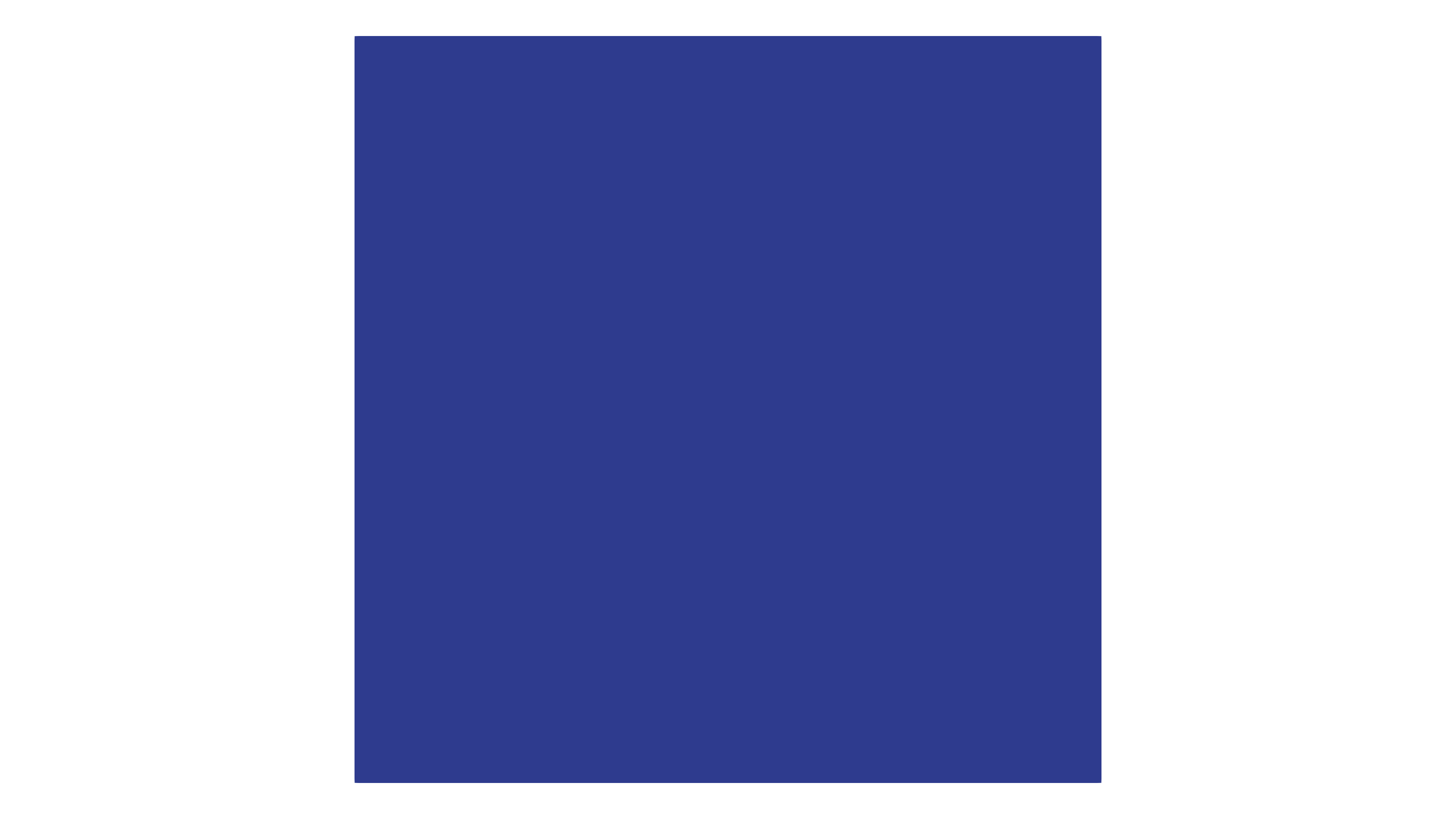} 
        \caption{}
        \label{fig:infillAnglea}
    \end{subfigure}
    \begin{subfigure}{0.45\textwidth}
        \includegraphics[trim={600pt 0pt 600pt 0pt}, clip=true, width=\linewidth]{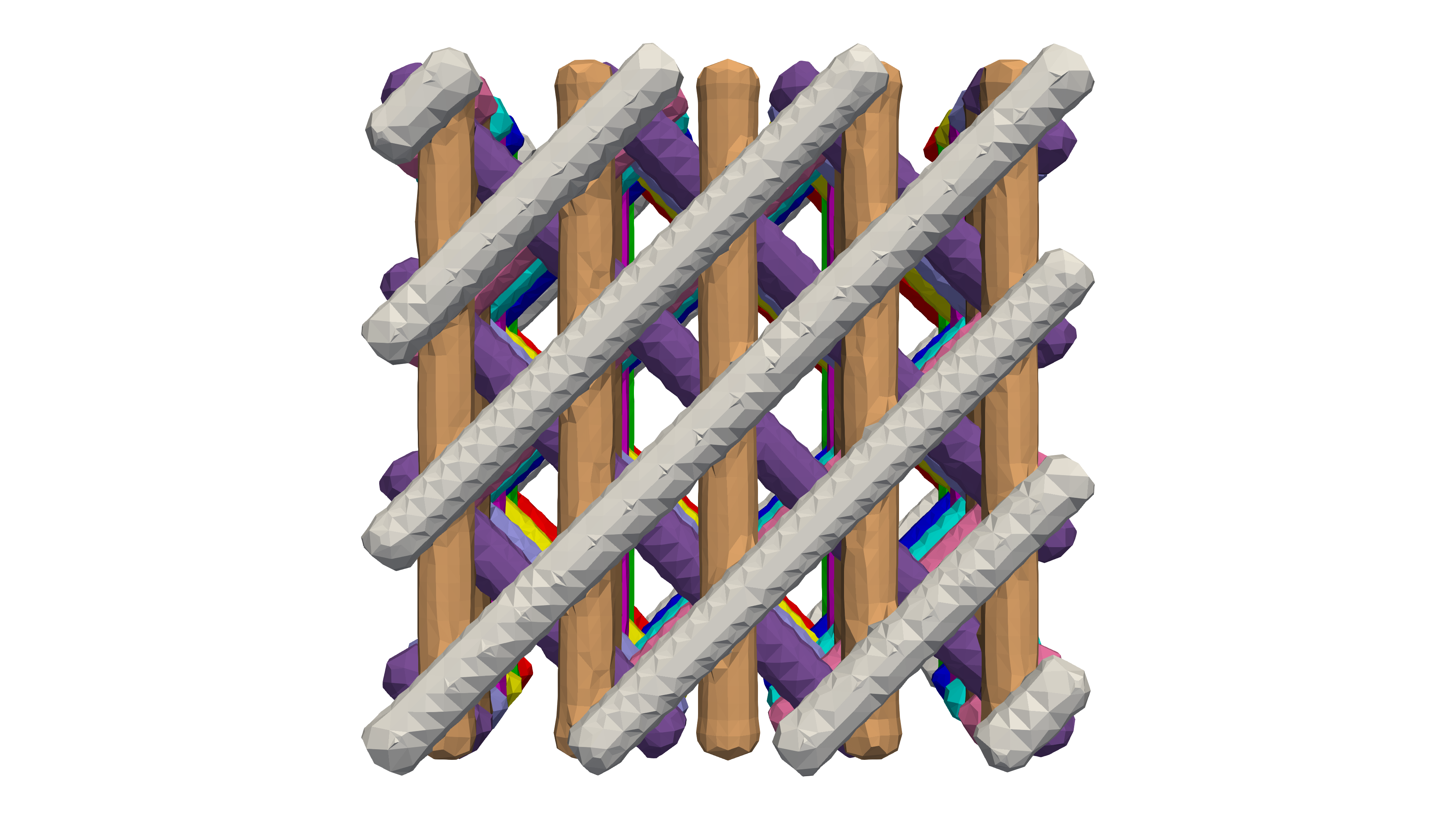} 
        \caption{}
        \label{fig:infillAngleb}
    \end{subfigure}
    \begin{subfigure}{0.45\textwidth}
        \includegraphics[trim={600pt 0pt 600pt 0pt}, clip=true, width=\linewidth]{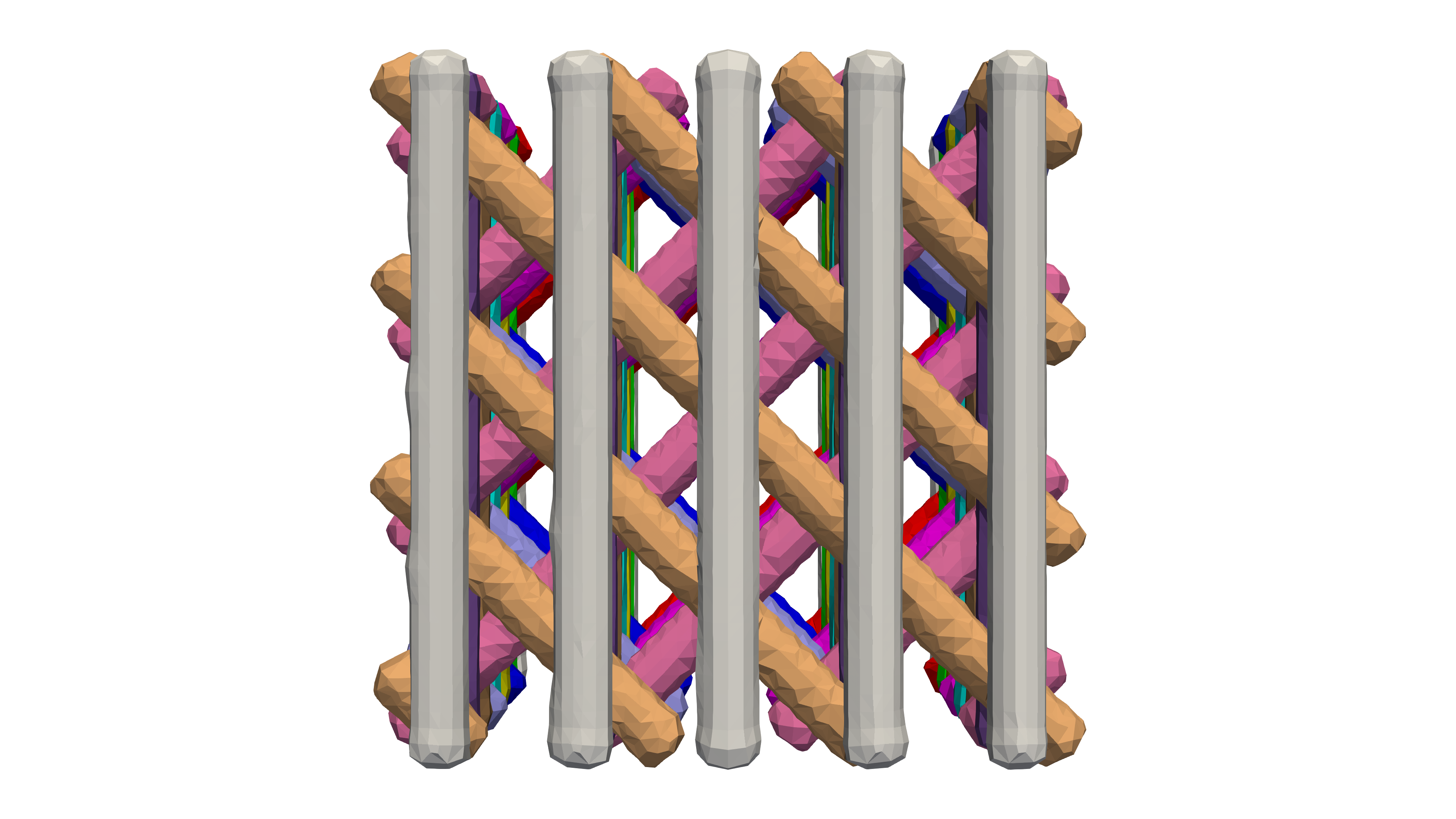}
        \caption{}
        \label{fig:infillAnglec}
    \end{subfigure}
    \caption{Example meshes showing the variation of infill angle; \ref{fig:infillAnglea} Geometry file, \ref{fig:infillAngleb} 45, 90, 135 degree infill angle , \ref{fig:infillAnglec} 90, 135, 45 degree infill angle.}
    \label{fig:infillAngle}
\end{figure}

The angle of the infill pattern also influences the mechanical properties and behavior of AM structures, where toolpaths oriented in the same direction of loading typically results in higher magnitudes of mechanical properties, \cite{QAMARTANVEER2022100, Suteja_2020}. The layer-by-layer angle can be specified as well, either for the whole model or in repeating patterns, allowing for unique anisotropic responses not typically seen in subtractively manufactured components, as shown in Figure~\ref{fig:infillAngle}. Here, the angles of the layers are alternating between 0, 45, 0, and 135 degrees. It should be noted that when the infill angle is referenced and used in future sections, it is structured in a repeating pattern of [0.0, angle, 0.0, 180 - angle] degrees, where the angle is user specified.

\begin{figure}[H]
    \centering
    \begin{subfigure}{0.45\textwidth}
        \includegraphics[trim={400pt 0pt 400pt 0pt}, clip=true, width=\linewidth]{figures/cube_STL_TOP.png} 
        \caption{}
        \label{fig:infillGeometrya}
    \end{subfigure}
    \begin{subfigure}{0.45\textwidth}
        \includegraphics[trim={400pt 0pt 400pt 0pt}, clip=true, width=\linewidth]{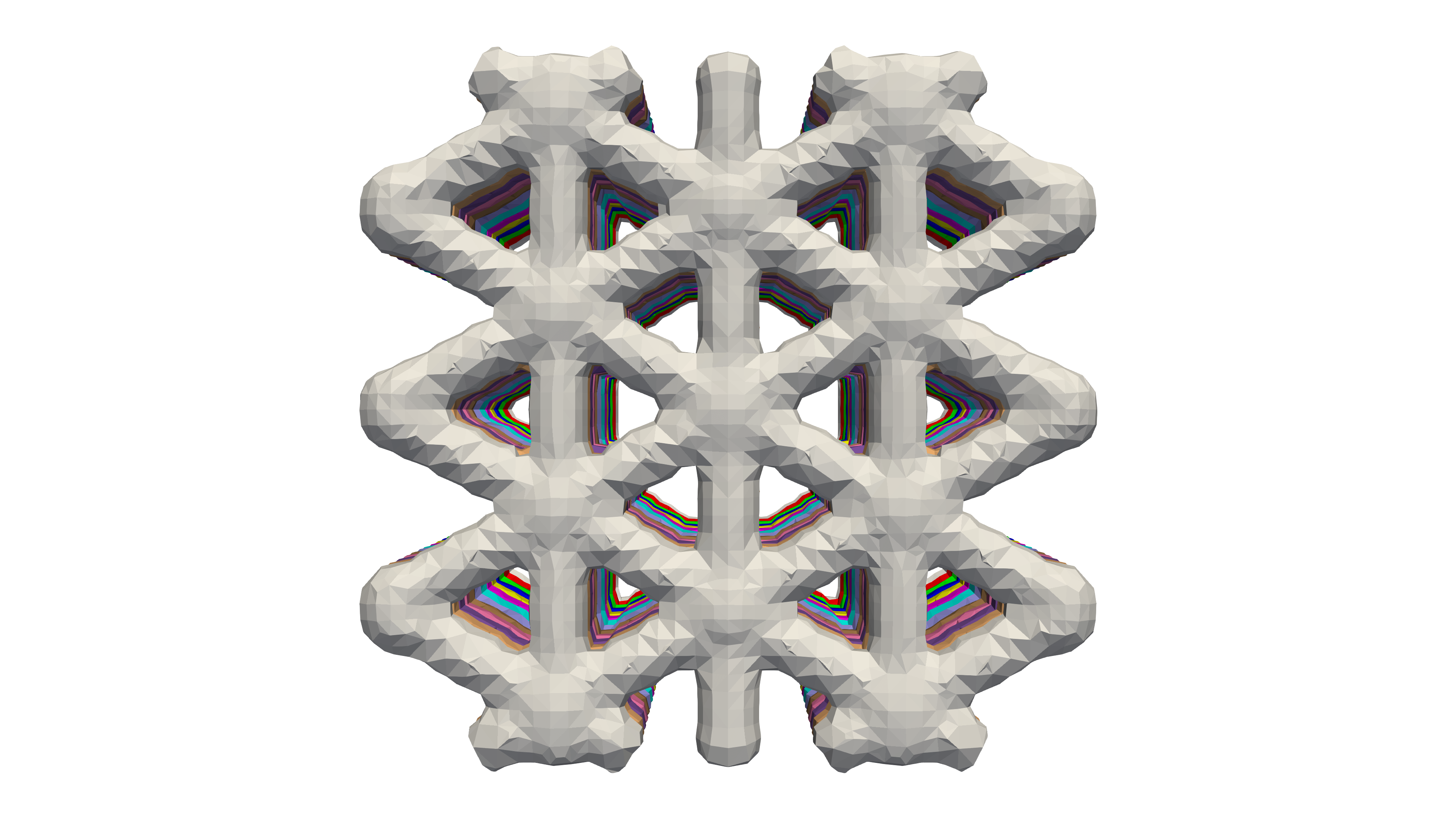} 
        \caption{}
        \label{fig:infillGeometryb}
    \end{subfigure}
    \begin{subfigure}{0.45\textwidth}
        \includegraphics[trim={400pt 0pt 400pt 0pt}, clip=true, width=\linewidth]{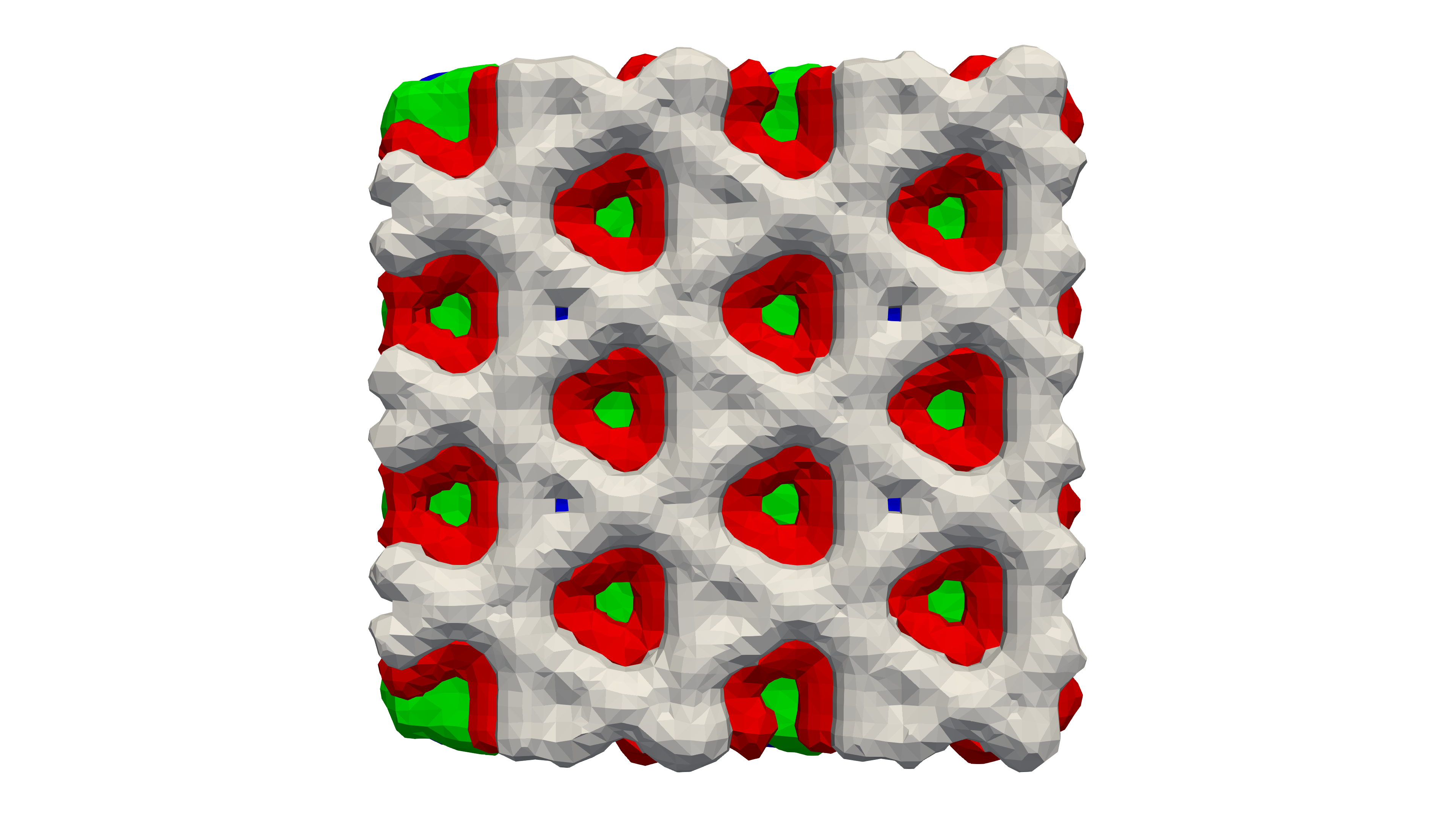}
        \caption{}
        \label{fig:infillGeometryc}
    \end{subfigure}
    \caption{Example meshes showing the variation of infill pattern; \ref{fig:infillGeometrya} top view of cube stl, \ref{fig:infillGeometryb} triangle infill pattern, \ref{fig:infillGeometryc} cubic infill pattern.}
    \label{fig:infillGeometry}
\end{figure}

As expected in a standard slicing software, the infill pattern itself is also variable, however it is limited to the user's selected slicing program, as shown in Figure~\ref{fig:infillGeometry}. Similarly to the infill spacing and angle, the pattern chosen will impact the mechanical properties of the finished part, though the exact relationship is less intuitive, \cite{QAMARTANVEER2022100, Suteja_2020}. a potential unique design space is available when considering the described print parameters in unison.

\subsection{Meshing Quality}
\begin{figure}[H]
    \centering
    \begin{subfigure}{0.45\textwidth}
        \includegraphics[width=\linewidth]{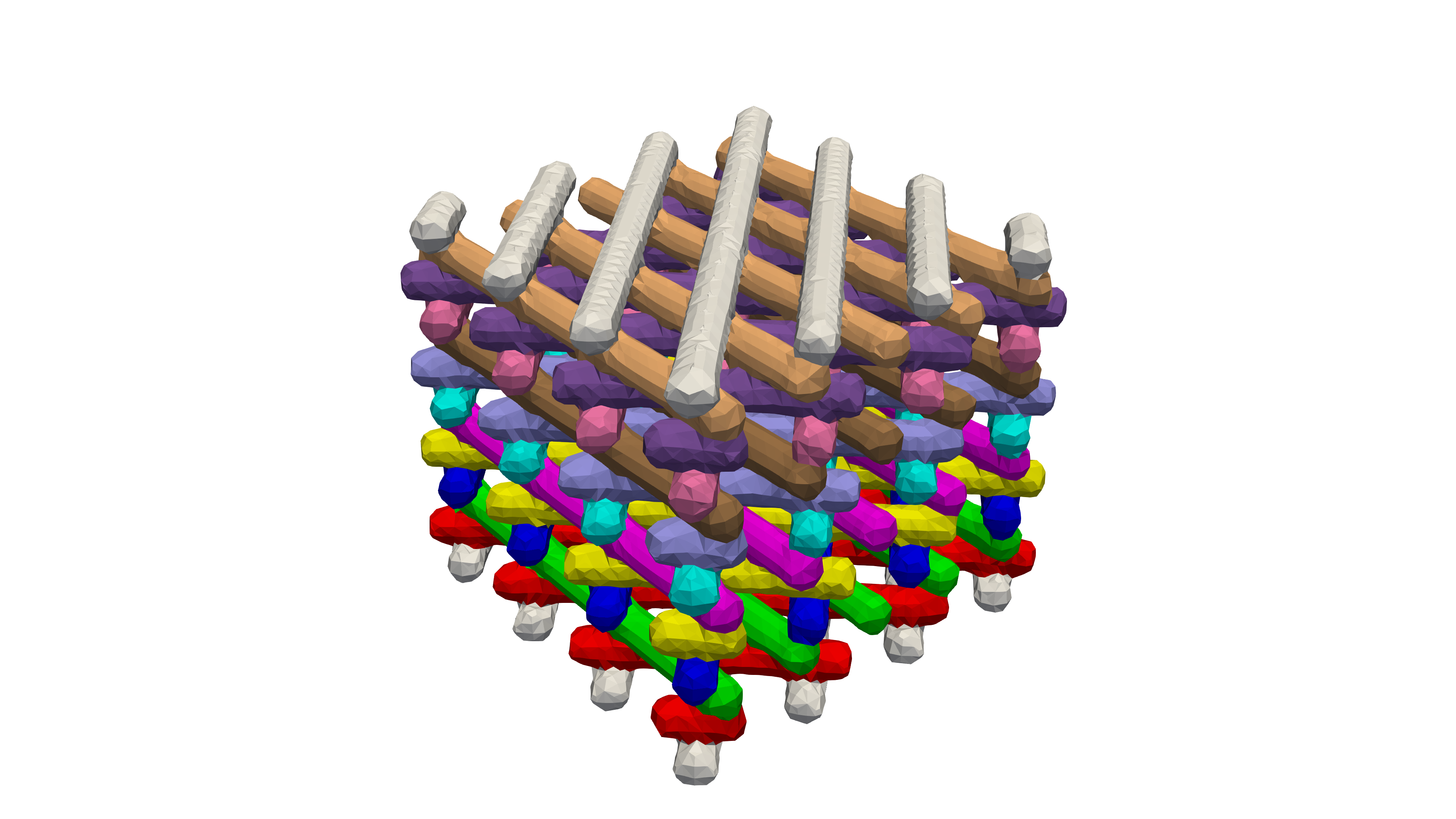} 
        \caption{}
        \label{fig:meshExamplea}
    \end{subfigure}
    \begin{subfigure}{0.45\textwidth}
        \includegraphics[width=\linewidth]{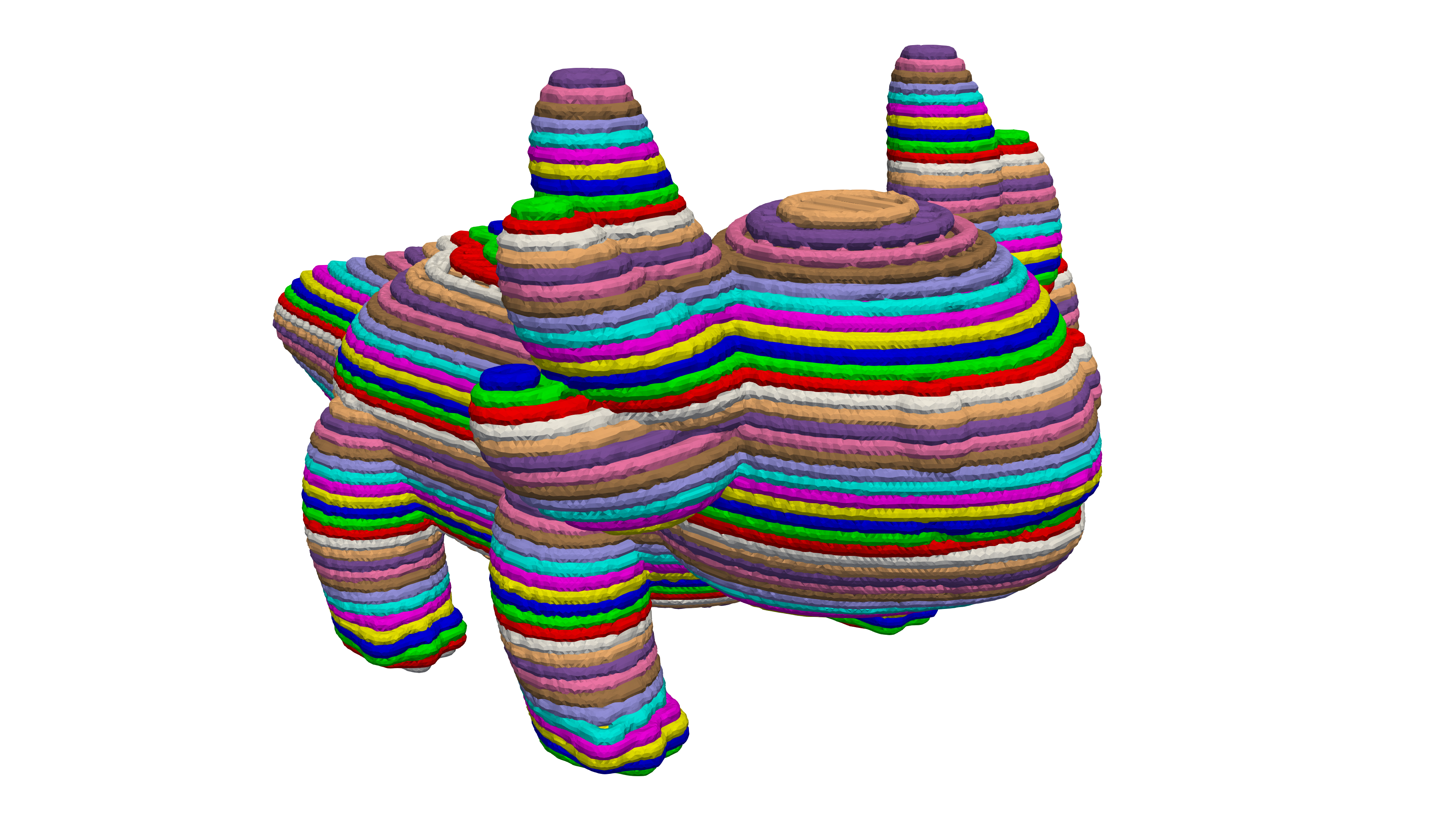}
        \caption{}
        \label{fig:meshExampleb}
    \end{subfigure}
    \begin{subfigure}{0.45\textwidth}
        \includegraphics[width=\linewidth]{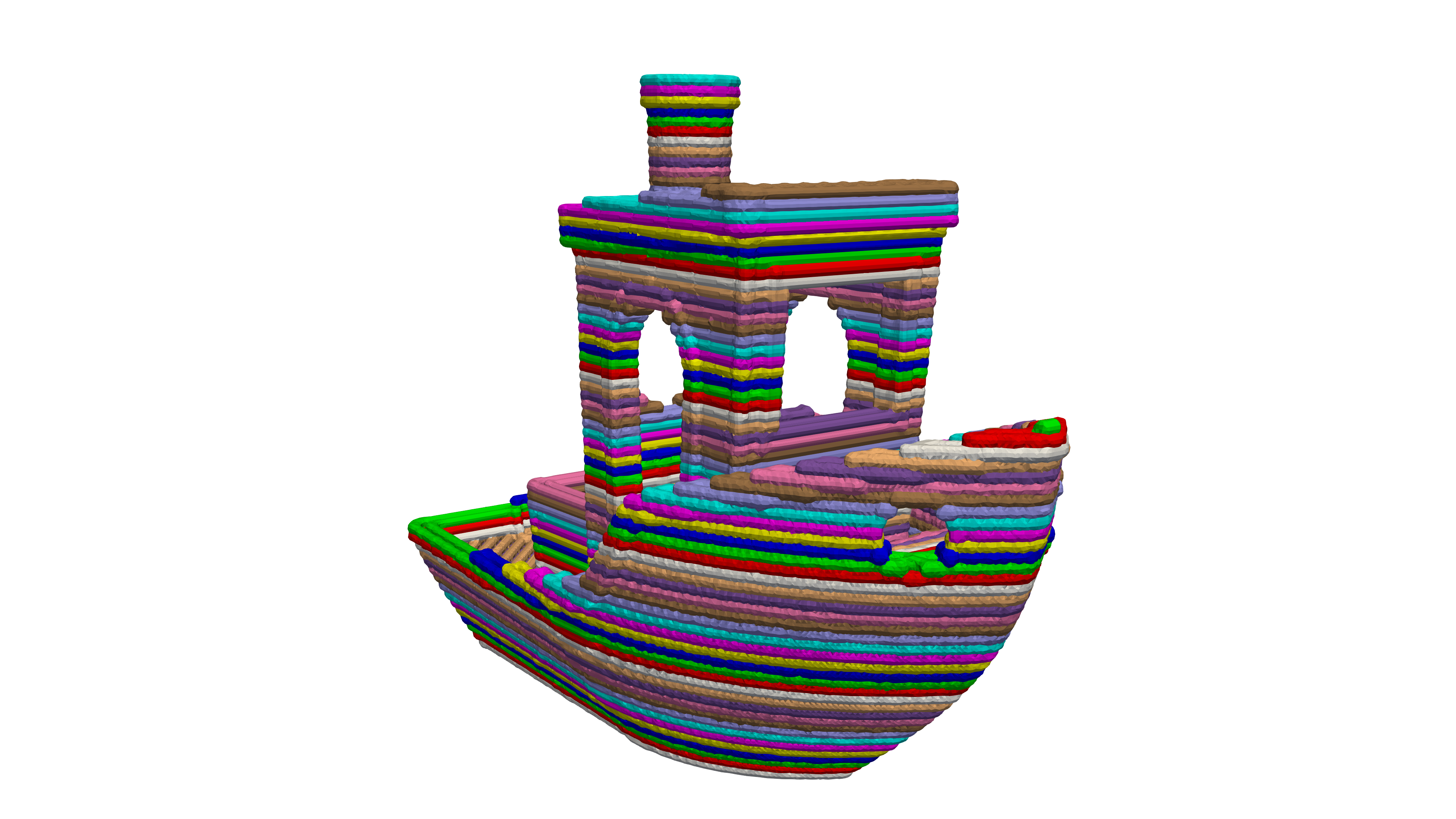}
        \caption{}
        \label{fig:meshExamplec}
    \end{subfigure}
    \begin{subfigure}{0.45\textwidth}
        \includegraphics[width=\linewidth]{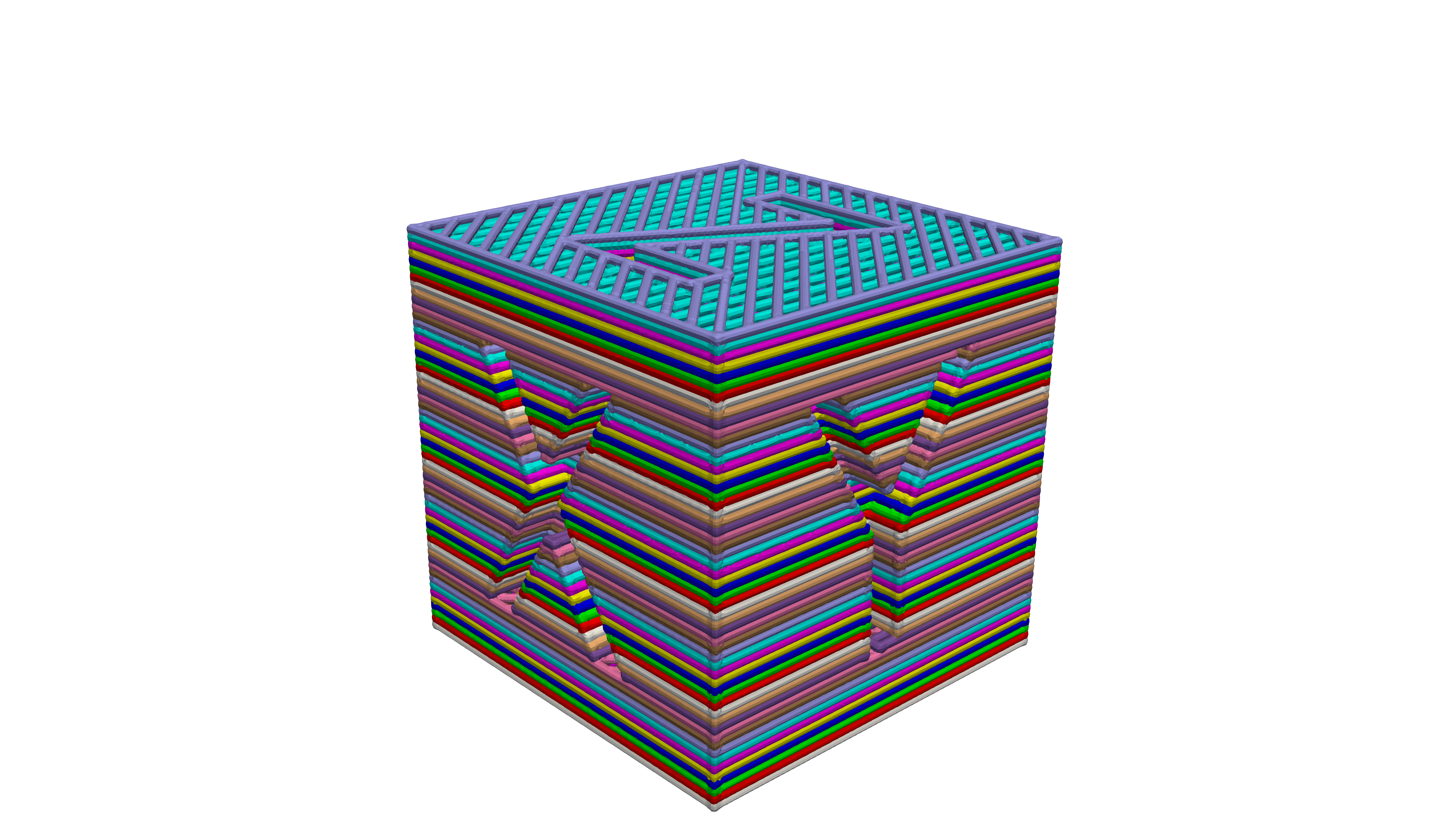}
        \caption{}
        \label{fig:meshExampled}
    \end{subfigure}
    \caption{Example meshes of standard AM structure benchmarks, \ref{fig:meshExamplea} cube, \ref{fig:meshExampleb} axolotl, \ref{fig:meshExamplec} Benchy, \ref{fig:meshExampled} and calibration cube.}
    \label{fig:meshExamples}
\end{figure}

The mesh composition in a numerical simulation has an impact on the efficiency and accuracy of the solution; it is therefore important for a suitable mesh to be used. Sculpt is capable of generating high-quality, hex-dominant meshes for both simple and complex geometries. Figure~\ref{fig:meshExamples} shows several example meshes generated for complex standard AM benchmark structures. Many different factors of the hex-dominant mesh are variable according to user specification, such as the approximate element size. 

\begin{figure}[H]
    \centering
    \begin{subfigure}{0.45\textwidth}
        \includegraphics[width=\linewidth]{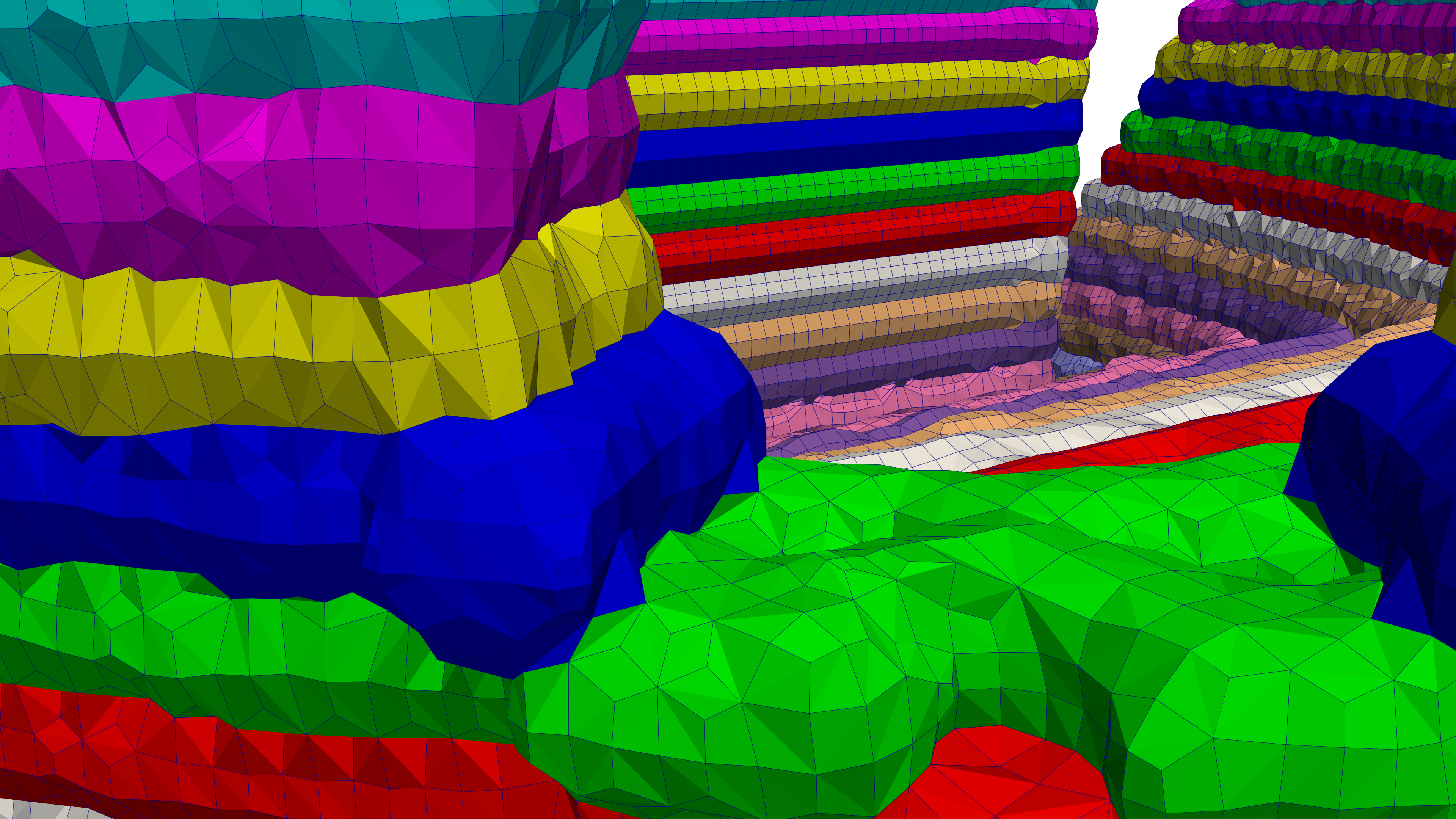} 
        \caption{}
        \label{fig:courseMesha}
    \end{subfigure}
    \begin{subfigure}{0.45\textwidth}
        \includegraphics[width=\linewidth]{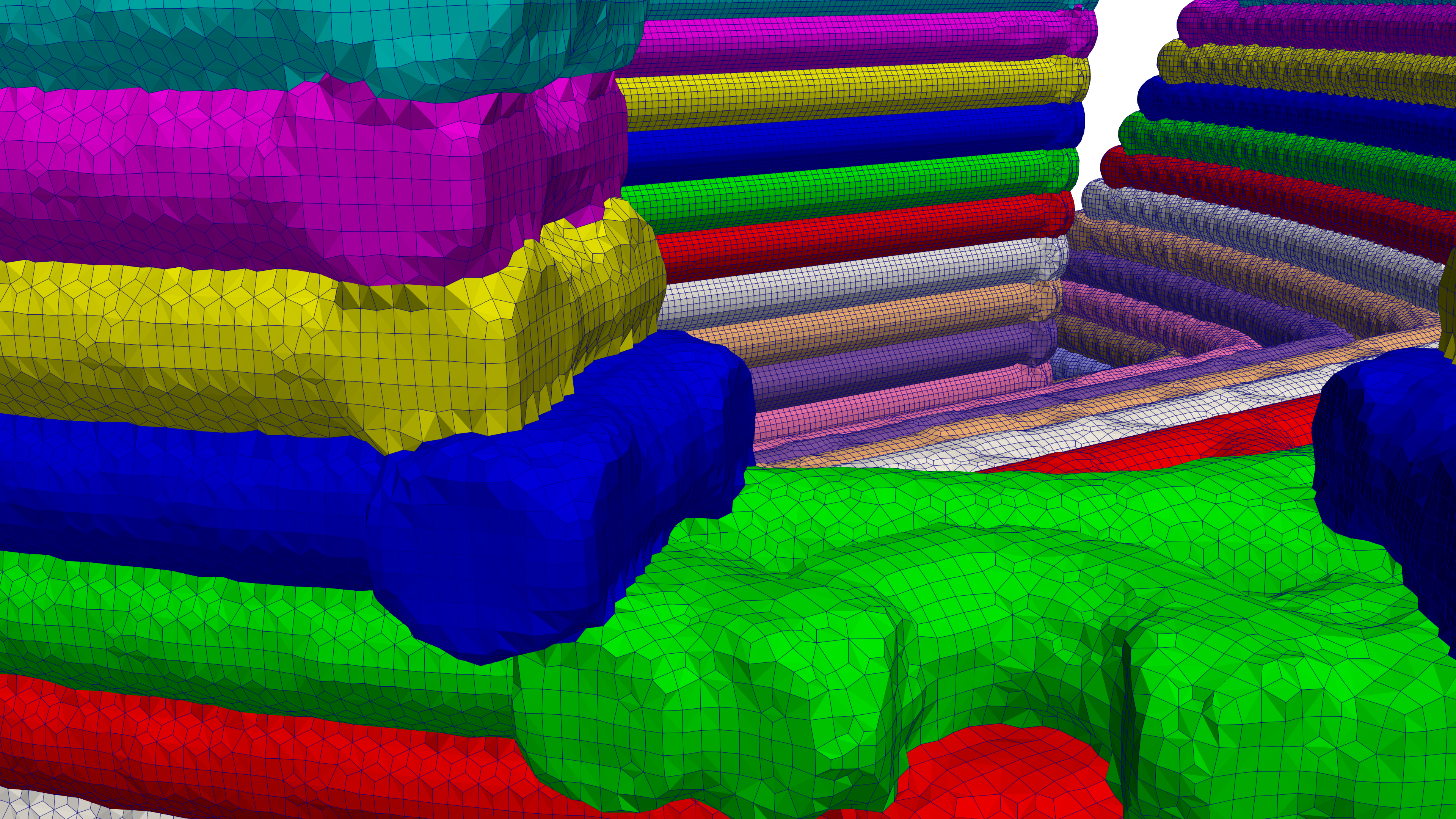} 
        \caption{}
        \label{fig:courseMeshb}
    \end{subfigure}
    \begin{subfigure}{0.45\textwidth}
        \includegraphics[width=\linewidth]{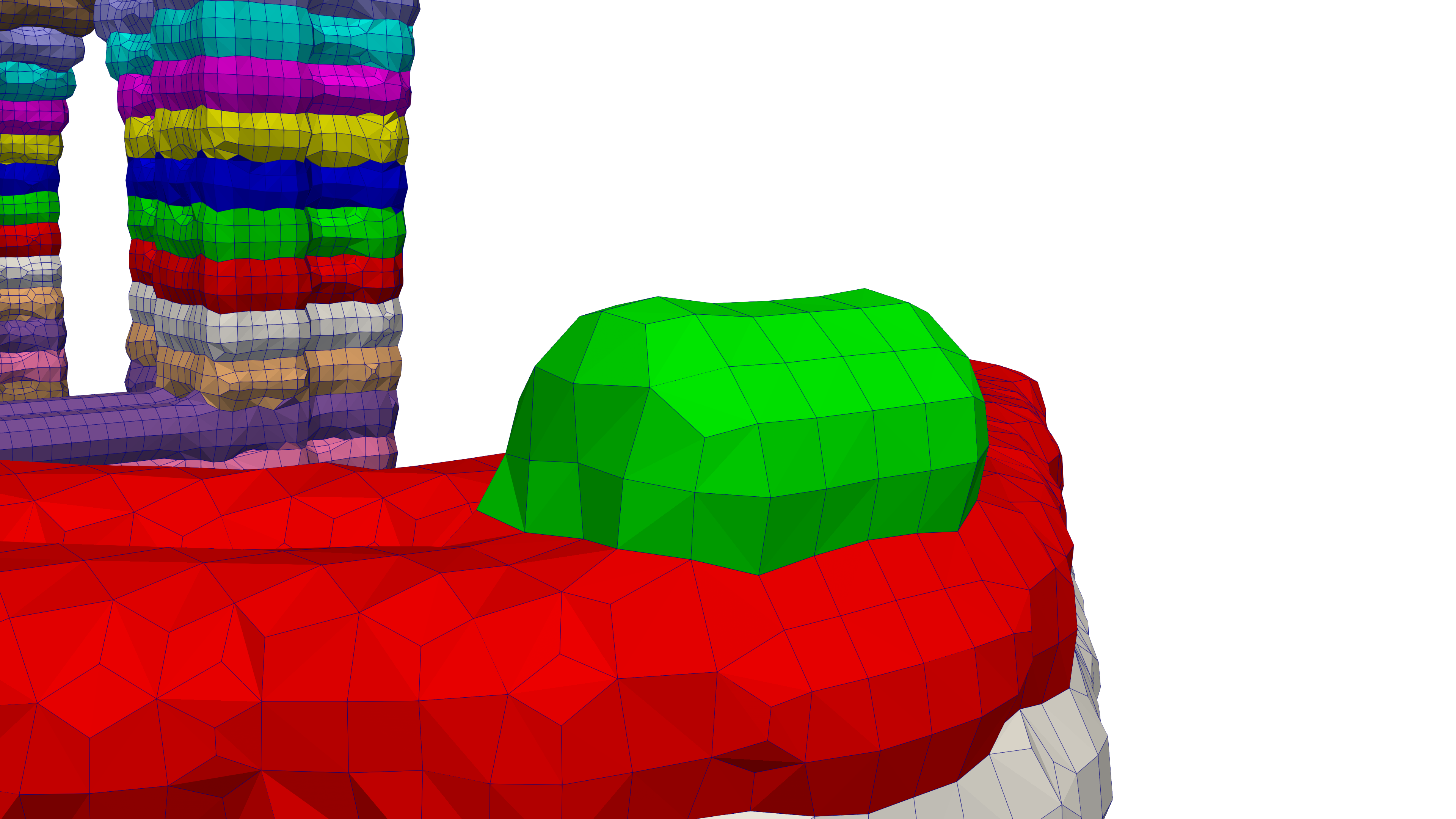}
        \caption{}
        \label{fig:courseMeshc}
    \end{subfigure}
    \begin{subfigure}{0.45\textwidth}
        \includegraphics[width=\linewidth]{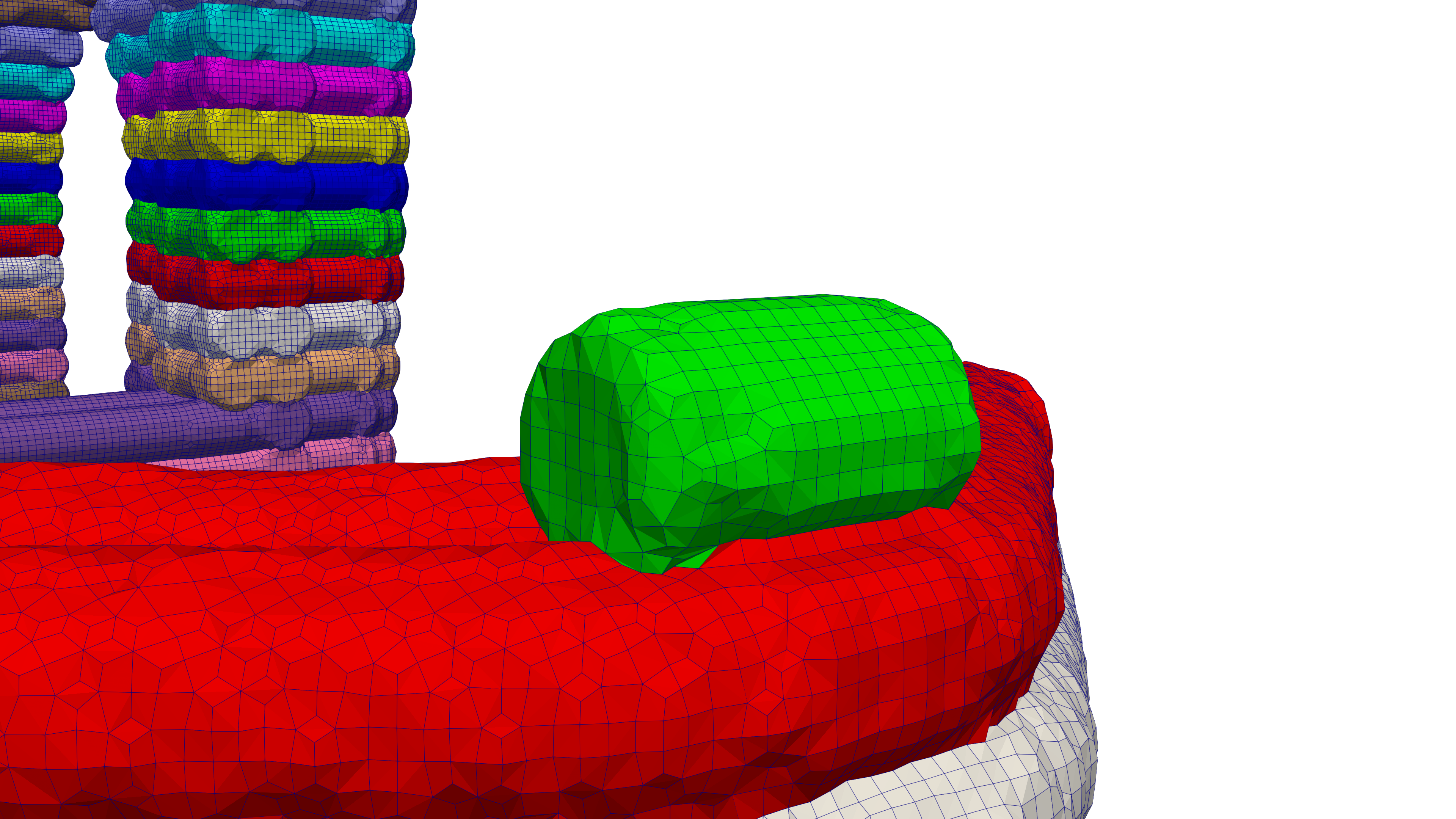}
        \caption{}
        \label{fig:courseMeshd}
    \end{subfigure}
    \caption{Examples of coarse and refined mesh examples, \ref{fig:courseMesha} varying print layer geometry of coarse mesh, \ref{fig:courseMeshb} varying print layer geometry of refined mesh, \ref{fig:courseMeshc} stand-alone print line of coarse mesh, \ref{fig:courseMeshd} stand-alone print line of refined mesh.}
    \label{fig:courseMesh}
\end{figure}

By decreasing the size of the elements, features of the extruded material are more clearly represented. For example, Figure~\ref{fig:courseMesh} shows a coarse mesh of the classic 3D printing model, Benchy, and a refined mesh. By reducing the element size, the fidelity of the mesh increases greatly, thus better representing the layer and print line interaction. Conversely, this will also increase element count and simulation time, highlighting the importance of sufficient mesh refinement in conjunction with execution time. The coarse meshed Benchy has a characteristic element size of 0.25 mm and is composed of 1,870,378 elements while the refined meshed Benchy has a characteristic element size of 0.1 mm and has 18,181,185 elements.

Finding the appropriate element size while maintaining necessary feature resolution and reasonable simulation time is key. The interaction of print lines becomes smoother and closer to the expected interaction between print layers as the element size decreases, however solution convergence of QOI's should be investigated to ensure that the element size isn't unnecessarily small.

\subsection{Direct Numerical Simulations}
\subsubsection{1D Compression}
The first exemplar application of our framework investigates the 1D compression response of soft AM lattices. We utilized Sierra/SM to conduct explicit dynamics calculations which solve the following partial differential equation,
\begin{equation}
    \rho\ddot{\bm{u}} = \nabla\cdot\bm{\sigma} + \rho\bm{b},
\end{equation}
where $\rho$ is the density, $\bm{u}$ is the displacement field, $\bm{\sigma}$ is the Cauchy stress, and $\bm{b}$ is a body force. An appropriate material model must be selected for the AM material of interest to close the above equation. For our examples, we utilized the Neo-Hookean model with the following specific model form,
\begin{equation}
    \bm{\sigma} = \frac{1}{2}K\left(J - \frac{1}{J}\right)\bm{I} + J^{-5/3}G\left[\bm{B} - \frac{1}{3}tr\left(\bm{B}\right)\bm{I}\right],
\label{eq:conservationOfMomentum}
\end{equation}
where $K$ is the bulk modulus, $J$ is the Jacobian $J = \det\bm{F}$ (where $\bm{F}$ is the deformation gradient), $\bm{I}$ is the second order identity tensor, and $\bm{B} = \bm{F}\bm{F}^T$ is the left Cauchy-Green Tensor, ~\cite{lame_2024}.

Several other simulation details were specifically chosen for robust evaluation. A uniform gradient hexahedral element formulation with hourglass control was utilized for reduced integration~\cite{Belytschko1981}. Additionally, the complex and often compact nature of an additively manufactured parts causes individual print lines to contact each other during loading. To account for this interaction, Coulomb friction was included when contact between nodes and faces occurred, ~\cite{sierra_sm_2024}. 

In the case of high fidelity FEA, complications with singular elements can cause the entirety of the simulation to slow or fail. To help mitigate this problem, element death was introduced. While there are many options to approach element death, we utilized two key criteria; element inversion and minimum critical time step. When an element is inverted, it is permanently removed from the analysis along with the associated mass from the corresponding nodes. Similarly, when elements become highly distorted, the explicit time step drops dramatically, therefore elements that cause the time step to drop below a specified limit are removed. 

\begin{figure}[H]
    \centering
    \begin{subfigure}{0.35\textwidth}
        \includegraphics[trim={700pt 0pt 700pt 0pt}, clip=true, width=\linewidth]{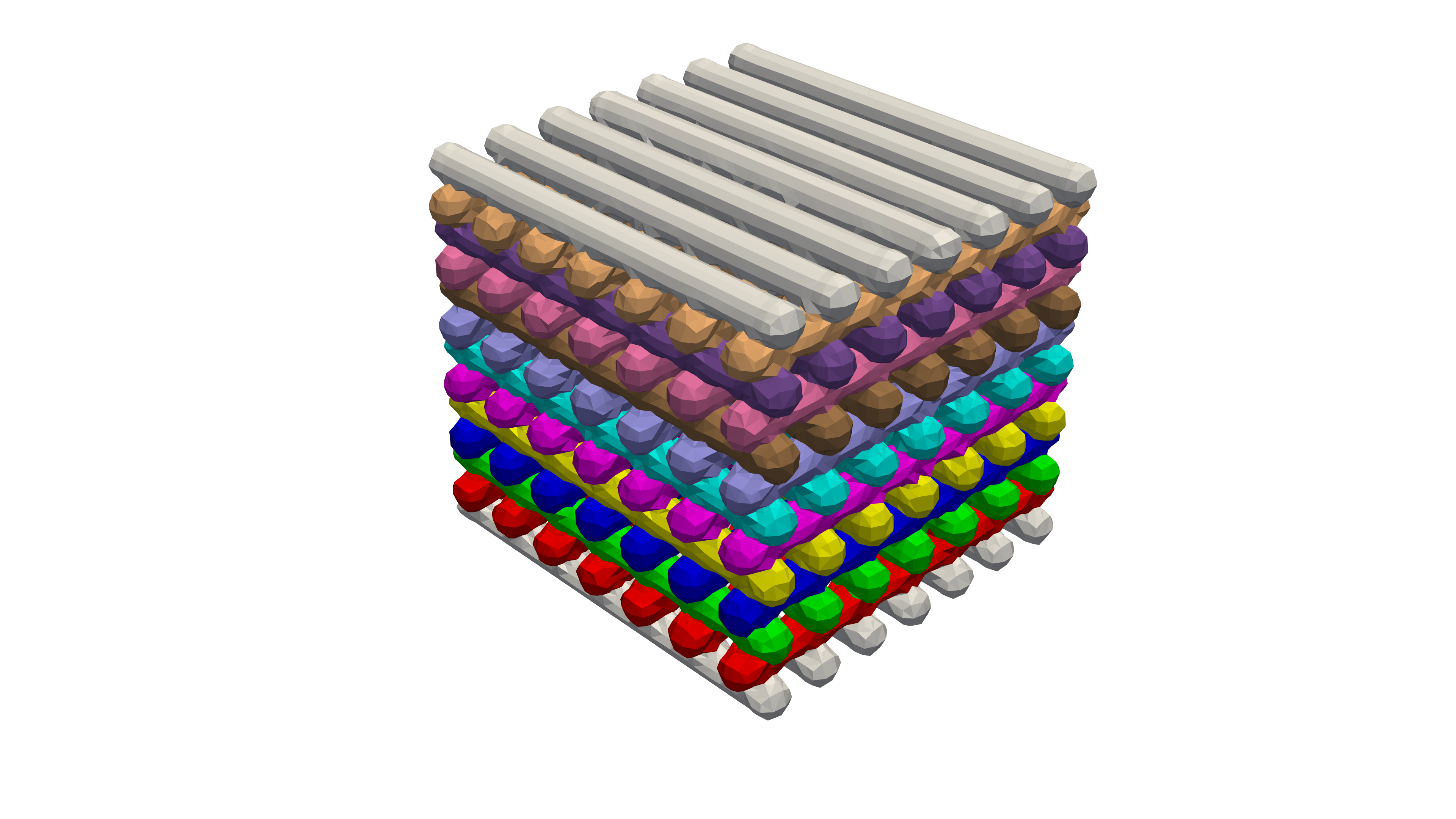} 
        \caption{}
        \label{fig:preloadExamplea}
    \end{subfigure}
    \begin{subfigure}{0.35\textwidth}
        \includegraphics[trim={700pt 0pt 700pt 0pt}, clip=true, width=\linewidth]{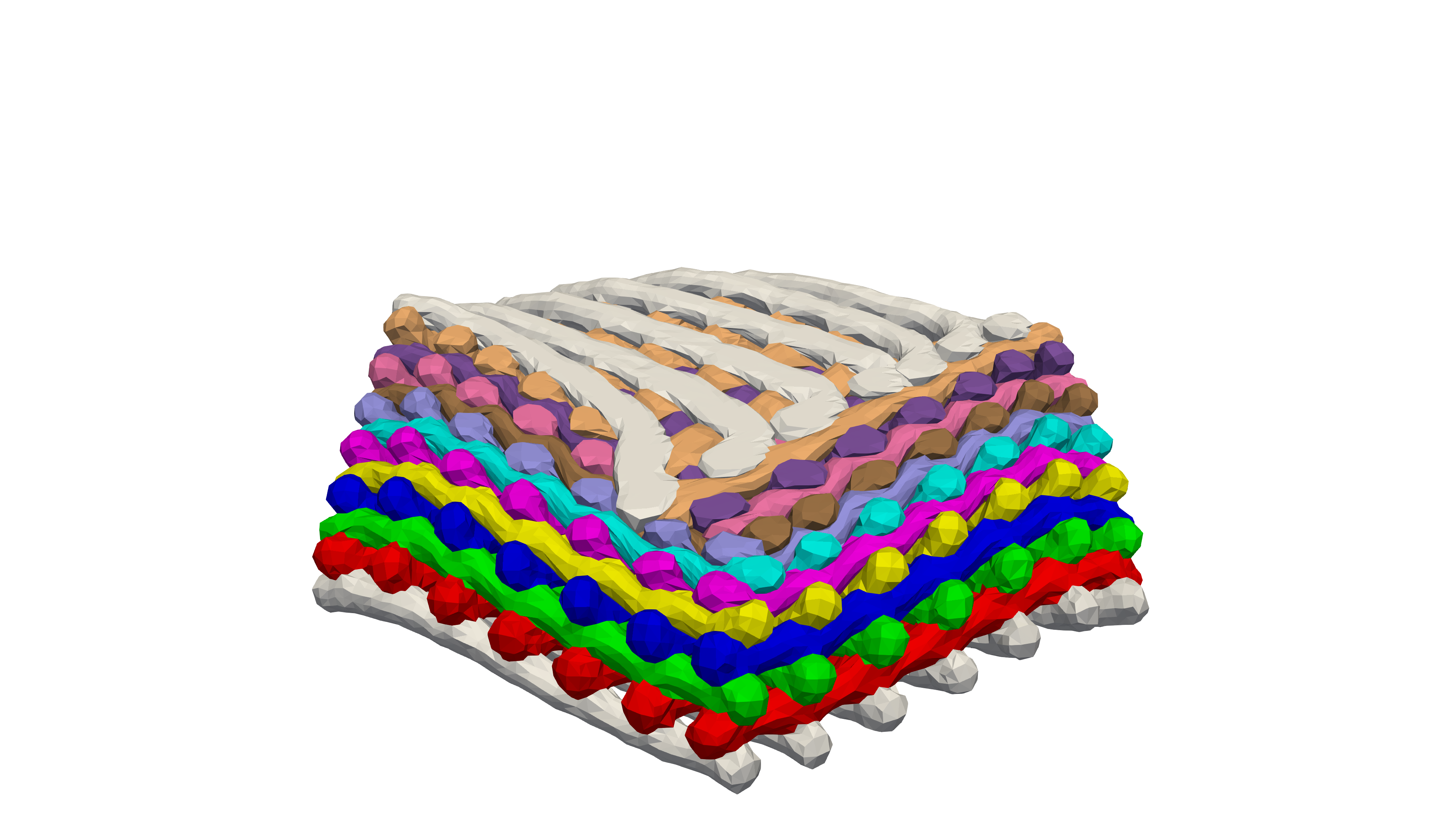}
        \caption{}
        \label{fig:preloadExampleb}
    \end{subfigure}
    \begin{subfigure}{0.35\textwidth}
        \includegraphics[trim={700pt 0pt 700pt 0pt}, clip=true, width=\linewidth]{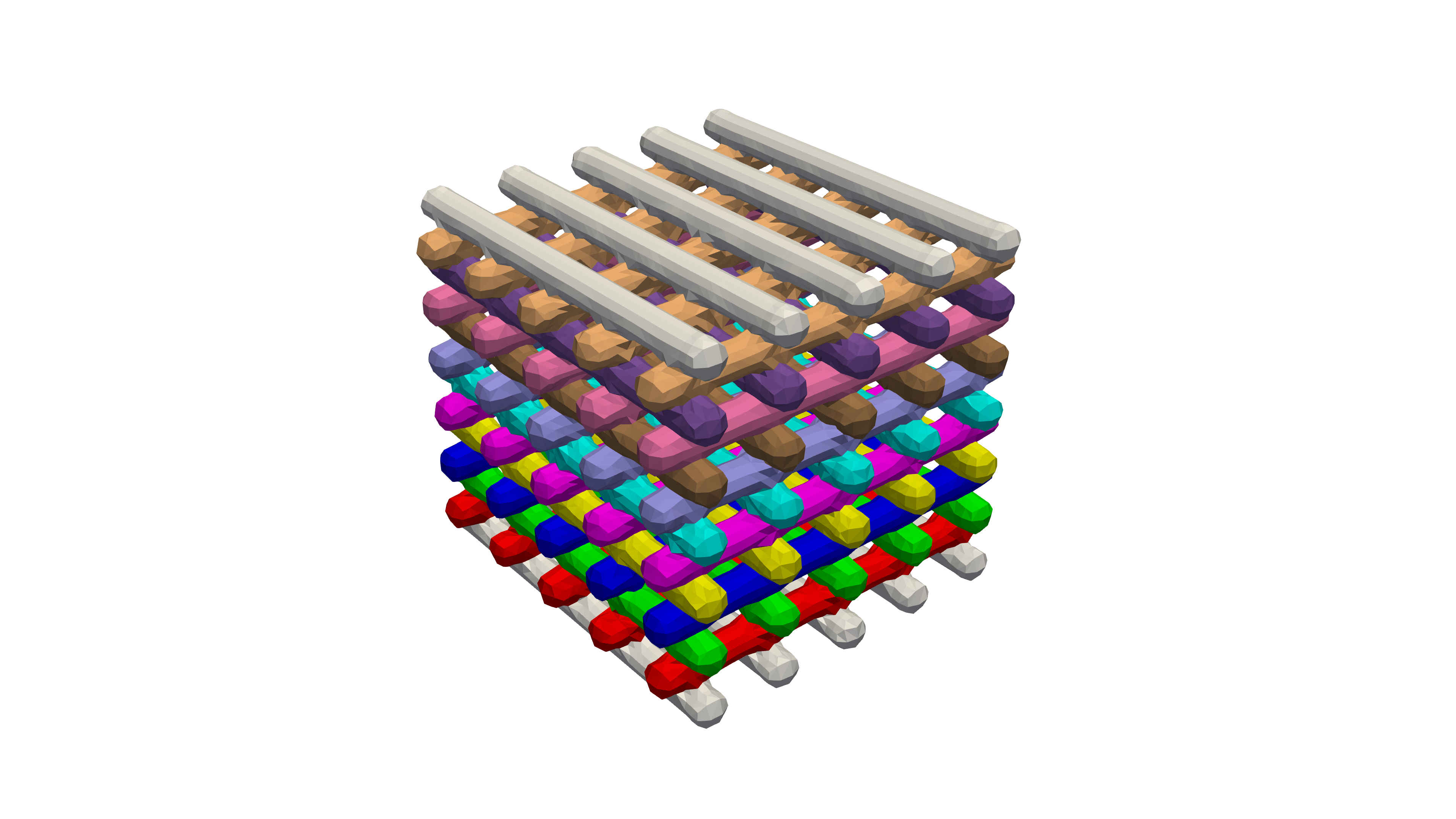}
        \caption{}
        \label{fig:preloadExamplec}
    \end{subfigure}
    \begin{subfigure}{0.35\textwidth}
        \includegraphics[trim={700pt 0pt 700pt 0pt}, clip=true, width=\linewidth]{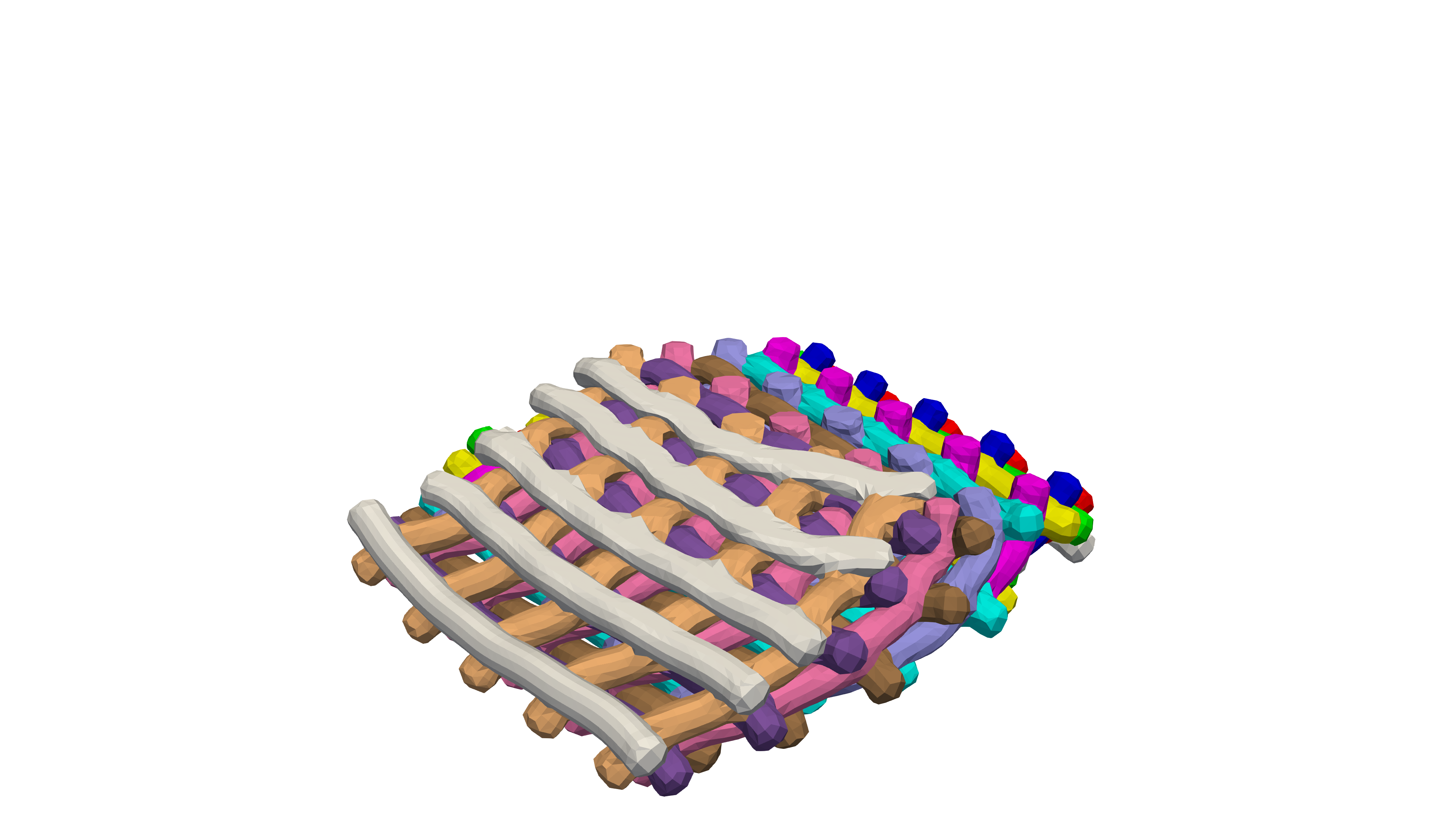}
        \caption{}
        \label{fig:preloadExampled}
    \end{subfigure}
    \begin{subfigure}{0.35\textwidth}
        \includegraphics[trim={700pt 0pt 700pt 0pt}, clip=true, width=\linewidth]{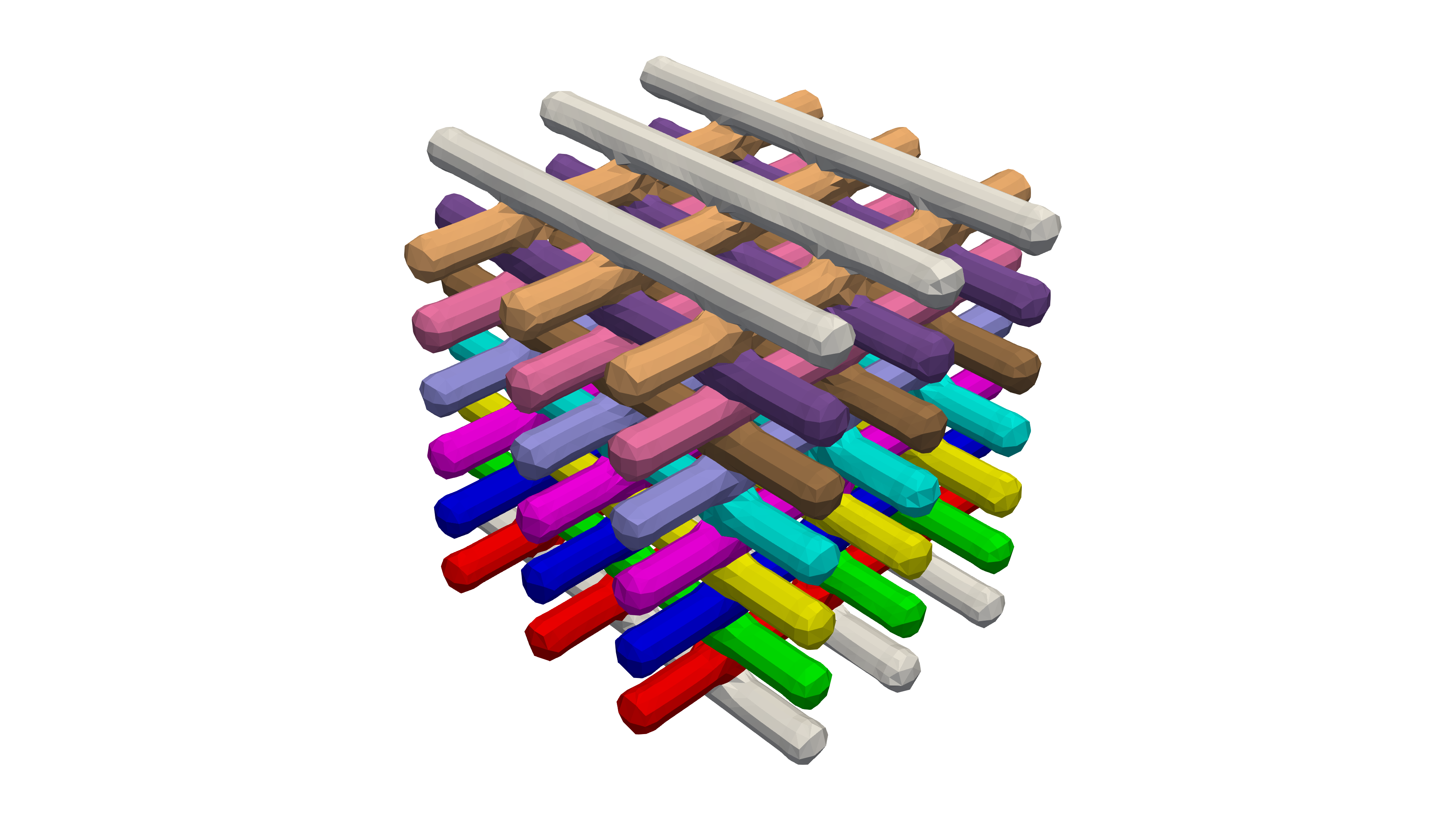}
        \caption{}
        \label{fig:preloadExamplee}
    \end{subfigure}
    \begin{subfigure}{0.35\textwidth}
        \includegraphics[trim={600pt 0pt 700pt 0pt}, clip=true, width=\linewidth]{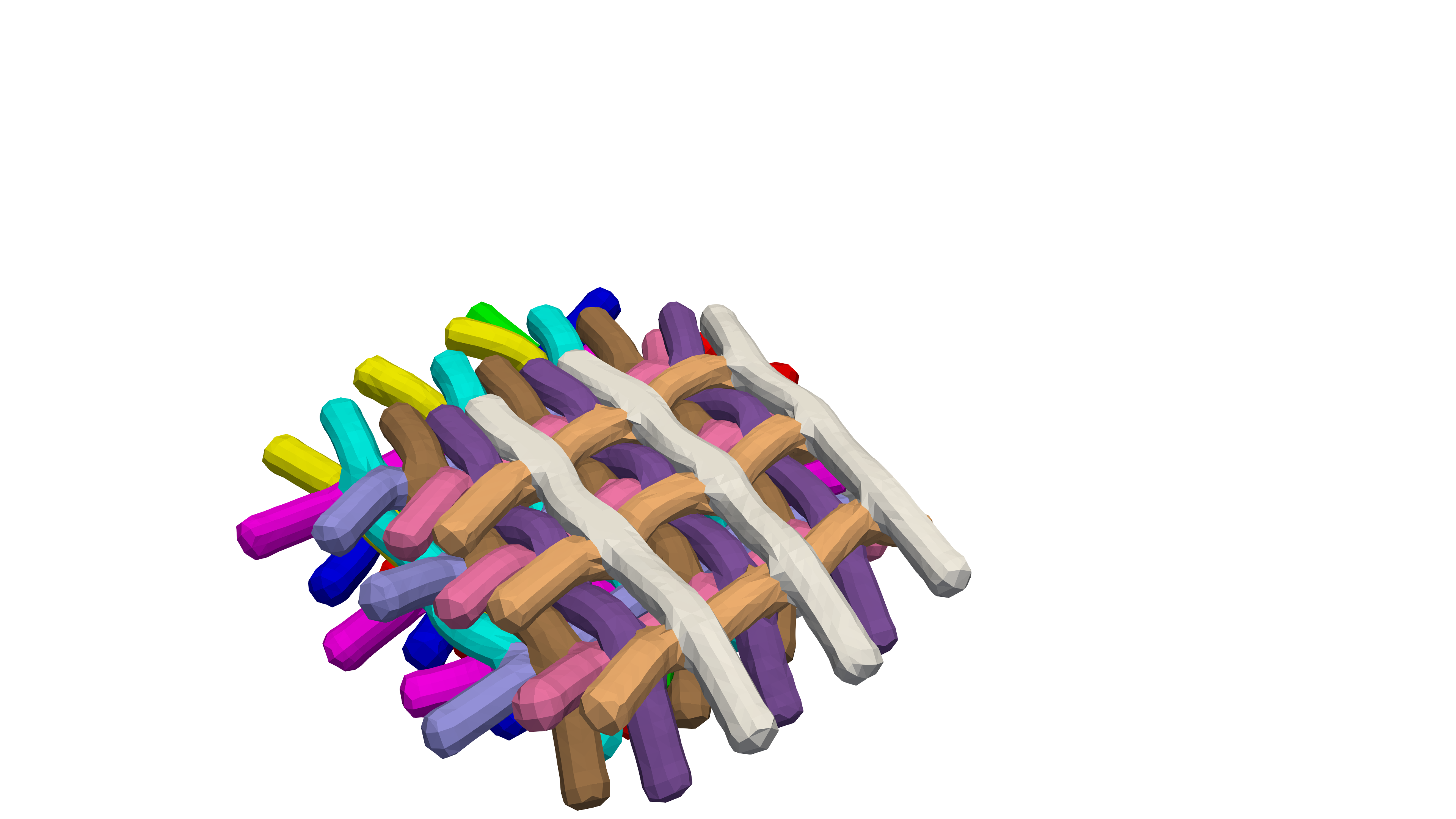}
        \caption{}
        \label{fig:preloadExamplef}
    \end{subfigure}
    \caption{Example 1D compression responses of high fidelity AM structure meshes with variable infill spacing between 0.0 and 0.65 compressive engineering strain, \ref{fig:preloadExamplea} 0.75 infill spacing, \ref{fig:preloadExampleb} 0.75 infill spacing, \ref{fig:preloadExamplec} 1.5 infill spacing, \ref{fig:preloadExampled} 1.5 infill spacing, \ref{fig:preloadExamplee} 2.25 infill spacing, \ref{fig:preloadExamplef} 2.25 infill spacing.}
    \label{fig:preloadExample}
\end{figure}

One of the simplest examples of the solid mechanics capabilities of the proposed framework is 1D compression loading; in this problem, a model is compressed between two platens along a single axis. A single, 5 mm cube subjected to 1D compressive loading was evaluated with our our framework, Figure~\ref{fig:preloadExample}, at varying infill spacing. The cube was subjected to a maximum compressive engineering strain of 0.65 in monotonic loading. 

\begin{figure}[H]
    \centering
    \begin{subfigure}{0.48\textwidth}
        \includegraphics[trim={15pt 8pt 10pt 40pt}, clip=true, width=\linewidth]{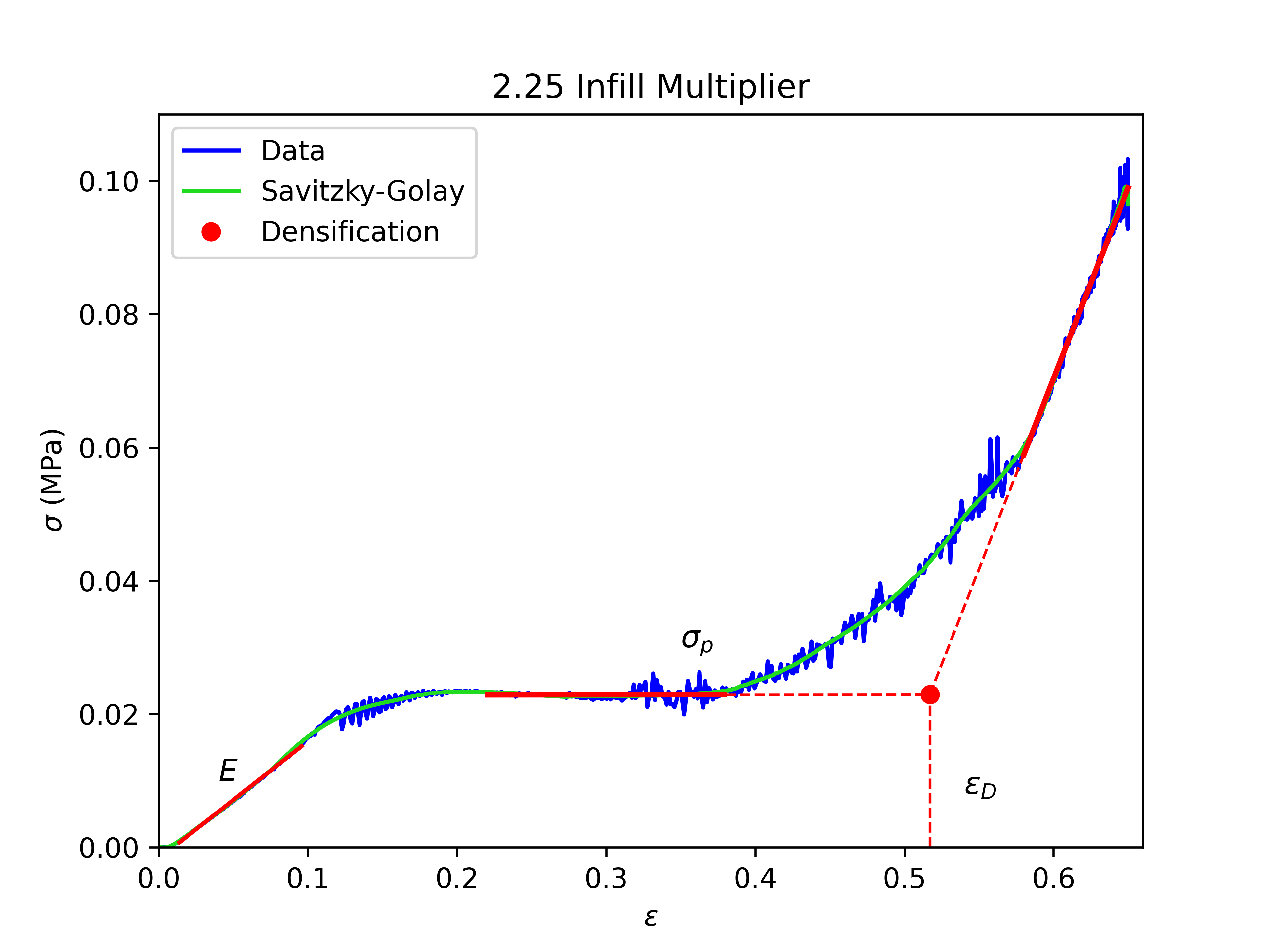} 
        \caption{}
        \label{fig:cubeDataa}
    \end{subfigure}
    \begin{subfigure}{0.48\textwidth}
        \includegraphics[trim={15pt 8pt 10pt 40pt}, clip=true, width=\linewidth]{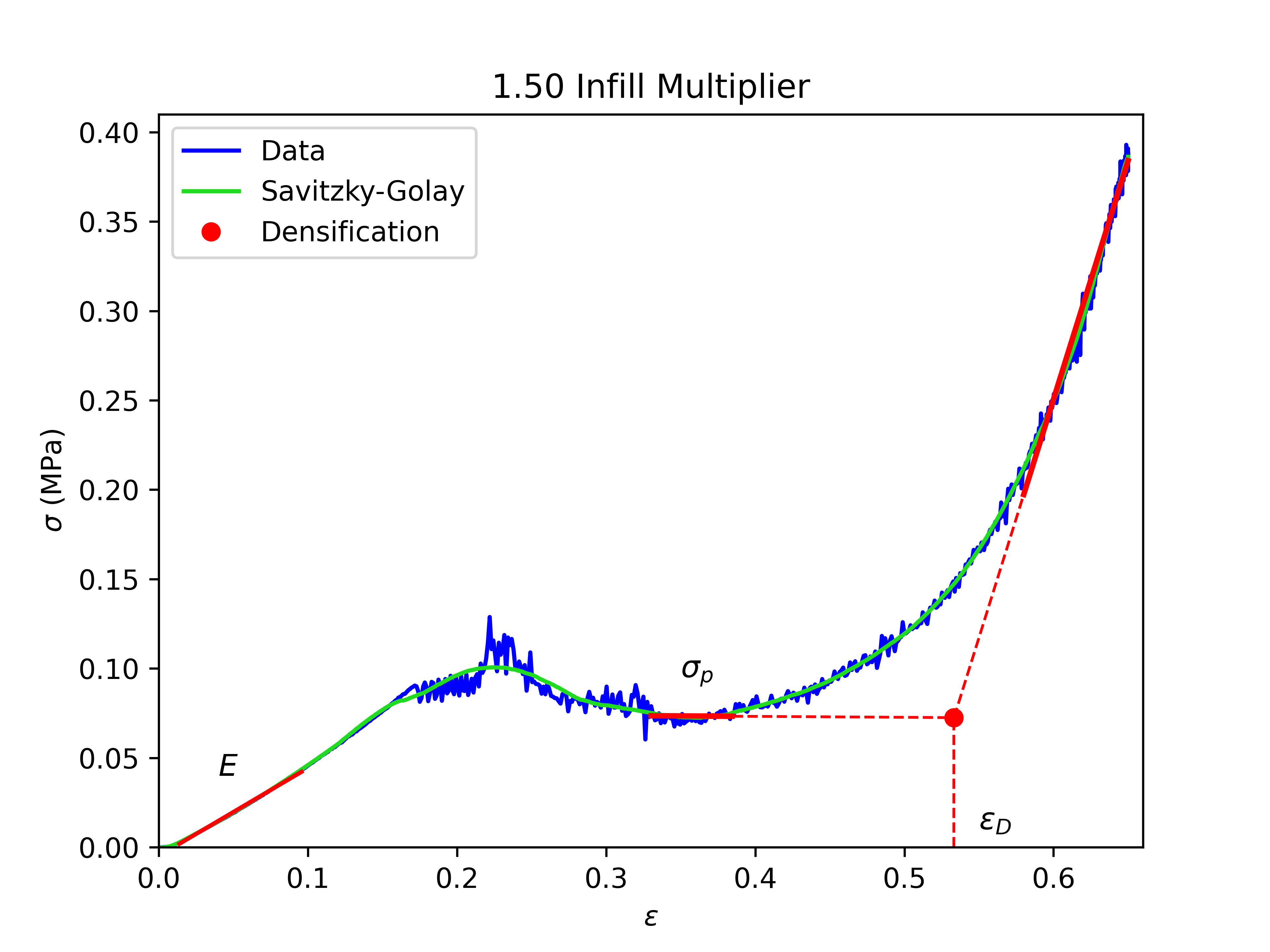} 
        \caption{}
        \label{fig:cubeDatab}
    \end{subfigure}
    \begin{subfigure}{0.48\textwidth}
        \includegraphics[trim={15pt 8pt 10pt 40pt}, clip=true, width=\linewidth]{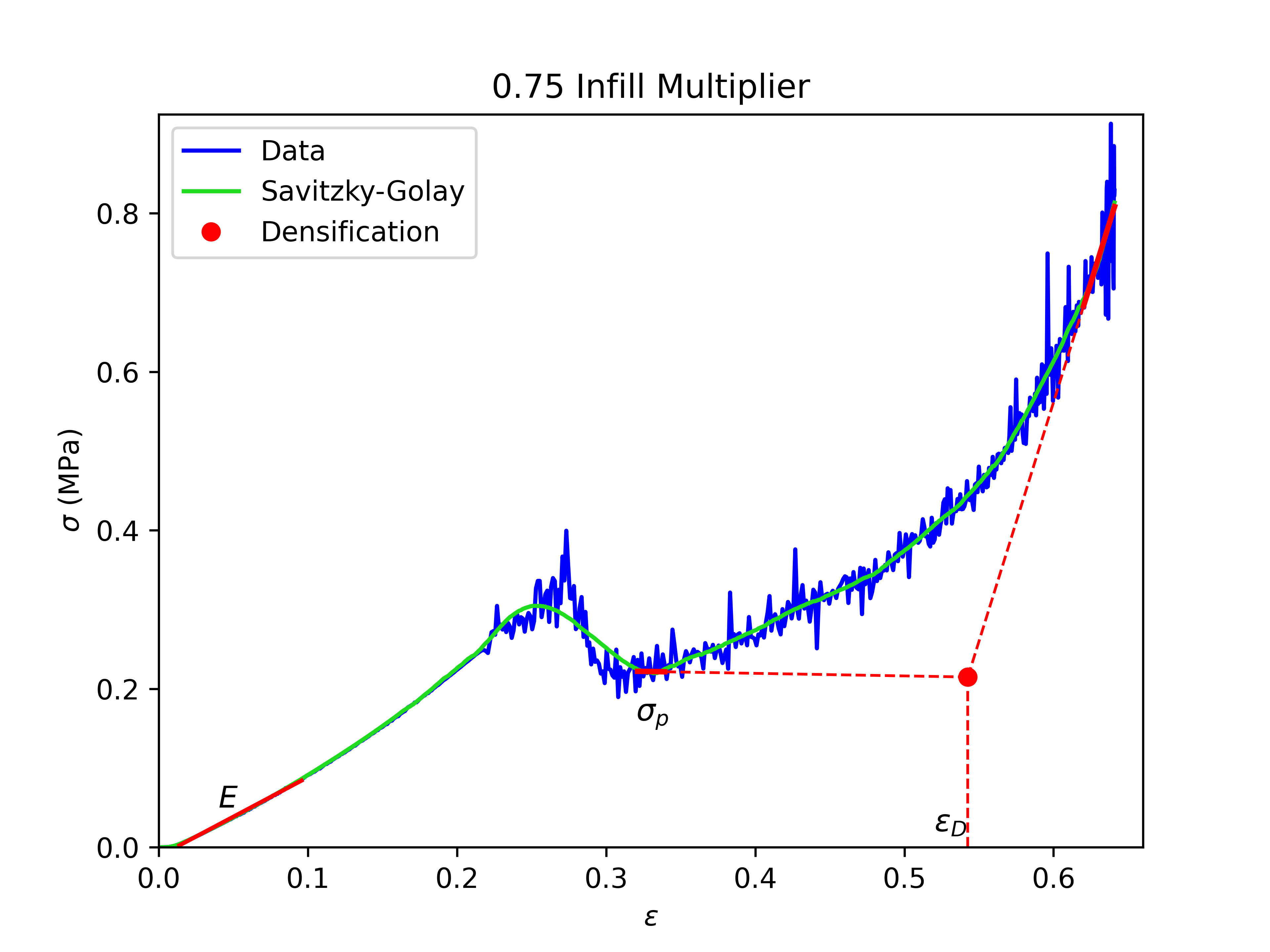}
        \caption{}
        \label{fig:cubeDatac}
    \end{subfigure}
    \caption{Uniaxial compression of 5 mm cube with varying infill spacing, explicit simulation data smoothed with Savitzky-Golay filter~\cite{Savitzky}; \ref{fig:cubeDataa} 0.923 mm spacing, \ref{fig:cubeDatab} 0.615 mm spacing, \ref{fig:cubeDatac} 0.308 mm spacing.}
    \label{fig:cubeData}
\end{figure}

Stress-strain curves were produced from the responses of the cubes. Overall, the curves exhibit the same compressive behavior; a linear region, followed by a buckling transition, then a plateau stress region where stress was relatively constant for a given range of strain, and finally a densification region where the cubes began to behave similar to a homogeneous, non-lattice structured model due to widespread self contact of the lattice struts. Figure~\ref{fig:cubeData} shows examples of these points on the stress-strain curve for the cube with varying infill spacing, where $E$ is the elastic modulus, $\sigma_{p}$ is the plateau stress, and $\varepsilon_{D}$ is the densification strain. The example cubes were fixed between two platens, with the top platen linearly lowering to 0.65 engineering strain.

Firstly, the Elastic Modulus was determined, 
\begin{equation}
    \bm{E} = \bm{\sigma_{E}} / \bm{\varepsilon_{E}},
\end{equation}
where $E$ is the Elastic Modulus, \bm{$\sigma_{E}$} and \bm{$\varepsilon_{E}$} are the engineering stress and engineering strain in the elastic region, respectively. 

The evaluated elastic moduli for the AM structures show in Figure~\ref{fig:preloadExample} are summarized in Table~\ref{ElasticModulus}.
\begin{table}[H]
\centering
\begin{tabular}{ | c | c | }
    \hline
    Infill Spacing & Elastic Modulus (MPa) \\
    \hline
    \hline
    0.75 & 0.99 \\
    \hline
    1.5 & 0.49 \\
    \hline
    2.25 & 0.17 \\
    \hline
\end{tabular}
\caption{Elastic modulus of the 5 mm cube with variable infill spacing.}
\label{ElasticModulus}
\end{table}

Next the plateau stress was calculated as the average stress in the region where the slope of the stress-strain curve is approximately zero. A first-order curve was fitted to this region to determine the error, similar to the Elastic Modulus. The plateau stress was calculated for all three cubes and are summarized in Table ~\ref{Plateau}.

\begin{table}[H]
\centering
\begin{tabular}{ | c | c | }
    \hline
    Infill Spacing & Plateau Stress (MPa)\\
    \hline
    \hline
    0.75 & 0.228 \\
    \hline
    1.5 & 0.072 \\
    \hline
    2.25 & 0.023 \\
    \hline
\end{tabular}
\caption{Plateau stress and mean residual difference of 5 mm cube with variable infill spacing.}
\label{Plateau}
\end{table}

Lastly, the densification strain was calculated, which we define here as the intersection of the densification region and the plateau stress. 

\begin{table}[H]
\centering
\begin{tabular}{ | c | c | }
    \hline
    Infill Spacing & Densification strain \\
    \hline
    \hline
    0.75 & 0.543 \\
    \hline
    1.5 & 0.533 \\
    \hline
    2.25 & 0.517 \\
    \hline
\end{tabular}
\caption{Densification strain and mean residual difference of 5 mm cube with variable infill spacing.}
\label{Densification}
\end{table}

\begin{figure}[H]
    \centering
    \begin{subfigure}{0.45\textwidth}
        \includegraphics[trim={600pt 0pt 600pt 0pt}, clip=true, width=\linewidth]{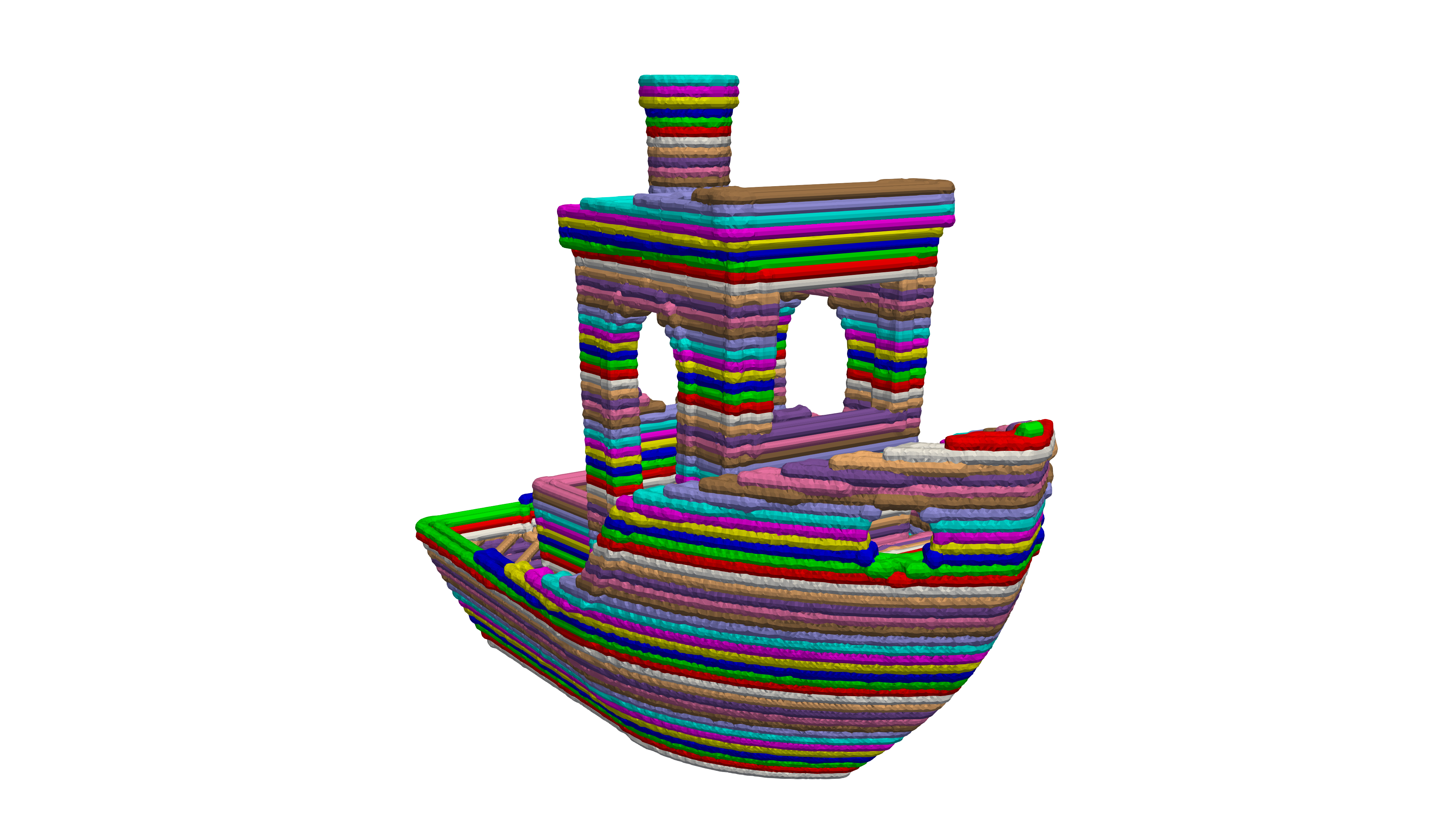} 
        \caption{}
        \label{fig:CrushBenchya}
    \end{subfigure}
    \begin{subfigure}{0.45\textwidth}
        \includegraphics[trim={600pt 0pt 600pt 0pt}, clip=true, width=\linewidth]{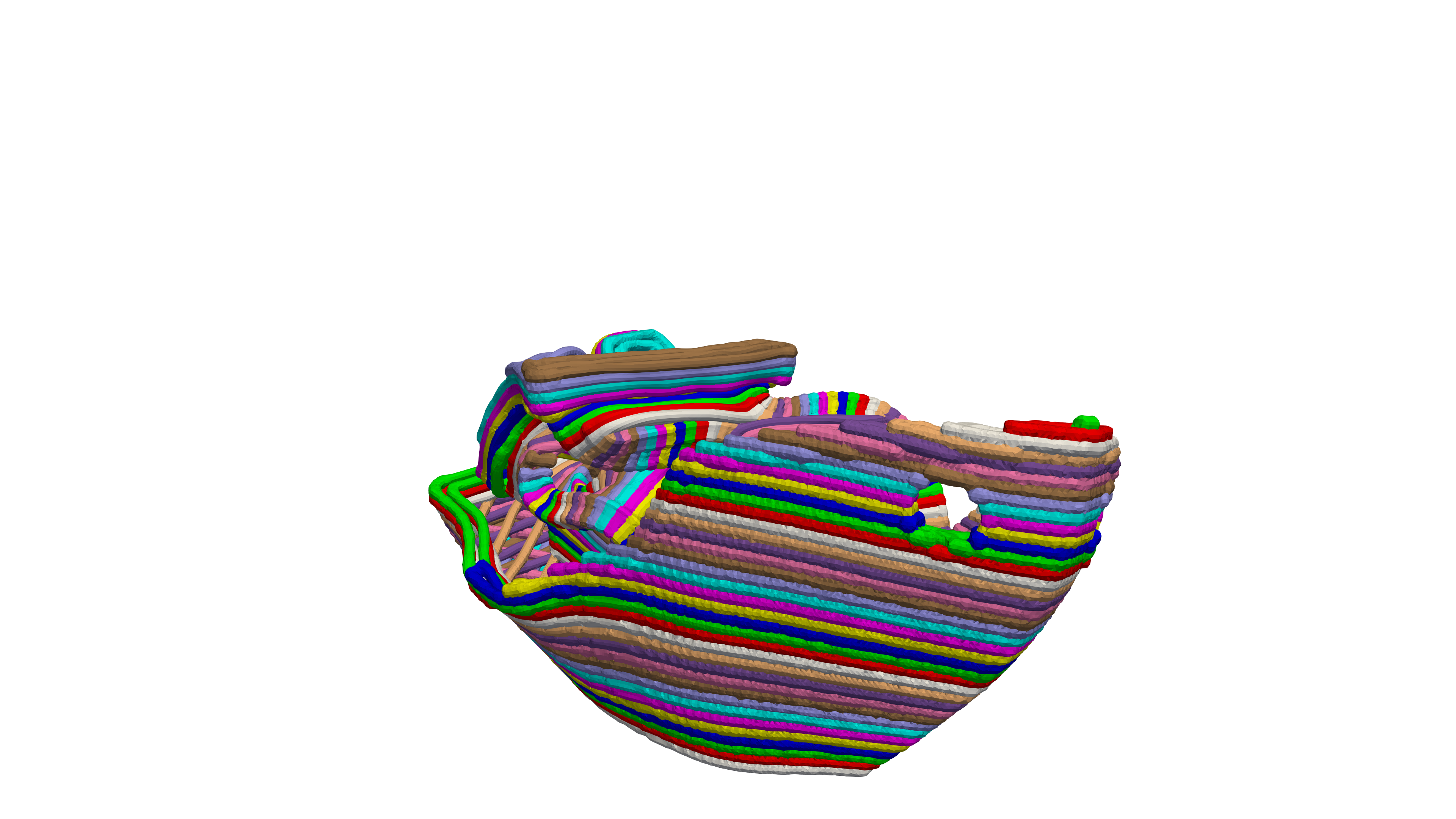}
        \caption{}
        \label{fig:CrushBenchyb}
    \end{subfigure}
    \caption{Example of a large, complex model under large, 1D compressive loading; \ref{fig:CrushBenchya} pre-loading, \ref{fig:CrushBenchyb} post-loading.}
    \label{fig:CrushBenchy}
\end{figure}

The example cubes show the relationship between the infill spacing and the mechanical properties. There is a consistent negative trend in elastic modulus, plateau stress, and densification strain as the infill spacing increases. Since higher infill spacings results in fewer toolpaths, it is expected that the magnitude of certain mechanical properties decreases, as seen in Tables \ref{ElasticModulus}, \ref{Plateau}, \ref{Densification}. Similar studies can be applied to more complicated and larger structures, as shown in Figure~\ref{fig:CrushBenchy}.

\subsubsection{Eigen Solutions}\label{subsectionEigenSolutions}
\begin{figure}[H]
    \centering
    \begin{subfigure}{0.25\textwidth}
        \includegraphics[width=\linewidth]{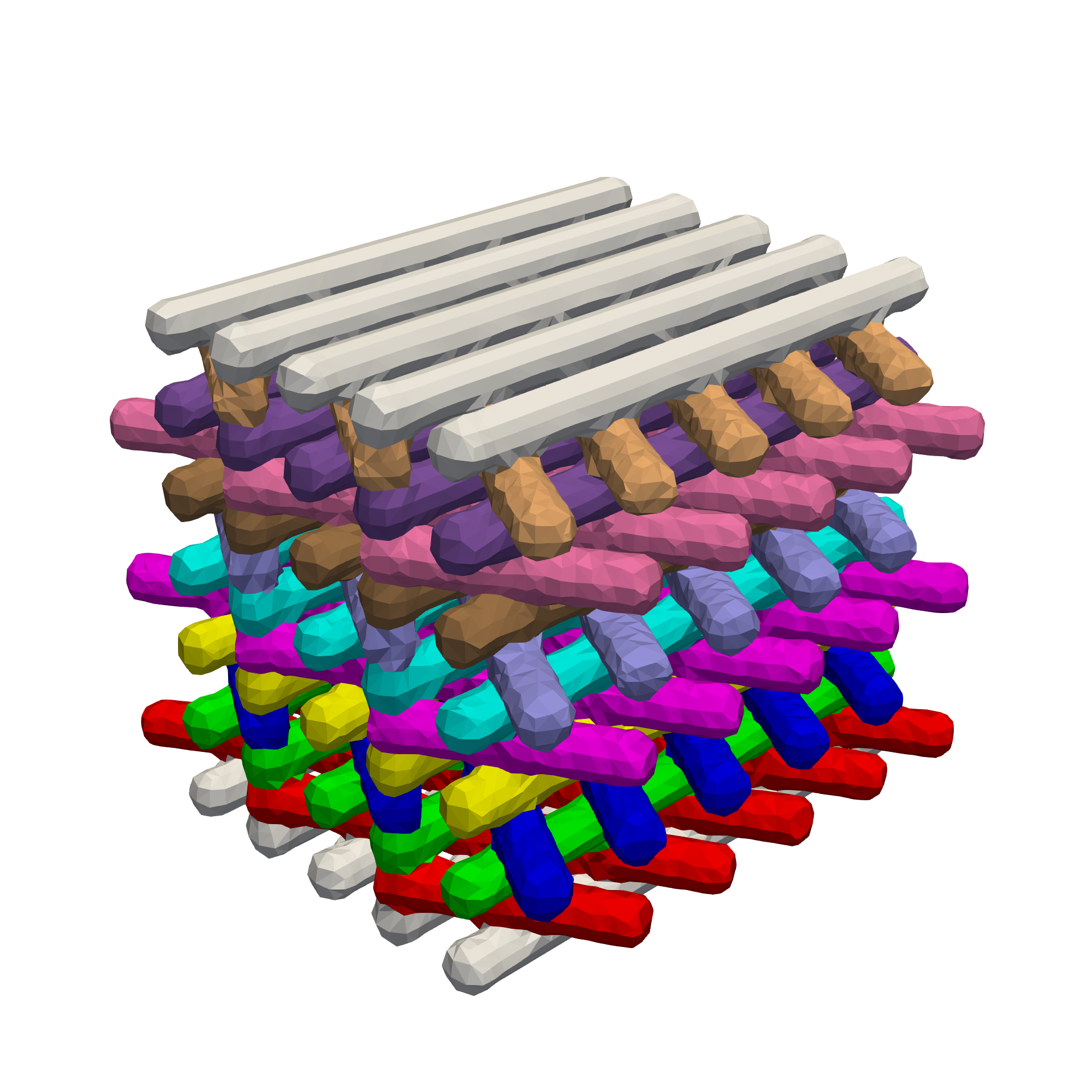} 
        \caption{}
        \label{fig:eigenExamplea}
    \end{subfigure}
    \begin{subfigure}{0.25\textwidth}
        \includegraphics[width=\linewidth]{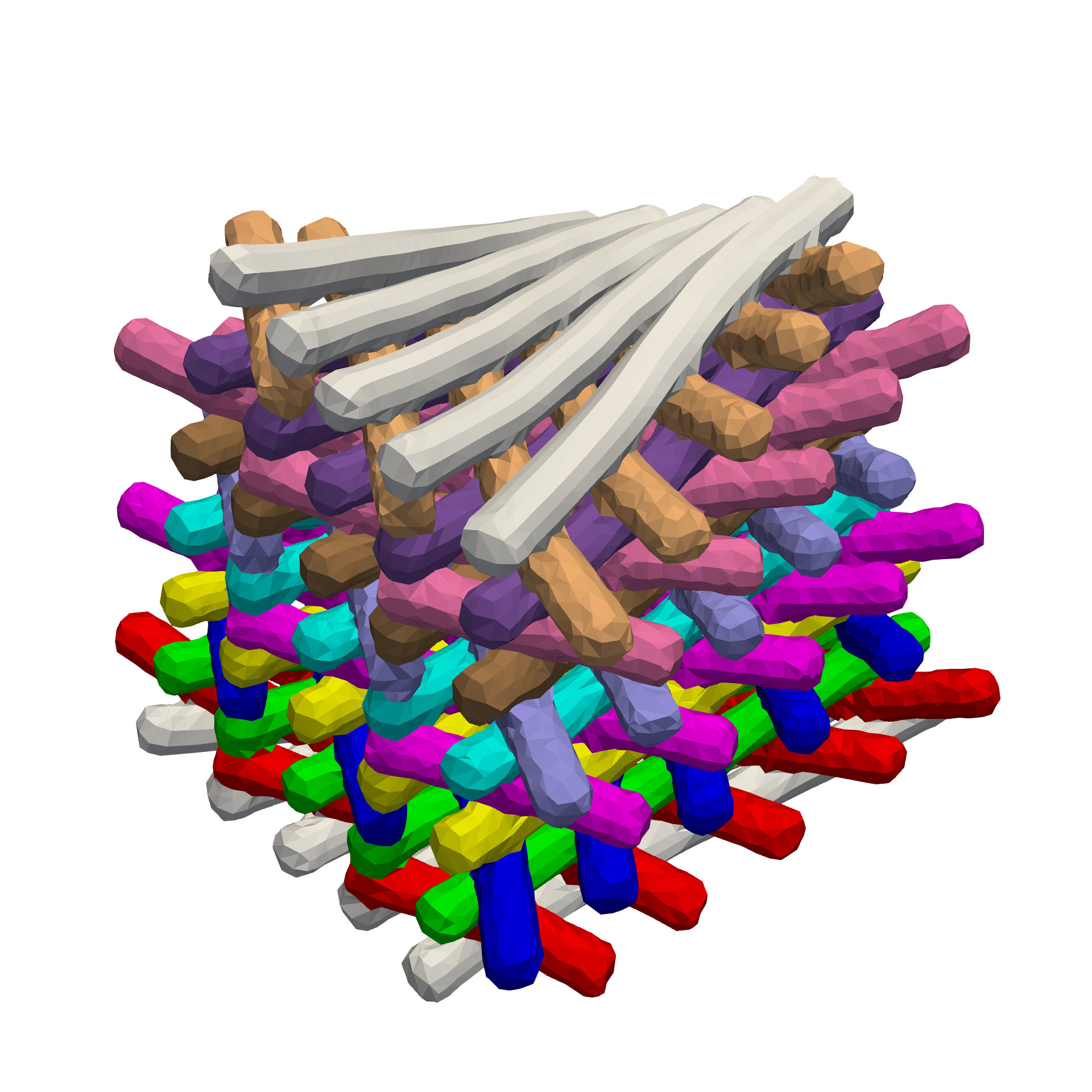}
        \caption{}
        \label{fig:eigenExampleb}
    \end{subfigure}
    \begin{subfigure}{0.25\textwidth}
        \includegraphics[width=\linewidth]{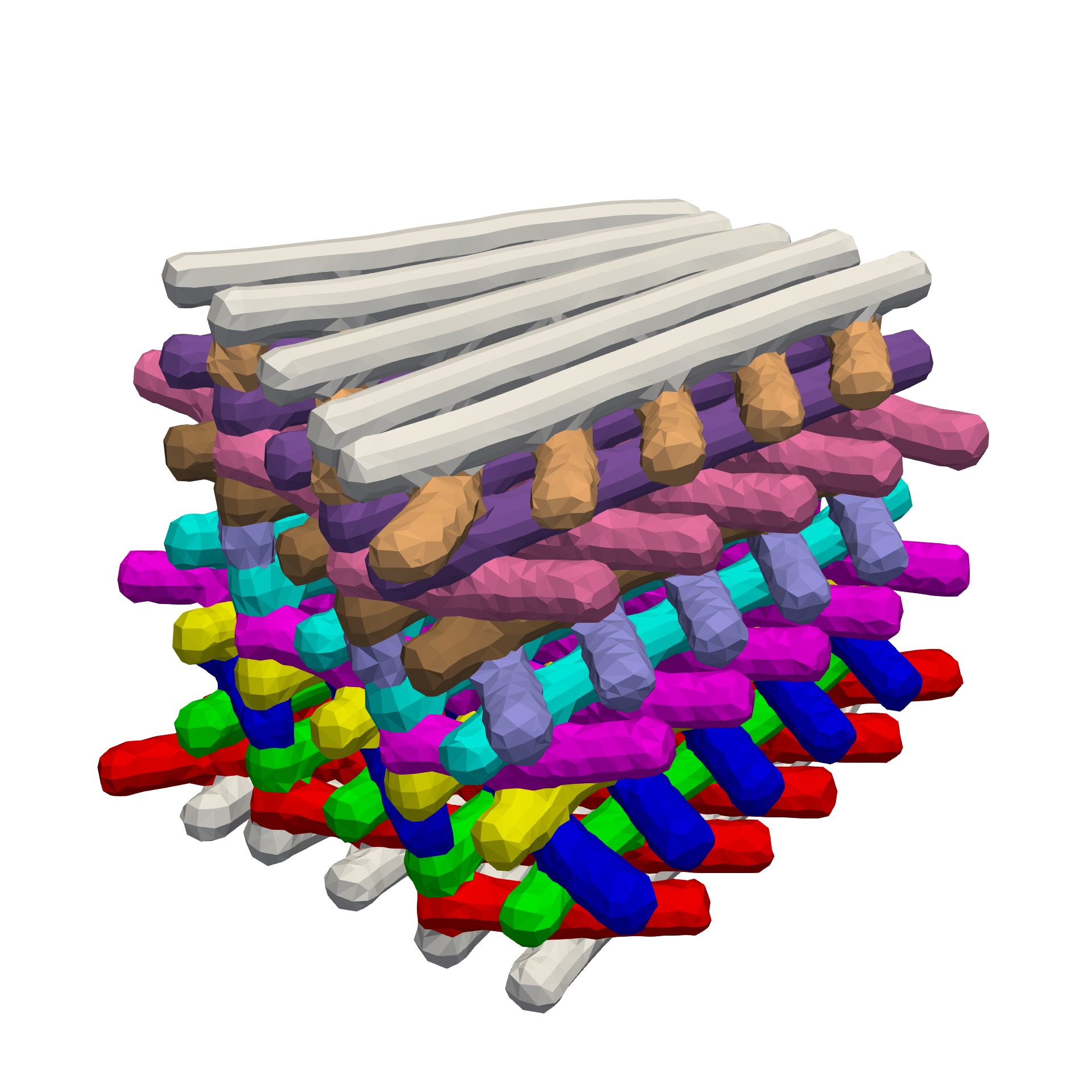}
        \caption{}
        \label{fig:eigenExamplec}
    \end{subfigure}
    \begin{subfigure}{0.25\textwidth}
        \includegraphics[width=\linewidth]{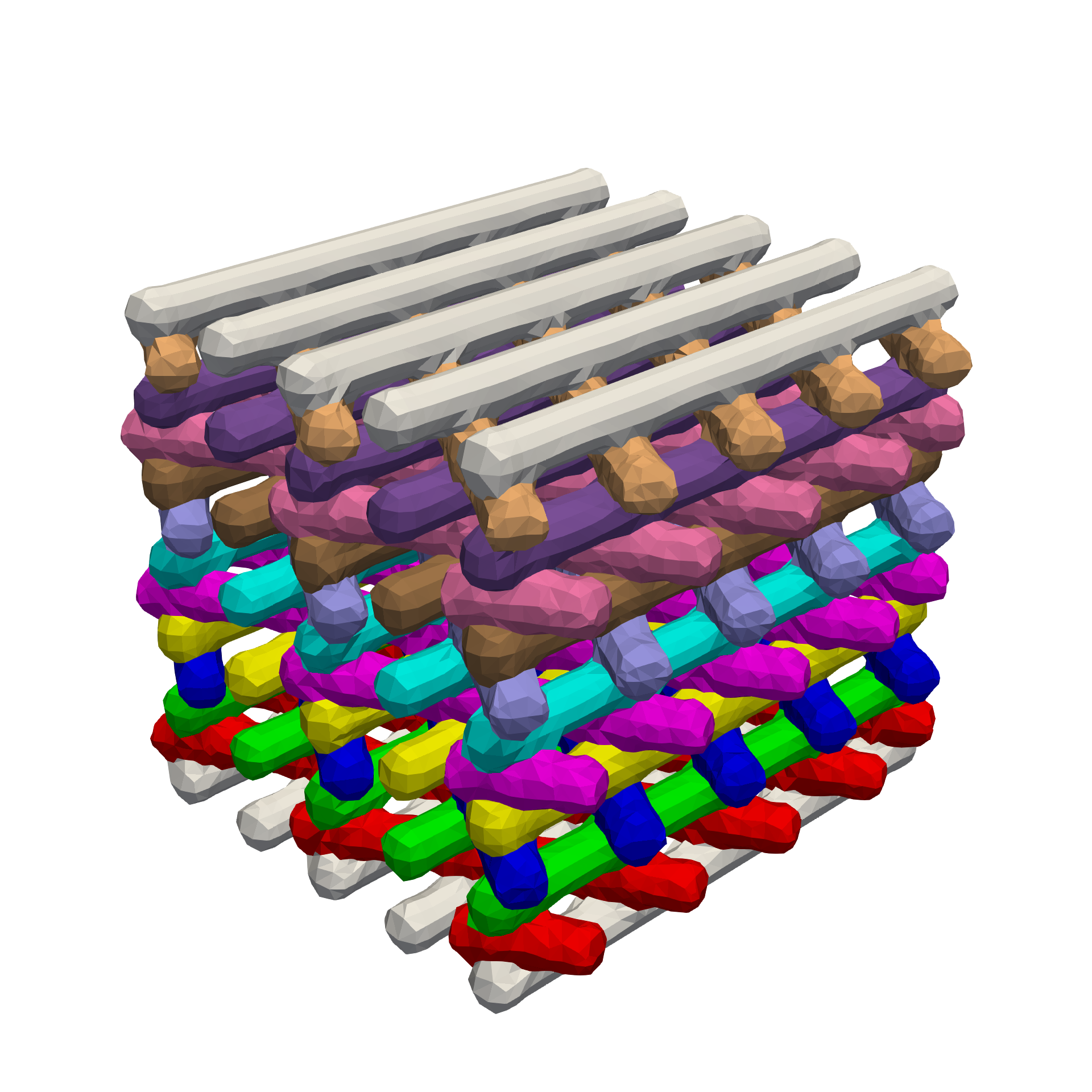}
        \caption{}
        \label{fig:eigenExampled}
    \end{subfigure}
    \begin{subfigure}{0.25\textwidth}
        \includegraphics[width=\linewidth]{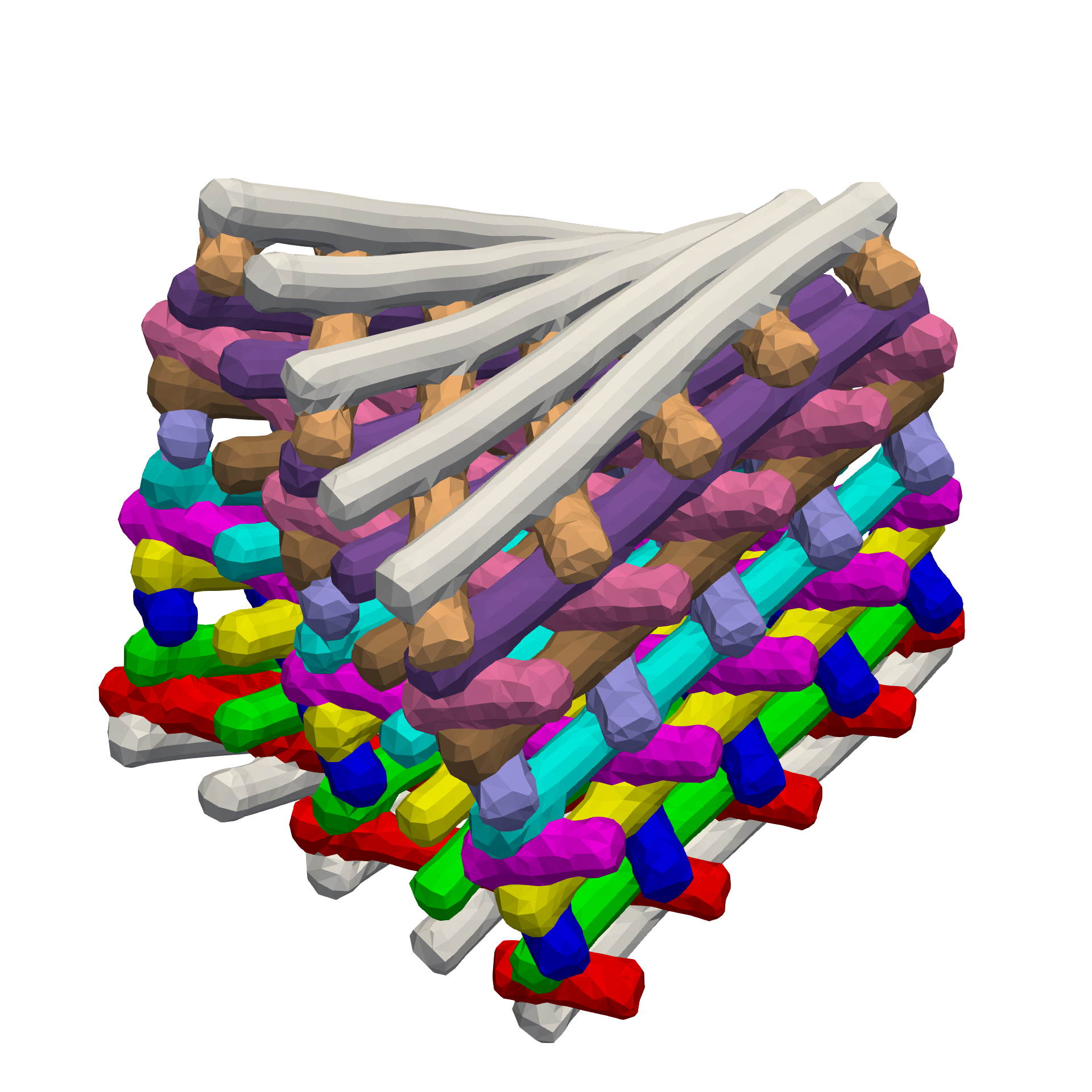}
        \caption{}
        \label{fig:eigenExamplee}
    \end{subfigure}
    \begin{subfigure}{0.25\textwidth}
        \includegraphics[width=\linewidth]{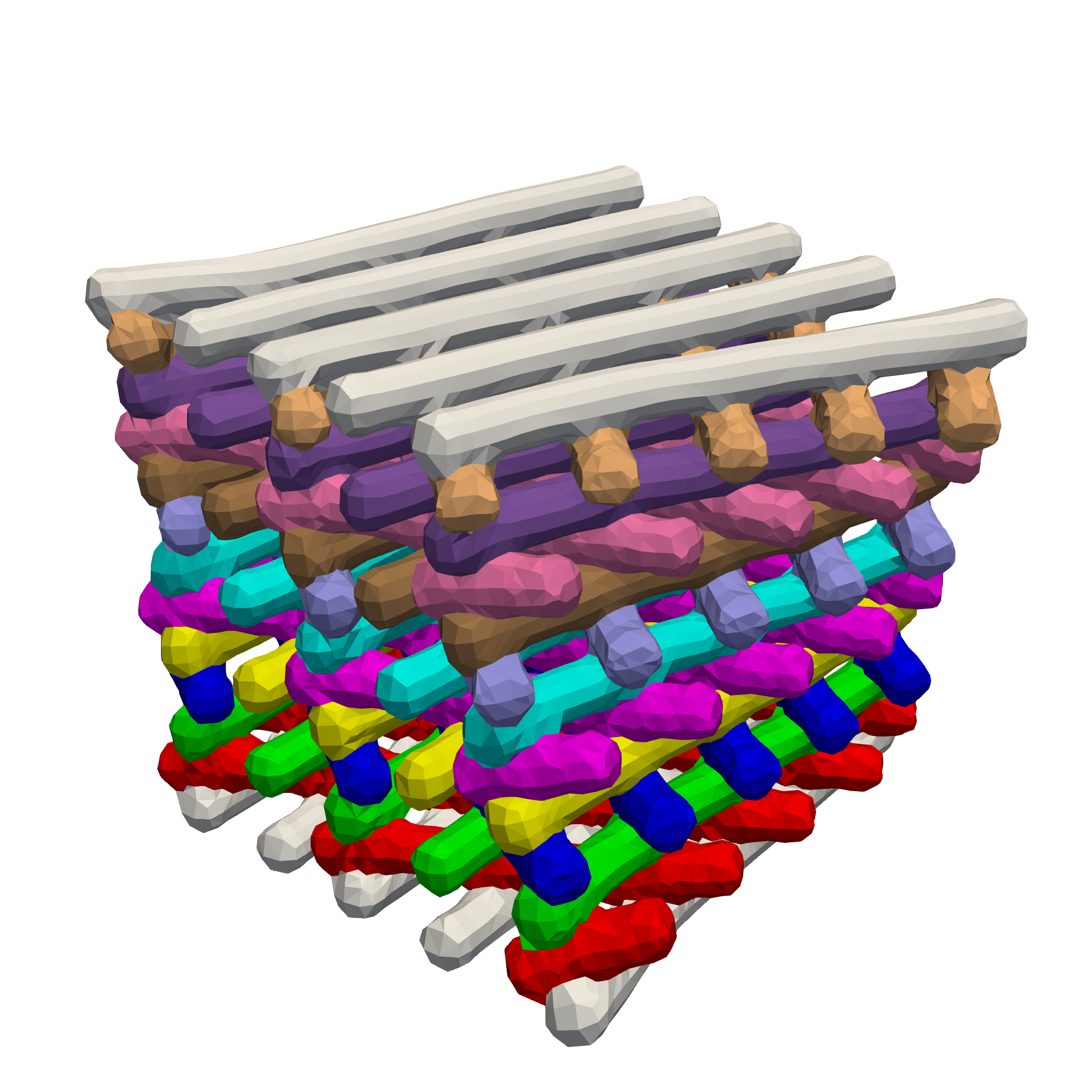}
        \caption{}
        \label{fig:eigenExamplef}
    \end{subfigure}
    \caption{Example eigenmodes of high fidelity AM structure mesh with variable infill spacing and angle, \ref{fig:eigenExamplea} 1.00 infill spacing, 60.0 degree angle, \ref{fig:eigenExampleb} first eigenmode 62.6 Hz, \ref{fig:eigenExamplec} second eigenmode 72.2 Hz, \ref{fig:eigenExampled} 1.35 infill spacing, 60.0 degree angle, \ref{fig:eigenExamplee} first eigenmode 53.0, \ref{fig:eigenExamplef} second eigenmode 63.4.}
    \label{fig:eigenExample}
\end{figure}

The proposed framework is also applicable for solving for the eigenvalues of the following linearized form of Equation~\ref{eq:conservationOfMomentum},
\begin{equation}
    \bm{M\ddot{x} + Kx = f}\left(t\right),
\end{equation}
where \bm{$M$} is the mass matrix, \bm{$K$} is the stiffness matrix, and \bm{$f$}$\left(t\right)$ is a forcing function. Understanding the eigenfrequencies and eigenmodes of an AM structure is essential for systems that experience a wide range of vibrational loading. While the magnitude of the responses are impacted by both the material and the structure of an AM component, this work focuses on the changes due to print structure. As an example, the first and second eigenmodes of a 5 mm cube with varied infill spacing were calculated with results shown in Figure~\ref{fig:eigenExample}. Given the same infill angle, 60 degrees, increasing the infill spacing from 1.00 to 1.35 causes the first eigenfrequency to decrease from 62.6 Hz to 52.9 Hz and the second to decrease from 72.2 Hz to 64.9 Hz suggesting that a lower density lattice structure will result in a lower eigenfrequency. By changing multiple printing parameters, the design space for eigenmodes and frequencies opens up significantly. 

\begin{figure}[H]
\centering
\includegraphics[scale=0.75]{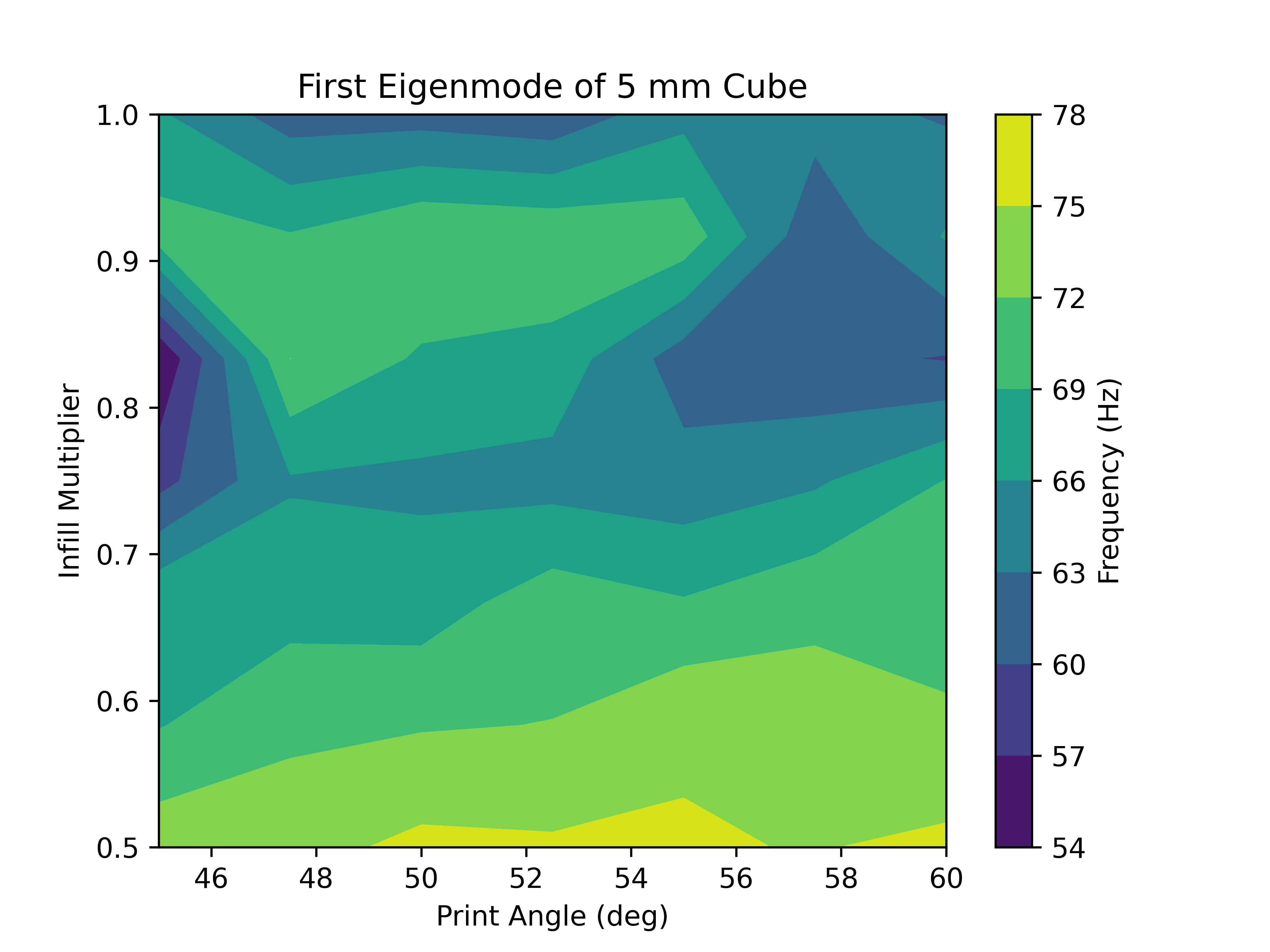} 
\caption{Design chart for the first eigenmode of a 5 mm cube with respect to infill spacing and print angle. In general, an increase in infill spacing decreases the frequency, though localized regions exist that cause an increase in frequency.}
\label{fig:EigenCountour}
\vspace{0cm}
\end{figure}

An example design space, shown in Figure~\ref{fig:EigenCountour}, shows the relationship between the first non-rigid eigenfrequency and the infill spacing and angle. In this case, the first mode is more sensitive to the infill spacing than the print angle. It should be noted that a minimization of the eigenmode occurs when the infill spacing is increased suggesting again that denser objects have higher eigenmodes than their less dense counterparts for the given property ranges. A design space can be readily made using the proposed framework, varying desired toolpath generation parameters by any range of viable values and further streamlining the PSPP linkage. 

\begin{figure}[H]
    \centering
    \begin{subfigure}{0.35\textwidth}
        \includegraphics[width=\linewidth]{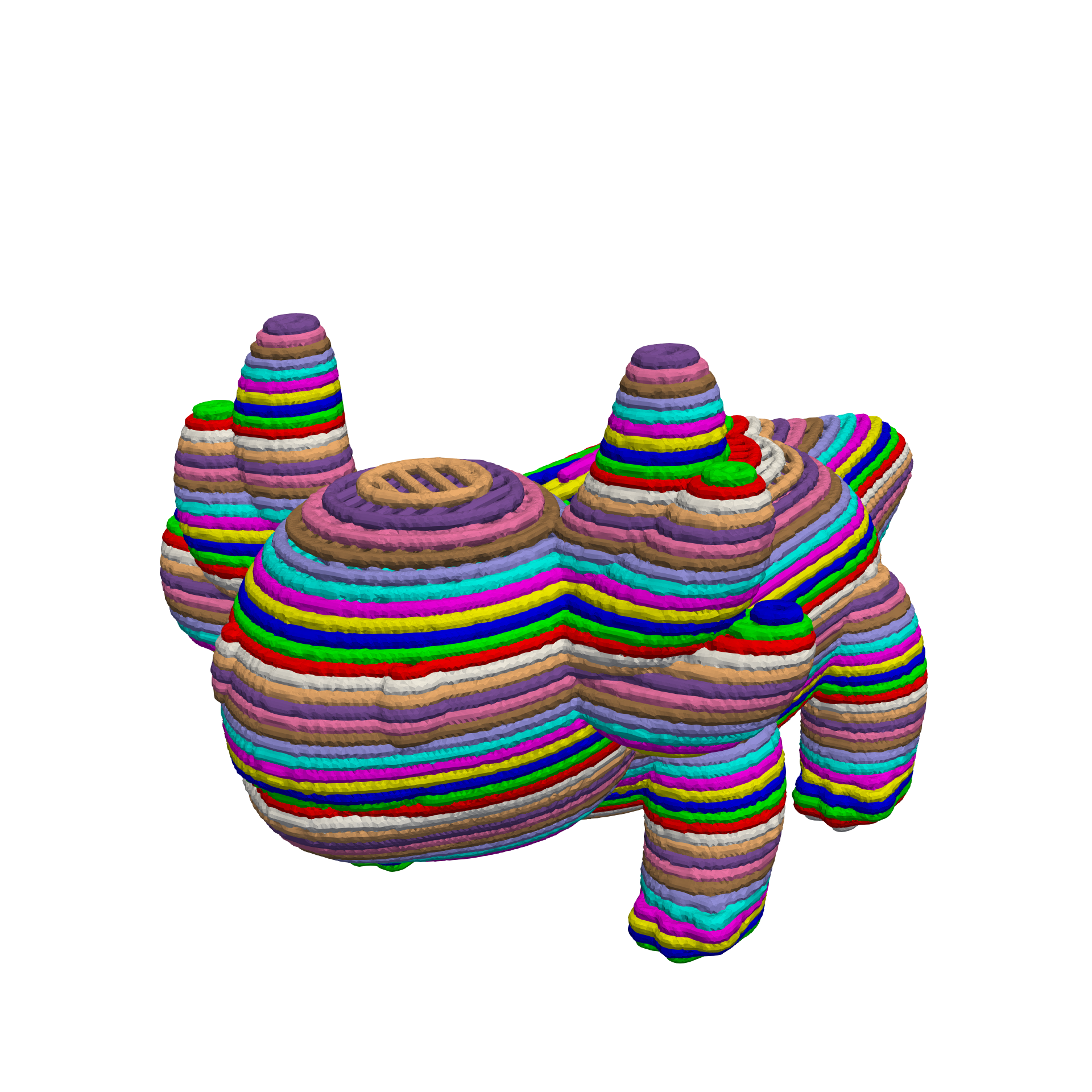}
        \caption{}
        \label{fig:eigenComplexa}
    \end{subfigure}
    \begin{subfigure}{0.35\textwidth}
        \includegraphics[width=\linewidth]{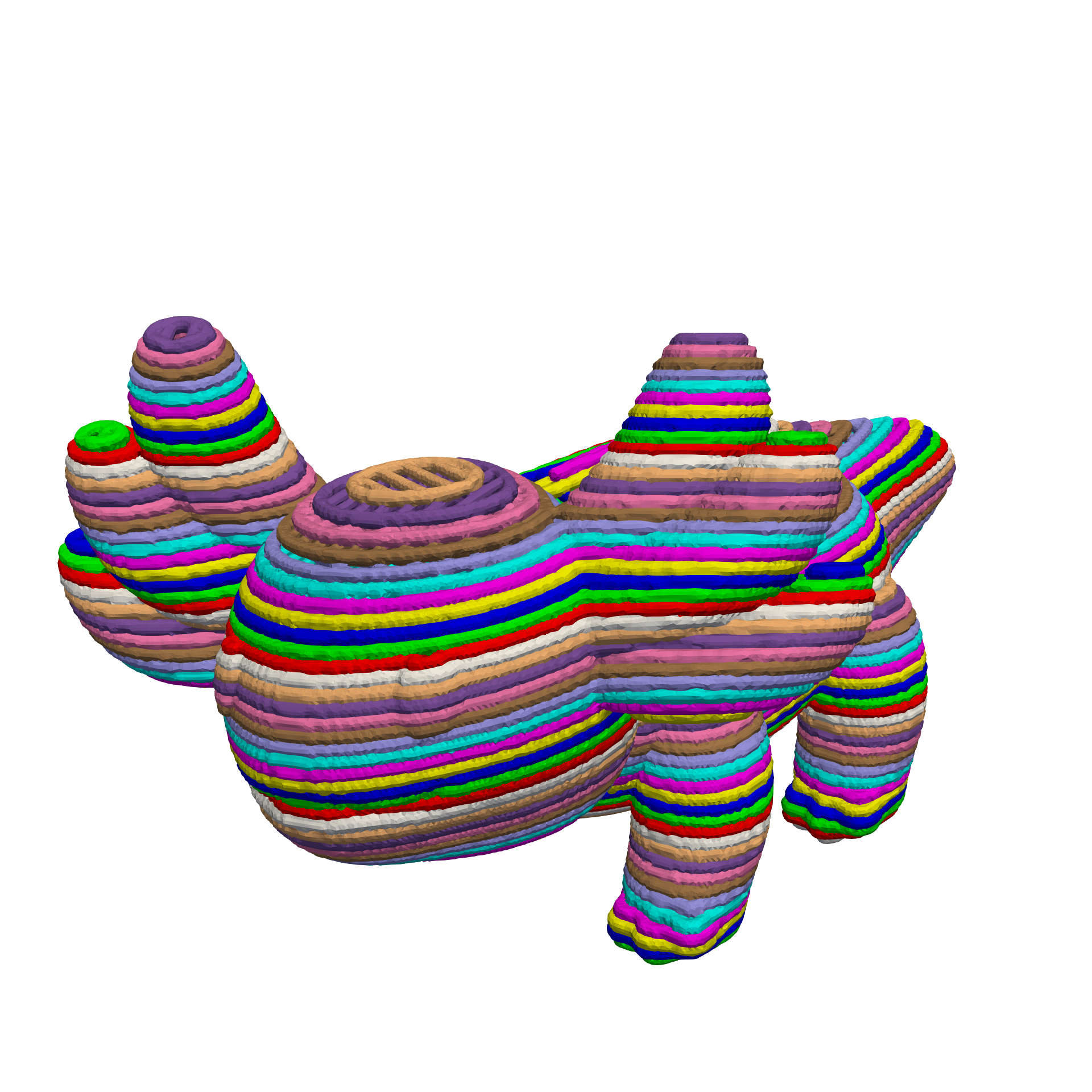}
        \caption{}
        \label{fig:eigenComplexb}
    \end{subfigure}
    \begin{subfigure}{0.35\textwidth}
        \includegraphics[width=\linewidth]{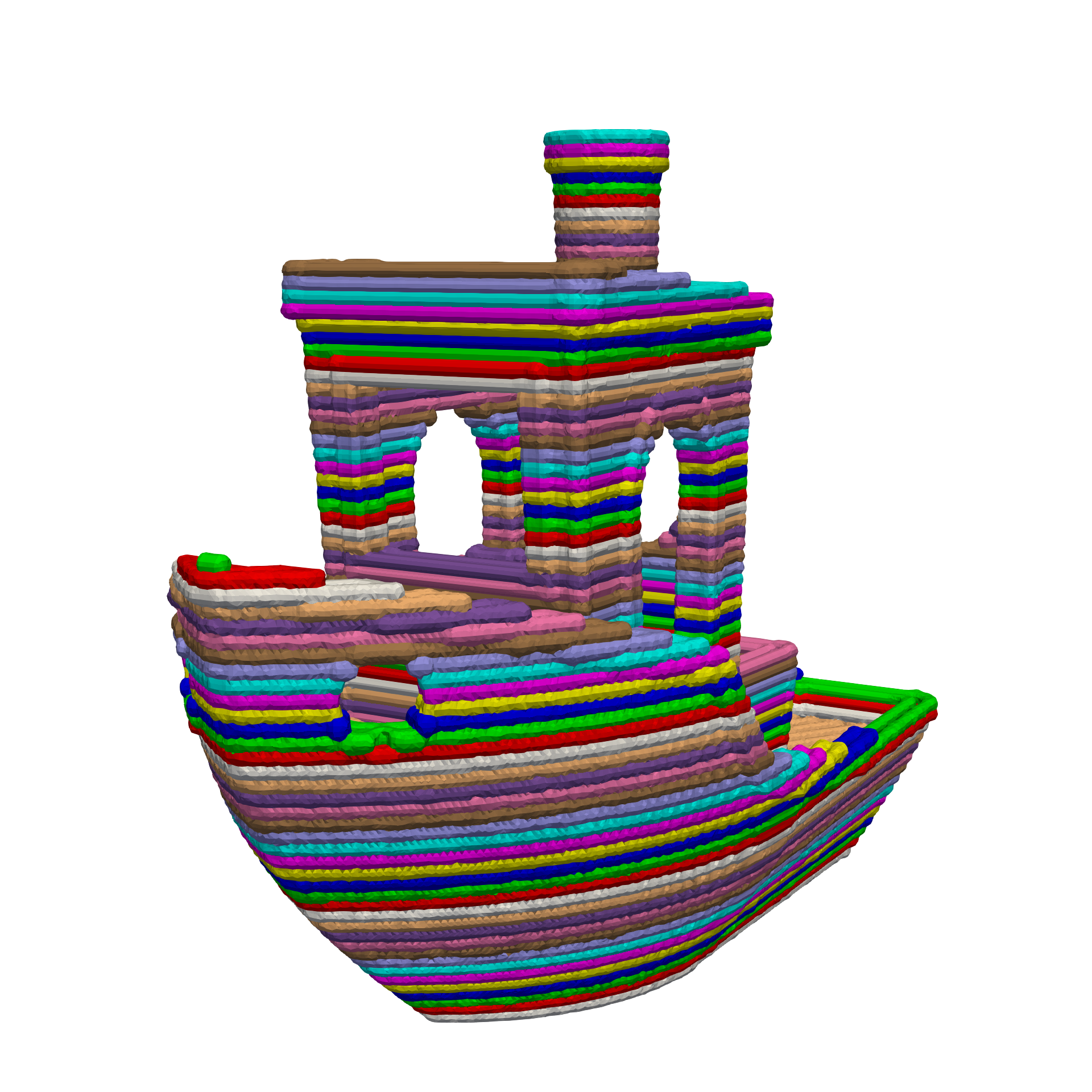}
        \caption{}
        \label{fig:eigenComplexc}
    \end{subfigure}
    \begin{subfigure}{0.35\textwidth}
        \includegraphics[width=\linewidth]{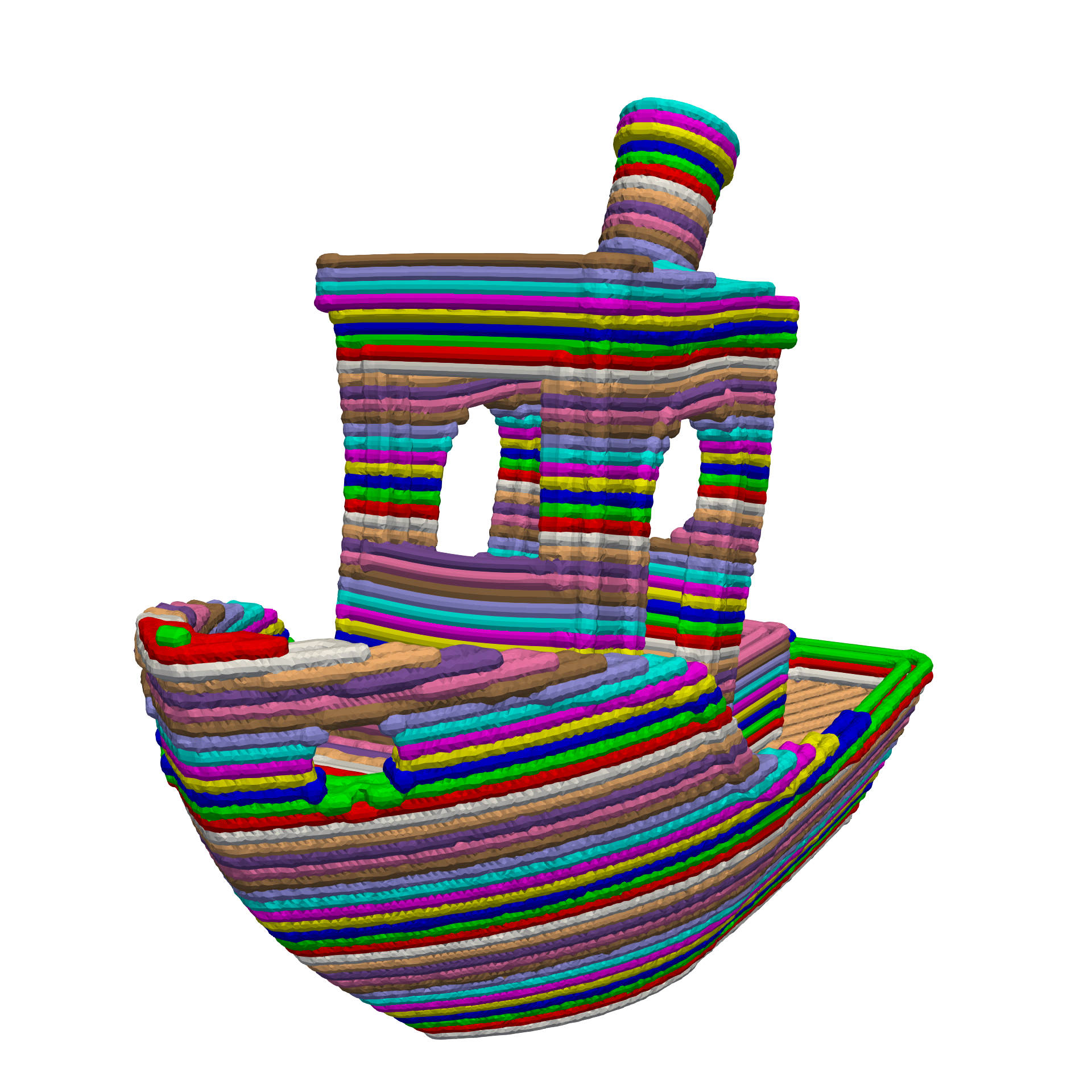}
        \caption{}
        \label{fig:eigenComplexd}
    \end{subfigure}
    \caption{Complex models (left) and the respective deformed eigenmodes (right), \ref{fig:eigenComplexa} axolotl model, \ref{fig:eigenComplexb} first mode 10.8 Hz, \ref{fig:eigenComplexc} Benchy model, \ref{fig:eigenComplexd} first mode 7.3 Hz.}
    \label{fig:eigenComplex}
\end{figure}

This framework readily processes both simple and complex geometries into high-fidelity extruded models, as shown in Figure~\ref{fig:eigenComplex}, for any applied study. In contrast to the previous example, the eigenfrequency of these larger models decreased in comparison to the smaller cube; the first eigenmode and frequency of an axolotl and a classic Benchy were also found, 10.8 Hz and 7.3 Hz respectively. This is more aligned with classical understanding of the natural frequencies of homogeneous materials, suggesting that at a certain size, the impact of varied print parameters on a printed part diminishes and the component behaves more homogeneously.

\subsection{Parameter Optimization for the Minimization of the First Eigen Frequency}

The modular nature of our proposed framework lends to its relatively straightforward insertion into other research interests. For example it may be of interest to minimize the eigenfrequency of an AM component, such as those introduced in Section \ref{subsectionEigenSolutions}, by tuning only the toolpath generation parameters. With the design parameters acting as the inputs, the output is designated to be the eigenfrequency. 

In this demonstration, the infill spacing, angle, and infill pattern were treated as constrained design variables, Table \ref{DesignParameters}. The implemented minimization algorithm provided by the Dakota toolkit, developed by Sandia National Laboratories to aid in the design and optimization of complex multi-physics systems, determines the minimum natural frequency depending on the toolpath generation parameters using the Fletcher-Reeves conjugate gradient (FRGC) algorithm, utilizing the finite difference method to compute gradients, \cite{dakota_2020}. The simple setup and low memory requirements of FRCG was beneficial to the simulation due to the large size of the microstructure model, as well as the quadratically converging nature of the method providing a near-exact solution, \cite{Fletcher1964}. While the spacing and angle were continuous, the infill pattern was treated as pseudo-continuous. Each pattern was assigned to a range of values to allow for the FRCG algorithm usage, e.g. lines pattern from 0$\leq$x$\leq$1, grid pattern from 1$<$x$\leq$1, etc. The pattern step size was set to a large enough size to ensure that a change in patter could occur.

\begin{table}[H]
\centering
\begin{tabular}{ | c | c | c | }
    \hline
    Design Parameter & Value Ranges & Initial Guess\\
    \hline
    \hline
    spacing & 0.5-2.0 & 1.0 \\
    \hline
    angle & 45.0-90.0 deg & 60.0 deg\\
    \hline
    pattern & lines, grid, gyroid, triangle, cubic & grid\\
    \hline
\end{tabular}
\caption{Design Parameters for Minimization of 1st Natural Frequency.}
\label{DesignParameters}
\end{table}

\begin{figure}[H]
    \centering
    \begin{subfigure}{0.45\textwidth}
        \includegraphics[trim={700pt 100pt 700pt 100pt}, clip=true, width=\linewidth]{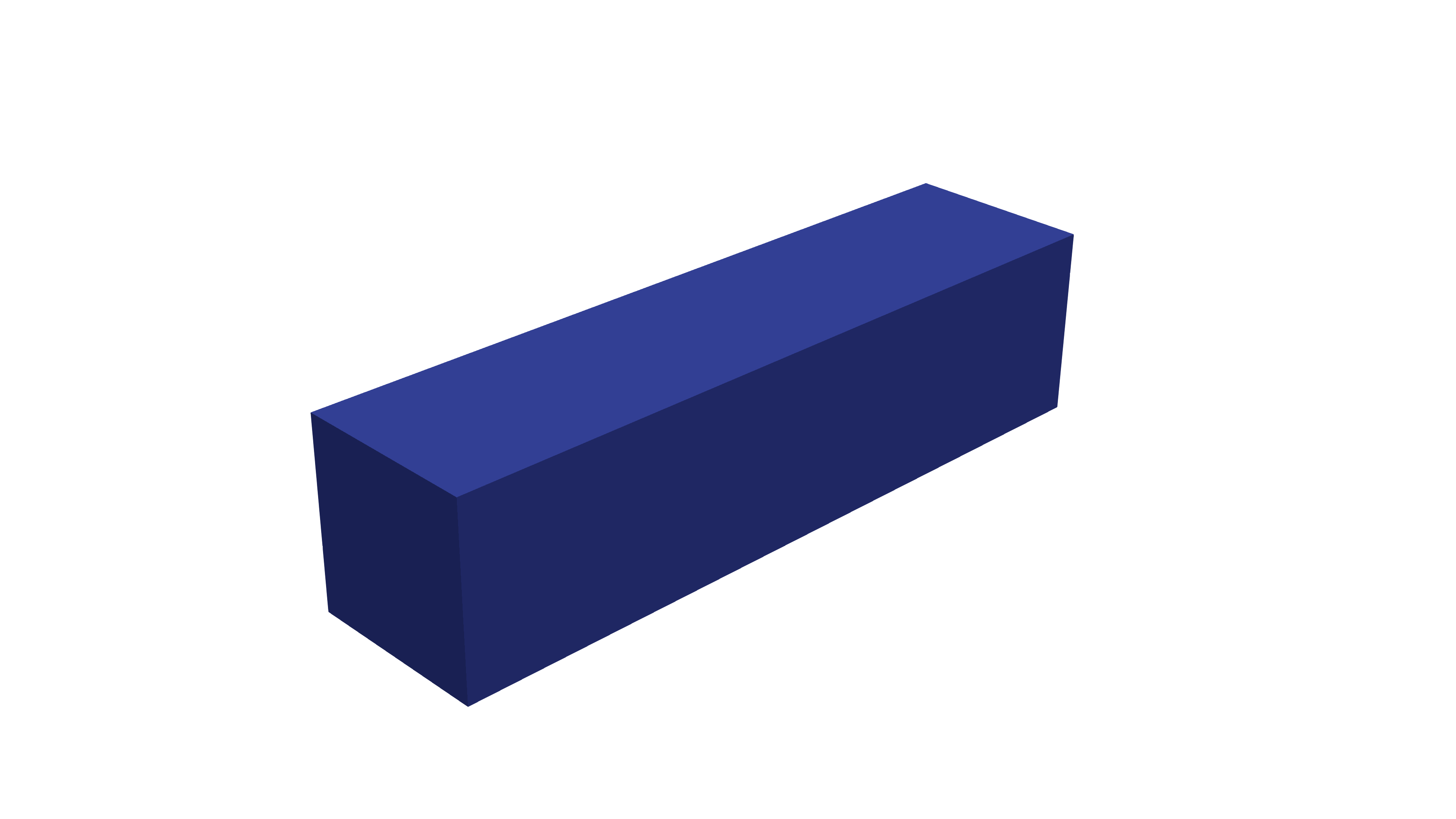} 
        \caption{}
        \label{fig:EigenExamplea}
    \end{subfigure}
    \begin{subfigure}{0.45\textwidth}
        \includegraphics[trim={700pt 100pt 700pt 100pt}, clip=true, width=\linewidth]{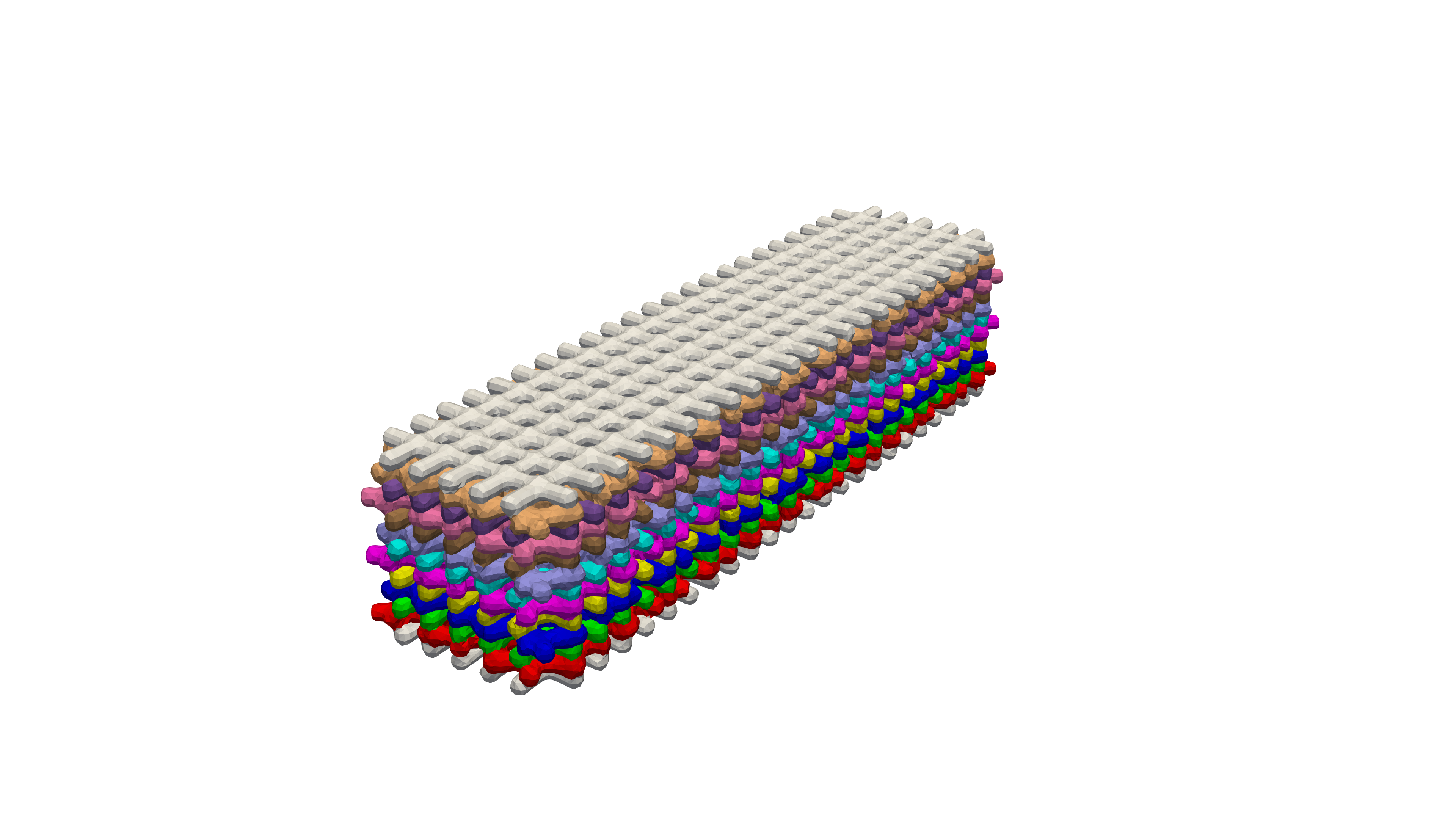}
        \caption{}
        \label{fig:EigenExampleb}
    \end{subfigure}
    \begin{subfigure}{0.45\textwidth}
        \includegraphics[trim={700pt 100pt 700pt 100pt}, clip=true, width=\linewidth]{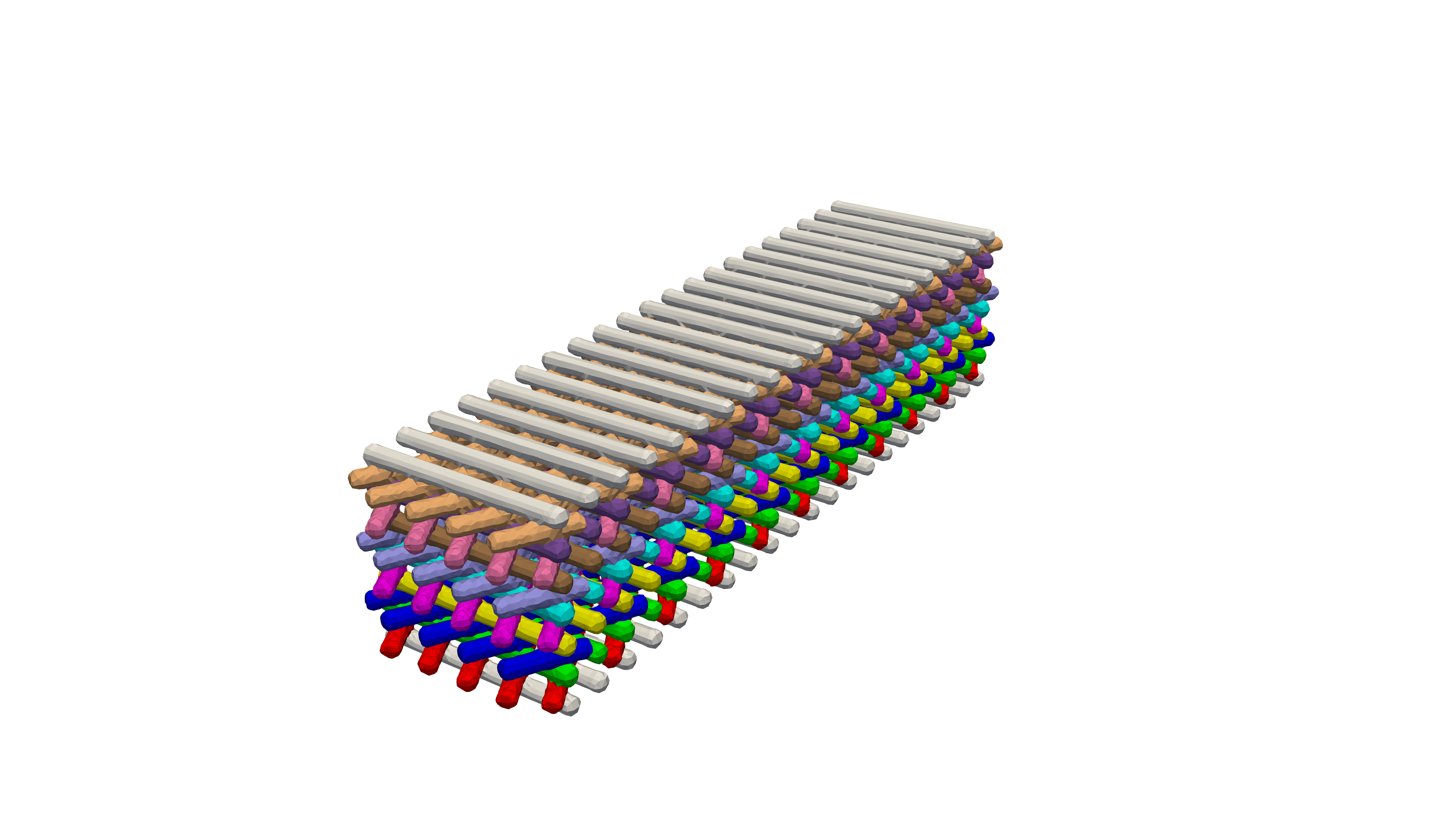}
        \caption{}
        \label{fig:EigenExamplec}
    \end{subfigure}
    \caption{Example of Eigenmode optimization, \ref{fig:EigenExamplea} .stl of example bar, \ref{fig:EigenExampleb} sliced model of bar from initial conditions, \ref{fig:EigenExamplec} optimized sliced model of bar.}
    \label{fig:EigenExample}
\end{figure}

A rectangular bar, with dimensions of 5 mm wide, 5 mm thick, and 20 mm long, was used as an example of the Eigenmode optimization, Figure~\ref{fig:EigenExample}. No degrees of freedom of the model were constrained in the study; therefore the first six resultant modes and corresponding frequencies calculated during the simulation describe the rigid body modes. The objective of the optimization was therefore to minimize the first non-rigid body mode.

\begin{figure}[H]
    \centering
    \begin{subfigure}{0.45\textwidth}
        \includegraphics[trim={700pt 100pt 700pt 100pt}, clip=true, width=\linewidth]{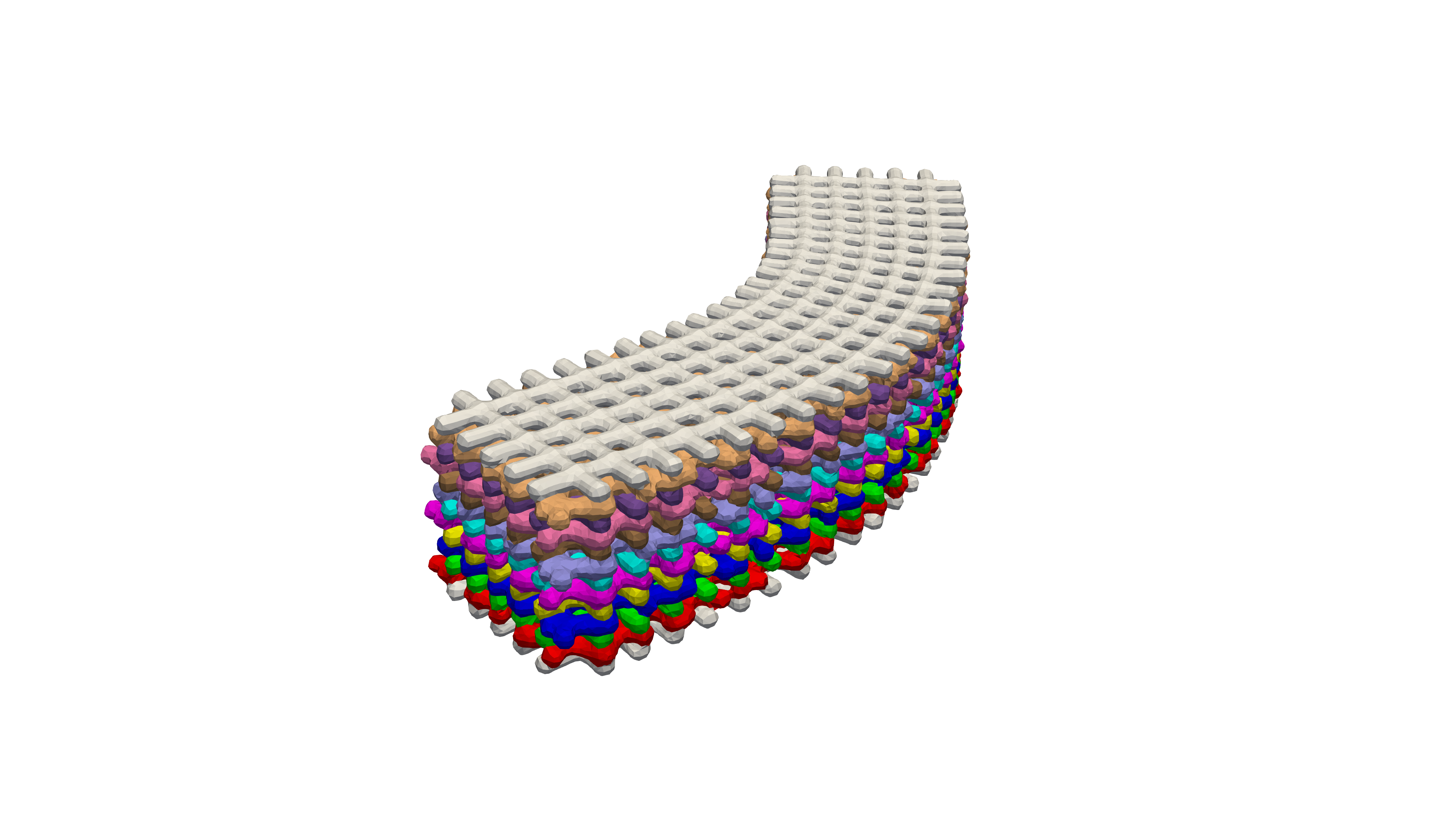} 
        \caption{}
        \label{fig:EigenOpta}
    \end{subfigure}
    \begin{subfigure}{0.45\textwidth}
        \includegraphics[trim={700pt 100pt 700pt 100pt}, clip=true, width=\linewidth]{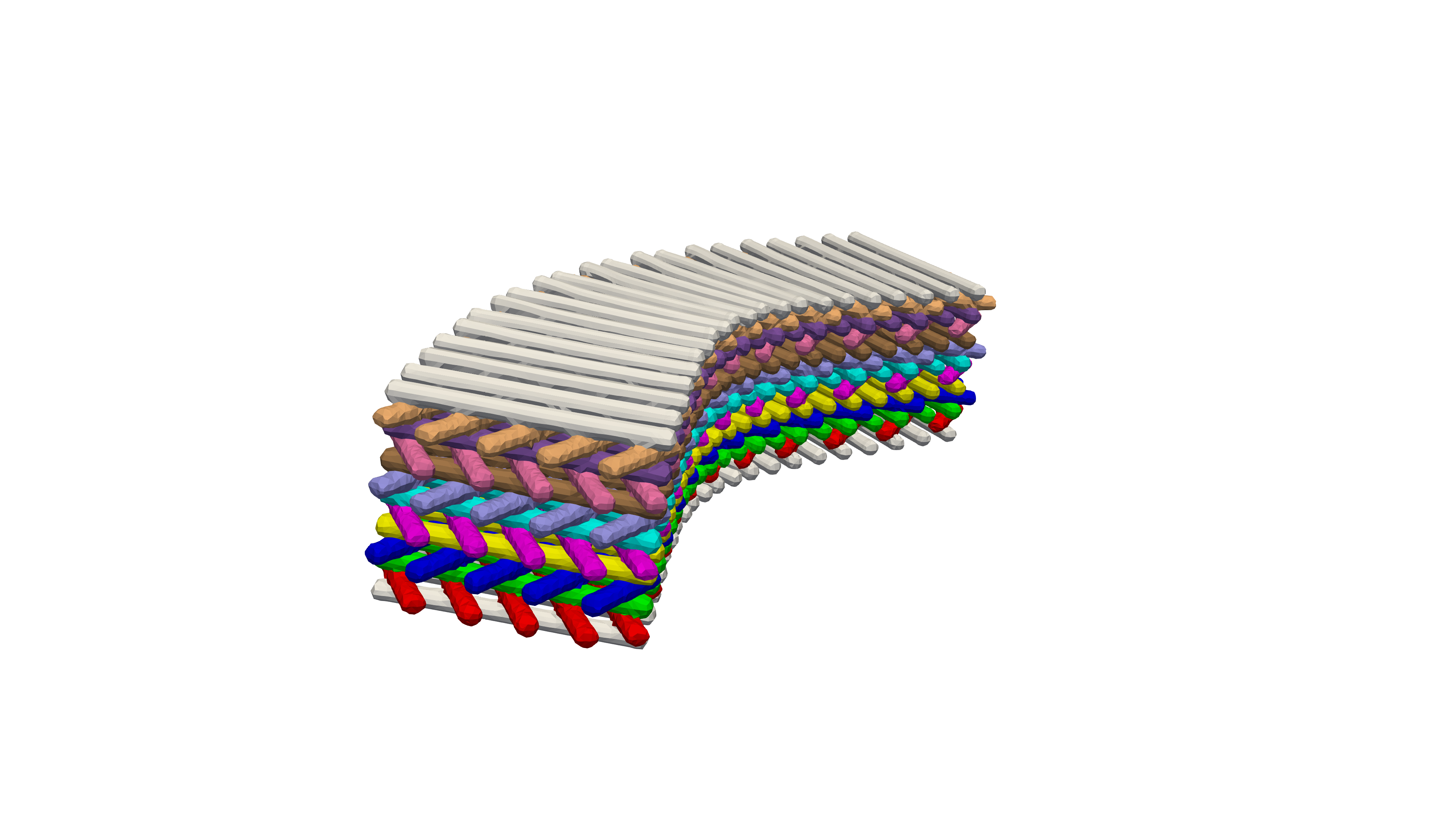}
        \caption{}
        \label{fig:EigenOptb}
    \end{subfigure}
    \caption{Eigenmode shape of optimized bar example, \ref{fig:EigenOpta} initial configuration, \ref{fig:EigenOptb} optimized configuration.}
    \label{fig:EigenOpt}
\end{figure}

The results of the optimization are shown in Table~\ref{optResults}. The 1st non-rigid mode was reduced by nearly a factor of 2 after optimization, demonstrating a significant drop in the natural frequency. This optimal mode was determined after 19 different simulation evaluations and 18 finite difference gradient evaluations, Figure~\ref{fig:EigenOpt}. The optimized beam utilized higher spacing between the printed lines, which is generally in agreement with the design space found in Section~\ref{subsectionEigenSolutions}. As the infill spacing increased, a general decrease in frequency was observed; less material in a printed object results in lower eigenfreqencies. 

\begin{table}[H]
\centering
\begin{tabular}{ | c | c | c | c | c |}
    \hline
    Iteration & Spacing & Angle (deg) & Pattern & 1st Mode (Hz)\\
    \hline
    \hline
    1 & 1.0 & 60.0 & grid & 19.63\\
    \hline
    Final & 1.21 & 60.0 & lines & 10.83\\
    \hline
\end{tabular}
\caption{Results of Optimization of 1st Natural Frequency.}
\label{optResults}
w\end{table}

Interestingly, the infill angle of the optimized configuration was the same as the infill angle of the initial guess. When comparing to Figure~\ref{fig:EigenCountour}, minima are possible with different infill spacings and print angles, which in conjunction with the provided constraints resulted in the initial guess of the angle being the optimized value for the angle.

\section{Conclusion}\label{sectionConclusion}

In this paper, a meshing framework for components fabricated with extrusion based additive manufacturing methods was presented. Our framework directly maps AM toolpaths to hexahedral meshes to allow for robust solid mechanics simulations, with the challenging motivation of considering the widespread self-contact inherent in extruded AM components. Other significant benefits are the modular nature of each stage allowing for the interchanging of software packages as well as the automatic processing of data hand-off between said stages of the framework. Exemplar simulations of fully meshed 3D structures were presented for uniaxial compression and eigen mode problems to show the wide range of physical phenomenon that can be replicated in-silico prior to the manufacturing and testing of designs. The provided examples are by no means exhaustive but are meant to be representative of quantities of interest AM part designers may be involved or interested in, such as the influence of toolpath parameters on the physical properties and responses of printed components. This modular framework considerably reduces the time and resources necessary to produce desirable, additively manufacturable components.

\section*{Acknowledgements}
Sandia National Laboratories is a multimission laboratory managed and operated by National Technology \& Engineering Solutions of Sandia, LLC, a wholly owned subsidiary of Honeywell International Inc., for the U.S. Department of Energy’s National Nuclear Security Administration under contract DE-NA0003525. This paper describes objective technical results and analysis. Any subjective views or opinions that might be expressed in the paper do not necessarily represent the views of the U.S. Department of Energy or the United States Government.


\end{document}